\newif\ifpaperdraft

\paperdrafttrue

\documentclass[nonacm,acmtog]{acmart}

\author{Chenghao Wu}
\email{chenghao.wu@manchester.ac.uk} 
\affiliation{%
  \institution{University of Manchester}
  \country{United Kingdom}
}

\author{Yuefan Shen}
\email{jhonve@zju.edu.cn} 
\affiliation{%
  \institution{LIGHTSPEED}
  \country{China}
}

\author{Tao Huang}
\email{tao_huang@ucsb.edu} 
\affiliation{%
  \institution{LIGHTSPEED}
  \country{China}
}

\author{Kai Yan}
\email{kyan8@uci.edu} 
\affiliation{%
  \institution{LIGHTSPEED}
  \country{China}
}

\author{Zahra Montazeri}
\email{zahra.montazeri@manchester.ac.uk} 
\affiliation{%
  \institution{University of Manchester}
  \country{United Kingdom}
}

\author{Kui Wu}
\email{walker.kui.wu@gmail.com} 
\orcid{0000-0003-3326-7943}
\affiliation{%
  \institution{LIGHTSPEED}
  \country{USA}
}

\usepackage{gensymb}
\usepackage{multirow}
\usepackage{prettyref}
\usepackage[labelfont=bf,format=plain,singlelinecheck=false]{caption}
\usepackage{subcaption}
\usepackage{bm,microtype}
\usepackage{overpic}
\usepackage{enumitem}

\definecolor{myblue}{RGB}{31, 119, 180}
\definecolor{myorange}{RGB}{255, 127, 14}
\definecolor{mygreen}{RGB}{44, 160, 44}
\definecolor{myred}{RGB}{214, 39, 40}

\usepackage{soul}
\AtBeginDocument{%
  }

\setcopyright{acmlicensed}
\acmJournal{TOG}

\usepackage{tikz}
\usetikzlibrary{angles, calc}
\tikzset{
  dot/.style={
    circle, fill=black, inner sep=1pt, outer sep=0pt
  },
  dot label/.style={
    circle, inner sep=0pt, outer sep=1pt
  },
  pics/right angle/.append style={
    /tikz/draw, /tikz/angle radius=5pt
  }
}
\usepackage{xcolor}

\acmSubmissionID{1083}

\usepackage{prettyref}
\newrefformat{fig}{Figure~\ref{#1}}
\newrefformat{par}{Section~\ref{#1}}
\newrefformat{appen}{Appendix~\ref{#1}}
\newrefformat{sec}{Section~\ref{#1}}
\newrefformat{sub}{Section~\ref{#1}}
\newrefformat{table}{Table~\ref{#1}}
\newrefformat{ass}{Assumption~\ref{#1}}
\newrefformat{alg}{Algorithm~\ref{#1}}
\newrefformat{def}{Definition~\ref{#1}}
\newrefformat{thm}{Theorem~\ref{#1}}
\newrefformat{cor}{Corollary~\ref{#1}}
\newrefformat{lem}{Lemma~\ref{#1}}
\newrefformat{step}{Step~\ref{#1}}
\newrefformat{ln}{Line~\ref{#1}}
\newrefformat{rem}{Remark~\ref{#1}}
\newrefformat{eq}{Equation~\ref{#1}}
\newrefformat{pb}{Problem~\ref{#1}}
\newrefformat{it}{Item~\ref{#1}}
\newrefformat{te}{Term~\ref{#1}}
\def\Eqref Eq:#1:{\eqref{eq:#1}}
\newrefformat{Eq}{Equation~\Eqref#1:}

\usepackage{rotating}

\newcommand{\qq}{\mathbf{q}}
\newcommand{\VV}{\mathbf{V}}
\newcommand{\cc}{\mathbf{c}}

\newcommand{\HH}{\mathbf{H}}

\newcommand{\ZZ}{\mathbf{Z}}

\newcommand{\CC}{\mathbf{C}}
\newcommand{\RR}{\mathbf{R}}

\newcommand{\DD}{\mathbf{D}}

\newcommand{\pp}{\mathbf{p}}
\newcommand{\LL}{\mathbf{L}}
\renewcommand{\SS}{\mathbf{S}}

\renewcommand{\ss}{\mathbf{s}}

\newcommand{\PP}{\mathbf{P}}
\newcommand{\XX}{\mathbf{X}}

\newcommand{\YY}{\mathbf{Y}}

\newcommand{\TT}{\mathbf{T}}
\newcommand{\II}{\mathbf{I}}

\newcommand{\UU}{\mathbf{U}}

\newcommand\restr[2]{{\left.\kern-\nulldelimiterspace{}#1\right|_{#2}}}

\usepackage{hyperref}
\usepackage{xr-hyper}

\usepackage[noend]{algpseudocode}
\usepackage{xcolor}
\usepackage[export]{adjustbox}
\usepackage{colortbl}
\usepackage[ruled,linesnumbered]{algorithm2e}

\def\BState{\State\hskip-\ALG@thistlm}
\makeatother

\newcommand{\algorithmicforeach}{\textbf{foreach}}

\algdef{SE}[FOREACH]{ForEach}{EndForEach}[1]{\algorithmicforeach\ #1\ \algorithmicdo}{\algorithmicend\ \algorithmicforeach}%
\algtext*{EndForEach}

\usepackage{svg}
\usepackage{xcolor}
\usepackage{wrapfig}
\usepackage[export]{adjustbox}
\usepackage{colortbl}
\usepackage{graphicx}
\usepackage{tikz}
\usetikzlibrary{angles, calc}
\tikzset{
  dot/.style={
    circle, fill=black, inner sep=1pt, outer sep=0pt
  },
  dot label/.style={
    circle, inner sep=0pt, outer sep=1pt
  },
  pics/right angle/.append style={
    /tikz/draw, /tikz/angle radius=5pt
  }
}

\makeatletter

\pgfkeys{
  /cornerimg/.is family,
  /cornerimg,
  inset scale/.initial=0.4,
  inset border/.initial=false,
  inset border color/.initial=white,
  inset border width/.initial=0.6pt,
  draw crop rect/.initial=false,
  rect color/.initial=red,
  rect width/.initial=0.6pt,
  image width/.initial=1,
  image height/.initial=1,
}

\NewDocumentCommand{\cornerimg}{O{0.09\linewidth} m m m m O{}}{%
  \pgfkeys{
    /cornerimg/.cd,
    inset scale=0.4,
    inset border=false,
    inset border color=white,
    inset border width=0.6pt,
    draw crop rect=false,
    rect color=red,
    rect width=0.6pt,
    image width=1,
    image height=1,
    #6
  }%

  \def\parseBaseTrim##1 ##2 ##3 ##4\relax{%
    \def\baseL{##1}%
    \def\baseB{##2}%
    \def\baseR{##3}%
    \def\baseT{##4}%
  }%
  \expandafter\parseBaseTrim#4\relax

  \def\parseInsetTrim##1 ##2 ##3 ##4\relax{%
    \def\cropL{##1}%
    \def\cropB{##2}%
    \def\cropR{##3}%
    \def\cropT{##4}%
  }%
  \expandafter\parseInsetTrim#5\relax

  \pgfmathsetmacro{\imgW}{\pgfkeysvalueof{/cornerimg/image width}}%
  \pgfmathsetmacro{\imgH}{\pgfkeysvalueof{/cornerimg/image height}}%

  \pgfmathsetmacro{\visL}{\baseL}%
  \pgfmathsetmacro{\visB}{\baseB}%
  \pgfmathsetmacro{\visR}{\imgW - \baseR}%
  \pgfmathsetmacro{\visT}{\imgH - \baseT}%

  \pgfmathsetmacro{\visW}{\visR - \visL}%
  \pgfmathsetmacro{\visH}{\visT - \visB}%

  \pgfmathsetmacro{\cropXA}{\cropL}%
  \pgfmathsetmacro{\cropYA}{\cropB}%
  \pgfmathsetmacro{\cropXB}{\imgW - \cropR}%
  \pgfmathsetmacro{\cropYB}{\imgH - \cropT}%

  \pgfmathsetmacro{\rectXA}{(\cropXA - \visL) / \visW}%
  \pgfmathsetmacro{\rectXB}{(\cropXB - \visL) / \visW}%
  \pgfmathsetmacro{\rectYA}{(\cropYA - \visB) / \visH}%
  \pgfmathsetmacro{\rectYB}{(\cropYB - \visB) / \visH}%

  \adjustbox{valign=m}{%
    \begin{tikzpicture}
      \node[anchor=south west, inner sep=0] (img) at (0,0)
        {\includegraphics[width=#1, trim=#4, clip]{#2}};

      \edef\drawinsetborder{\pgfkeysvalueof{/cornerimg/inset border}}%
      \def\trueval{true}%

      \ifx\drawinsetborder\trueval
        \node[
          anchor=south west,
          inner sep=0pt,
          draw=\pgfkeysvalueof{/cornerimg/inset border color},
          line width=\pgfkeysvalueof{/cornerimg/inset border width}
        ] at (img.south west)
        {%
          \scalebox{\pgfkeysvalueof{/cornerimg/inset scale}}{%
            \includegraphics[width=#1, trim=#5, clip]{#3}%
          }%
        };
      \else
        \node[
          anchor=south west,
          inner sep=0pt
        ] at (img.south west)
        {%
          \scalebox{\pgfkeysvalueof{/cornerimg/inset scale}}{%
            \includegraphics[width=#1, trim=#5, clip]{#3}%
          }%
        };
      \fi

      \edef\drawrect{\pgfkeysvalueof{/cornerimg/draw crop rect}}%
      \ifx\drawrect\trueval
        \begin{scope}[
          shift={(img.south west)},
          x={($(img.south east)-(img.south west)$)},
          y={($(img.north west)-(img.south west)$)}
        ]
          \draw[
            color=\pgfkeysvalueof{/cornerimg/rect color},
            line width=\pgfkeysvalueof{/cornerimg/rect width}
          ] (\rectXA,\rectYA) rectangle (\rectXB,\rectYB);
        \end{scope}
      \fi

    \end{tikzpicture}%
  }%
}

\makeatother

\definecolor{smGrey}{rgb}{0.8, 0.8, 0.8}  
\definecolor{smOrange}{rgb}{1.0, 0.64, 0.0}  
\definecolor{smBlue}{rgb}{0.0, 0.64, 1.0}

\begin{document}

\title{Real-Time Neural Hair G-Buffer Anti-Aliasing}

\begin{abstract}
We propose a lightweight real-time method for reconstructing strand-based hair G-Buffers from severely undersampled rasterized inputs. Our pipeline first applies neural spatial reconstruction and temporal accumulation to recover hair coverage, i.e., fractional hair visibility within a pixel, and tangent. It then uses a tangent-guided reconstruction step to complete the position, which is subsequently used for physically based deferred hair shading. We evaluate our method across a diverse set of hairstyles, including straight, wavy, afro, and ponytail styles, under both static and dynamic scenarios. Our method achieves higher hair reconstruction quality than general industrial neural reconstruction solutions such as DLSS and FSR. 


\end{abstract}


%
%



\begin{teaserfigure}
\centering
\includegraphics[width=\textwidth]{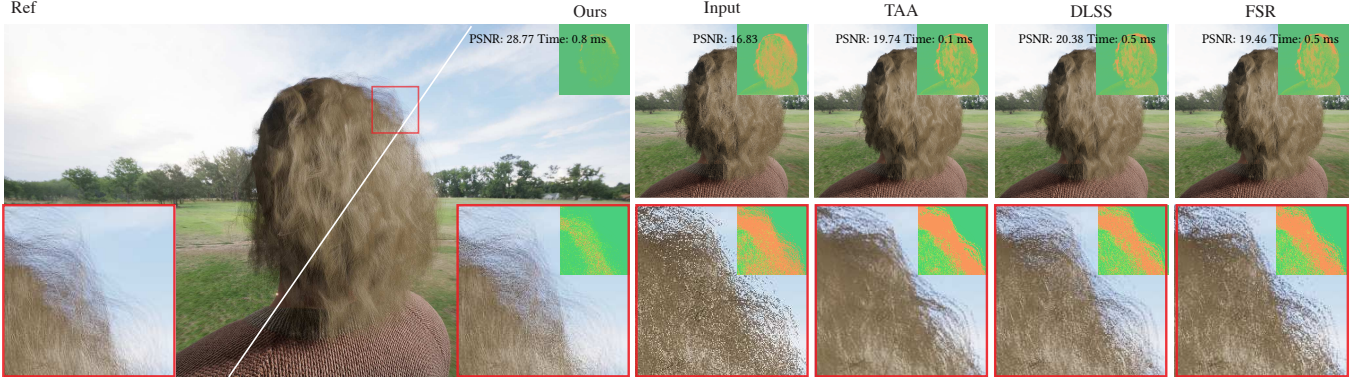}
\caption{We showcase the anti-aliasing results of our method given severely undersampled strand-based hair G-buffers. 
Compared with TAA, DLSS, and FSR, our approach better preserves sparse hair silhouettes and fine strand details, producing results closer to the high-sample reference. 
Insets show the corresponding error maps, and red boxes indicate zoomed regions.}
\Description{}
\label{fig:teaser}
\end{teaserfigure}


\maketitle
\section{Introduction}

Hair is one of the most important visual cues for digital characters, and high-quality hair rendering has become essential in modern real-time applications, including games, virtual production, AR/VR, and social avatars. Despite its importance, hair remains challenging to render efficiently. A typical hairstyle contains tens of thousands of strands, each of which is often much thinner than a pixel at common viewing distances. As a result, hair exhibits extreme subpixel complexity and requires a large number of samples per pixel to adequately suppress noise and aliasing.
For general scenes, rasterization is usually efficient, and shading tends to be the dominant bottleneck. Hair rendering, however, is fundamentally different: rasterization itself is already expensive due to the large number of thin geometric primitives. Moreover, the extremely small screen-space footprint of individual hairs makes them highly susceptible to severe undersampling, resulting in noisy renderings. The thin and highly detailed structure of hair also makes it difficult for general anti-aliasing methods, including recent learning-based approaches such as DLSS4~\cite{nvidia2025dlss4} and FSR~\cite{amd_fsr}, to handle effectively. Although recent approaches have explored hair filtering through multisample anti-aliasing on the visibility buffer~\cite{tafuri2019strand}, hair linking~\cite{Huang2025filter}, anisotropic filtering~\cite{fu2023}, and neural filtering~\cite{Currius2022}, their runtime cost still falls short of the constraints of modern game production, where the filtering stage is typically limited to only a few milliseconds.

Unlike prior work that denoises the final shaded image, where errors are already entangled with lighting, visibility, and the hair shading model, our method reconstructs low-sample rasterized hair G-buffers prior to shading. Operating in G-buffer space preserves the geometric and directional information required for physically based hair shading, enabling better temporal coherence and stronger geometric fidelity.
More specifically, undersampled rasterized hair G-buffers exhibit two major failure modes. First, attributes such as coverage and tangent are sparse, aliased, and noisy. Second, geometric buffers, such as position and depth, are often incomplete, making deferred shading unstable and prone to flickering artifacts. 
In this paper, we address these issues by proposing a novel \textit{lightweight real-time pipeline} for \textit{high-quality hair G-buffer reconstruction for deferred shading}. 
Our contributions include:
\begin{itemize}[leftmargin=*, itemsep=0.3em]
    \item A lightweight \textit{spatial-temporal neural reconstruction} framework for low-sample coverage and tangent buffers, improving both reconstruction quality and temporal stability (\S\ref{sec:NeuralPrediction}).
    \item A \textit{tangent-guided position completion method} that combines reconstructed coverage and tangent with noisy position and depth inputs to produce a shading-ready hair G-buffer (\S\ref{sec:position_reconstruction}).
\end{itemize}

We evaluate our method on a diverse set of hairstyles, including straight, wavy, afro, and ponytail styles, under both static and dynamic motion (\S\ref{sec:results}). 
Our method outperforms existing hair-specific denoising, conventional anti-aliasing, and industrial neural reconstruction baselines in hair reconstruction quality while maintaining comparable real-time performance. Code and data for this paper are at https://theo-wu.github.io/NeuralHairGAA/.

\section{Related work}


\paragraph{Hair Rendering}

\citet{KajiyaK89} introduced the first physically based analytic hair shading model. This was later extended by \citet{MarschnerJCWH03} into the widely used hair BCSDF model, which decomposes scattering inside dielectric fibers into the $R$, $TT$, and $TRT$ modes. Since then, many works have extended this model, including the addition of a diffuse term for improved data fitting~\cite{ZinkeRLWAK09}, artist-friendly lobe decompositions~\cite{Sadeghi2010}, additional lobes for improved energy conservation and high-order scattering~\cite{dEonFHLA11, ChiangBTB16}, and extensions to elliptical hairs, animal fur, and microfacet-based scattering~\cite{Yan2015, Khungurn2017, Huang2022hair}. More recently, wave-optics models have been proposed to capture the colorful glints of hair and fur~\cite{Xia2020, Xia2023}.
Efficiently approximating multiple scattering is critical for realistic strand-based rendering. Prior work has explored photon mapping with volumetric radiance storage~\cite{MoonM06, moon08}, neural prediction of high-order scattering~\cite{kt2023}, and dual scattering for real-time rendering~\cite{ZinkeYWK08, Yuksel2008}. \citet{Huang2025Knit} further enable efficient multiple-scattering approximation with textured strand details via precomputed dual-scattering corrections.
To further accelerate rendering, aggregation models represent clusters of fibers or hairs using coarser primitives. \citet{Koh2001} pioneered the use of polygon strips with alpha maps to represent clusters of hair strands for real-time applications; this approach remains widely used in the game industry~\cite{YIbing2016}. To better approximate the aggregated scattering behavior within fiber bundles, \citet{MontazeriGZJ20, Montazeri2021} modeled woven and knitted structures using normal mapping and a specialized BSDF with separate surface and body components. \citet{ZhuZWXY22, ZhuZJYA23} later introduced neural and analytical aggregated BCSDF models for fiber bundles and yarns, respectively. \citet{Bhokare2024} generated hair strand geometry on-the-fly based on \emph{hair mesh}~\cite{Yuksel2009, wu2016}, greatly reducing storage and memory bandwidth requirements for real-time hair rendering. More recently, \citet{Huang2025HairLoD} presented a real-time strand-based rendering framework with seamless level-of-detail transitions, using an aggregated BCSDF to preserve both single- and multiple-scattering effects within hair clusters. 
While many approaches have been proposed to improve the performance of strand-based rendering, including hair level-of-detail (LoD), aggregated models, and dual scattering, the intrinsic geometric nature of hair remains challenging: its width is often smaller than a screen pixel, which requires a large number of samples per pixel to achieve denoised results.

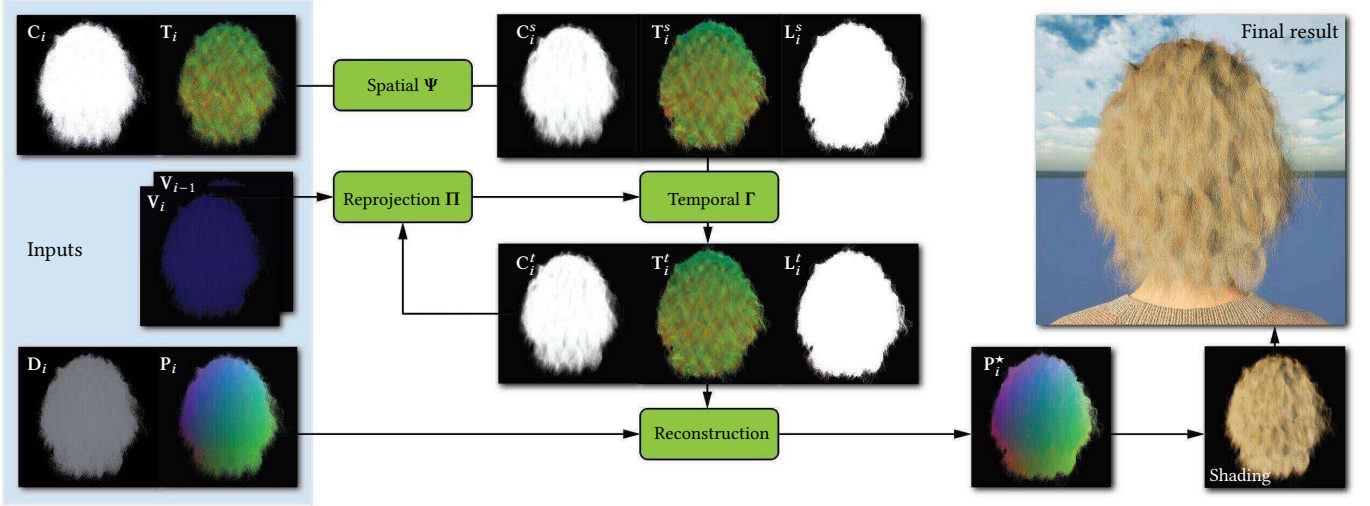
\begin{figure*}[ht!]
    \centering
    \includegraphics[width=\textwidth]{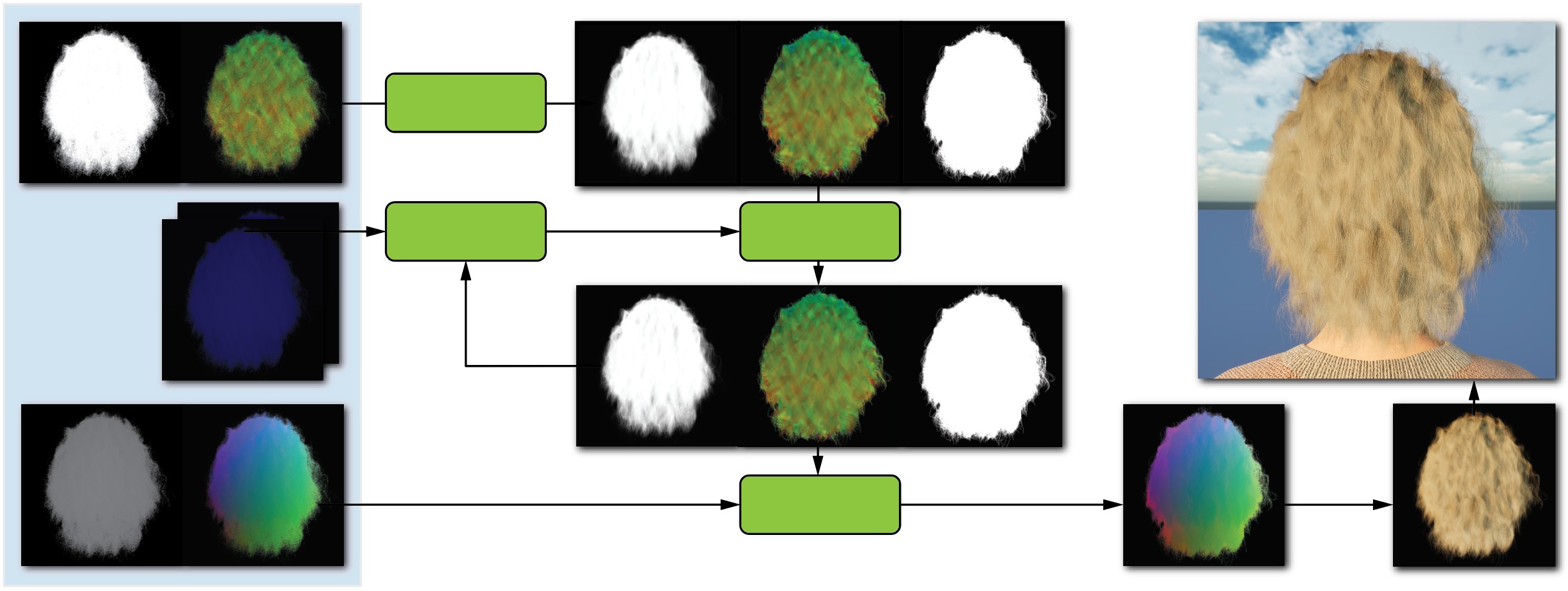}
    \put(-500,177){\footnotesize \color{white}{$\CC_i$} }
    \put(-450,177){\footnotesize \color{white}{$\TT_i$} }
    \put(-316,177){\footnotesize \color{white}{$\CC^s_i$} }
    \put(-265,177){\footnotesize \color{white}{$\TT^s_i$} }
    \put(-215,177){\footnotesize \color{white}{$\LL^s_i$} }
    \put(-316,90){\footnotesize \color{white}{$\CC^t_i$} }
    \put(-265,90){\footnotesize \color{white}{$\TT^t_i$} }
    \put(-215,90){\footnotesize \color{white}{$\LL^t_i$} }
    \put(-500,52){\footnotesize \color{white}{$\DD_i$} }
    \put(-450,52){\footnotesize \color{white}{$\PP_i$} }
    \put(-455,113){\footnotesize \color{white}{$\VV_i$} }
    \put(-450,120){\footnotesize \color{white}{$\VV_{i-1}$} }
    \put(-500,95){\small \color{black}{Inputs} }
    \put(-372,157){\footnotesize \color{black}{Spatial $ \boldsymbol\Psi$} }
    \put(-259,114){\footnotesize \color{black}{Temporal $ \boldsymbol \Gamma$} }
    \put(-381,114){\footnotesize \color{black}{Reprojection $ \boldsymbol \Pi$} }
    \put(-264,26){\footnotesize \color{black}{Reconstruction} }
    \put(-140,52){\footnotesize \color{white}{$\PP_i^\star$} }
    \put(-55,10){\footnotesize \color{white}{Shading} }
    \put(-43,177){\small \color{black}{Final result} }
    \caption{The pipeline of our neural hair filter. Starting from low-sample, aliased hair renderings, we extract noisy hair G-buffers, including coverage $\CC_i$, tangent $\TT_i$, position $\PP_i$, and depth $\DD_i$. A spatiotemporal neural filter is applied to tangent and coverage through a spatial module $\boldsymbol{\Psi}$ and a temporal module $\boldsymbol{\Gamma}$, while the noisy depth and position are directly fed into the reconstruction stage. Using consecutive noisy tangent and coverage inputs and motion vectors $\VV_i$ and $\VV_{i-1}$, the neural filter provides a recurrent feedback loop from the current frame to the next. With the filtered tangent and coverage, together with the original noisy depth and position, our method reconstructs the hair position $\PP_i^\star$ and completes the hair G-buffer. The reconstructed G-buffer is then used for deferred hair shading, producing a high-quality final image from sparse input samples.}
    \vspace{-10pt}
    \label{fig:pipeline}
    \Description{}
\end{figure*}

\paragraph{Hair Denoising}
Anti-aliasing and reconstruction are fundamental problems in real-time rendering. Supersample anti-aliasing (SSAA) mitigates aliasing by full-image supersampling at high cost, whereas multisample anti-aliasing (MSAA) improves geometric edge quality while reducing shading noise. Post-process filters, such as fast approximate anti-aliasing (FXAA)~\cite{Lottes2009FXAA}, provide efficient edge smoothing but may blur fine details. G-Buffer attribute filters, such as aggregate G-Buffer anti-aliasing (AGAA)~\cite{Crassin2015AGAA}, and normal distribution anti-aliasing~\cite{Kaplanyan2016normalantialiasing}. Temporal methods such as temporal anti-aliasing (TAA)~\cite{Yang2020taa} improve both anti-aliasing and denoising through cross-frame accumulation, but can introduce ghosting under motion or disocclusion. More recently, neural reconstruction methods such as deep learning super sampling (DLSS) 4~\cite{nvidia2025dlss4} and AMD FidelityFX Super Resolution (FSR)~\cite{amd_fsr} have combined temporal information with learned super-resolution to improve image quality. However, while these techniques work well for general geometry, they are less effective for high-frequency structures such as hair, where aliasing artifacts can persist even after edge detection and filtering.
In strand-based hair rendering, simply enabling MSAA is insufficient to eliminate aliasing and also increases rendering cost due to severe overdraw. To address this, the industrial solution~\cite{tafuri2019strand} adopts a pipeline in which all hair strands are first rasterized in a lightweight MSAA pass into the visibility buffer and then processed by sample deduplication, so that only distinct samples within each pixel are shaded to produce the anti-aliased result. This pipeline has been widely adopted in modern game engines~\cite{unrealengine, Kulikov2025strand}. \citet{fu2023} proposed an elliptical kernel designed to suppress noise while preserving fine strand detail.
In both industry and academia, general Monte Carlo denoising has been extensively studied using learning-based approaches, including robust average networks~\cite{kalojanov2023robust}, and optimization-based approaches, including ensemble denoising~\cite{Zheng2021ensembledenoising}. For hair-specific reconstruction, \citet{Currius2022} presented a CNN-based method that filters rendered color images produced from a small number of stochastic transparency samples, but it suffers from blurring, energy variation, and high inference cost.
In contrast, our method employs a lightweight network to directly filter the tangent and coverage, and fill position based on them in the G-buffer, yielding better visual quality with lower runtime overhead.

\section{Overview}

Because errors in the final shaded image are entangled with lighting, visibility, and material response, they are difficult to correct consistently; our goal is to recover a temporally stable, geometrically consistent hair representation from severely undersampled rasterized G-buffers that preserves the underlying geometric and directional structure needed for physically based hair shading. By restoring reliable intermediate attributes before shading, we can better preserve both temporal coherence and geometric fidelity.

At low sample counts, rasterized hair G-buffers exhibit two major failure modes. First, per-pixel coverage and tangent are sparse, aliased, and noisy. Second, geometric buffers such as position and depth are often incomplete, which in turn makes deferred shading unstable and prone to flickering artifacts. Among these attributes, coverage and tangent are relatively amenable to neural prediction, since they are locally correlated and can often be inferred from nearby strand patterns in screen space. In contrast, position and depth are more difficult to predict directly, as they require stronger geometric consistency and are highly sensitive to occlusion and missing samples. To address these issues, we propose a hybrid neural-analytic reconstruction pipeline that restores reliable hair attributes before deferred shading.
As shown in \autoref{fig:pipeline}, the pipeline consists of four stages:
neural spatial reconstruction of hair coverage and tangent (\S\ref{sec:spatial_prediction}),
neural temporal accumulation of the reconstructed coverage and tangent (\S\ref{sec:temporal_accumulation}),
analytic tangent-guided reconstruction of position and completion of the G-buffer (\S\ref{sec:position_reconstruction}), and
physically-based deferred hair shading. Our whole pipeline can be seamlessly integrated into the current industrial game engine. 



\section{Neural Prediction}
\label{sec:NeuralPrediction}

The neural prediction restores the two most critical hair attributes for subsequent geometric completion: coverage and tangent. We decompose this into a per-frame spatial reconstruction module and a recurrent temporal accumulation module, each with distinct characteristics: the spatial module learns to repair aliasing and local corruption within a single frame, while the temporal module learns to accumulate information over time and suppress flicker under motion and disocclusion. Both modules are trainable components. The first is a lightweight dual-branch U-Net, denoted by $\boldsymbol\Psi$, which operates on a single frame. The second is a compact recurrent convolutional module, denoted by $\boldsymbol\Gamma$, which refines the spatial predictions using reprojected temporal history.

\subsection{Spatial Reconstruction} \label{sec:spatial_prediction}

For each frame $i$, the spatial module takes as input the 4-channel tensor $\XX_i = [\CC_i, \TT_i] \in \mathbb{R}^{H \times W \times 4}$, where $\CC_i \in \mathbb{R}^{H \times W \times 1 }$ is the noisy coverage buffer and $\TT_i \in \mathbb{R}^{H \times W \times 3}$ is the noisy tangent buffer encoded in three channels. The spatial network predicts
$ 
\SS_i = \boldsymbol\Psi(\XX_i)
= [\CC^{s}_i, \TT^{s}_i, \LL^{s}_i] \in \mathbb{R}^{H \times W \times 5},
$
where $\CC^{s}_i$ is the reconstructed coverage, $ \TT^{s}_i$ is the reconstructed tangent, and $\LL^{s}_i$ is a foreground support-mask logit, used to suppress unreliable predictions outside the hair region. We demonstrate the network architecture of our neural filter in the supplementary material.

\paragraph{Dual-branch Spatial Encoder.}
As coverage and tangent have different statistics: coverage is scalar and sparse, while tangent is vector-valued and locally directional, we process them using separate shallow encoder branches before fusion. A $3 \times 3$ convolution maps the coverage channel to $N=32$ feature channels, and a second $3 \times 3$ convolution maps the 3-channel tangent input to $N$ feature channels. Each branch then applies two encoder stages. Each stage consists of residual convolutional blocks followed by a stride-2 $3 \times 3$ convolution for downsampling. The feature dimensions follow $N \rightarrow 2N \rightarrow 4N,$ with corresponding spatial resolutions $H \times W, \frac{H}{2} \times \frac{W}{2}, \frac{H}{4} \times \frac{W}{4}$.
Each residual block comprises two standard $3 \times 3$ convolutions with batch normalization and ReLU activations, followed by an identity skip connection. We use standard convolutions rather than depthwise separable convolutions, since the relatively small channel counts in our setting are more efficiently handled by dense convolution kernels on modern GPUs.

\paragraph{Bottleneck Fusion and Self-attention.}
At the bottleneck, the coverage and tangent features are concatenated to form an $8N$-channel representation at resolution $\frac{H}{4} \times \frac{W}{4}$. This fused representation is processed by additional residual blocks followed by a window-based multi-head self-attention module. The attention block uses GroupNorm pre-normalization, $1 \times 1$ convolutional projections for queries, keys, and values, four attention heads, and a feed-forward network composed of two $1 \times 1$ convolutions with GELU activation and expansion ratio $2$. We use $8\times8$ windows and additionally apply shifted-window attention with a half-window cyclic shift to enable interactions across neighboring windows.
Window-based attention keeps the computational and memory overhead low while still enabling interactions over larger hair structures through the shifted windows. This is particularly useful for sparse hair input, where valid local evidence may be weak, but consistent larger-scale flow patterns remain. When available, we use FlashAttention~\cite{dao2022flashattention} to reduce memory overhead.

\paragraph{Decoder and Skip Connections.}
The decoder mirrors the encoder with two upsampling stages. Each stage performs bilinear upsampling followed by a $1 \times 1$ convolution that halves the number of channels. The encoder features from the corresponding coverage and tangent branches are concatenated and fused into the decoder through skip connections. This design reintroduces high-frequency spatial detail while avoiding the larger memory cost of fully concatenating all decoder features. A final $3 \times 3$ convolution produces the output tensor.

\paragraph{Hierarchical Filtering Branch.}
In parallel with the main residual decoder, we introduce a lightweight multi-scale filtering branch that provides a low-frequency denoised estimate of the input. This branch is inspired by multi-scale kernel-predicting denoisers~\cite{vogels2018denoising}, but instead of predicting dense per-pixel spatial kernels, it predicts compact channel-wise filtering coefficients and blending weights, to achieve the fast inference speeds required for real-time performance.
At the bottleneck and the first decoder level, a $1 \times 1$ convolution predicts four sigmoid-normalized filtering coefficients $\boldsymbol{\kappa}^{l}$, one for each channel of the coverage-tangent input, together with a sigmoid-normalized blending coefficient $\beta^{l}$:
\begin{align}
    [\boldsymbol{\kappa}^{l}, \beta^{l}] = \text{\fontfamily{lmss}\selectfont K}^{l}(\mathbf{F}^{l}), l \in \{4\times, 2\times\}   \; ,
\end{align}
where $\mathbf{F}^{l}$ is the corresponding bottleneck or decoder feature map, and $\text{\fontfamily{lmss}\selectfont K}^{l}$ is implemented as a $1{\times}1$ convolution.
These coefficients are applied element-wise to average-pooled input buffers:
\begin{align}
\HH^{l} = \boldsymbol{\kappa}^{l} \odot \XX^{l} \;,
\end{align}
where \(\odot\) denotes element-wise multiplication, $\XX^{l}$ is the 4-channel input at scale $l$. 
The coarse \(4\times\) estimate is first upsampled and blended with the finer \(2\times\) estimate:
\begin{align}
    \HH^{2\times}_{\mathrm{fused}}
    =
    \bigl(\mathbf{1}-\text{\fontfamily{lmss}\selectfont up}(\beta^{4\times})\bigr)
    \odot \HH^{2\times}
    +
    \text{\fontfamily{lmss}\selectfont up}(\beta^{4\times})
    \odot
    \text{\fontfamily{lmss}\selectfont up}(\HH^{4\times}) \; .
\end{align}
The fused estimate is then upsampled to full resolution and modulated by the decoder-level blending coefficient:
\begin{align}
    \HH
    =
    \text{\fontfamily{lmss}\selectfont up}
    \bigl(\HH^{2\times}_{\mathrm{fused}}\bigr)
    \odot
    \text{\fontfamily{lmss}\selectfont up}(\beta^{2\times}) \; .
\end{align}
This auxiliary branch provides an adaptive low-frequency prior that stabilizes the residual reconstruction path. Operator $\text{\fontfamily{lmss}\selectfont up}(\cdot)$ denotes bilinear upsampling. 

\paragraph{Spatial Output Head.}
Let $\RR_i \in \mathbb{R}^{H \times W \times 4}$ denote the residual predicted for coverage and tangent, and $\HH_i$ denote the filtered estimate from the hierarchical branch. The spatial output is formed as
\begin{align}
\ZZ_i = \XX_i + s \RR_i + \HH_i \;,
\end{align}
where $s$ is a learned scalar residual scale. 

\subsection{Temporal Accumulation} \label{sec:temporal_accumulation}

Although the spatial module improves per-frame quality, its predictions may still flicker because sparse rasterization artifacts vary across frames. We therefore introduce a recurrent temporal module that refines the current spatial prediction using reprojected history.
Let $\YY_{i-1}$ denote the previous temporal output. For frame $i$, we reproject $\YY_{i-1}$ into the current frame using the current motion vector $\VV_i$. We also compute a warped motion vector difference
$
\Delta \VV_i
=
\VV_i - \boldsymbol\Pi(\YY_{i-1}, \VV_i),
$
where $\boldsymbol\Pi(\cdot)$ denotes bilinear reprojection using motion vectors. This term provides a local cue for motion inconsistency, acceleration, and disocclusion~\cite{edison17acceleartionmotion}. 
Note that obtaining the motion vectors incurs no additional computational cost, as they are already provided by the standard rendering pipeline by default.
%
The temporal input is
\begin{align}
\UU_i = [\SS_i,\,\boldsymbol\Pi(\YY_{i-1}, \VV_i),\, \VV_i,\, \Delta \VV_i] \in \mathbb{R}^{H \times W \times 14} \;.
\end{align}
The channel count is therefore $14 = 5 + 5 + 2 + 2$. On the first frame of a sequence, or whenever valid history is unavailable, we set the reprojected history and the motion-vector difference to zero.

The temporal module $\boldsymbol\Gamma$ is implemented as a compact residual CNN. A $3 \times 3$ convolution maps the 14-channel input to 32 hidden channels, followed by batch normalization and ReLU. This is followed by 4 residual blocks and a final $3 \times 3$ convolution that predicts a 5-channel residual correction:
\begin{align}
[\Delta \CC_i, \Delta \TT_i, \Delta \LL_i] = \boldsymbol\Gamma(\UU_i).
\end{align}
The final temporal prediction is formed as 
\begin{align}
[\CC^t_i, \TT^t_i, \LL^t_i]&= [\CC^{s}_i, \TT^{s}_i, \LL^{s}_i] + \alpha [\Delta \CC_i, \Delta \TT_i, \Delta \LL_i],
\end{align}
where $\alpha$ is a learned residual scale initialized to a small value. If $\II_i$ indicates that the current frame is the first frame in a sequence, the temporal module directly returns the spatial prediction, preventing invalid history from corrupting the output. During inference, we threshold the support mask derived from $\LL_i$ and use it to suppress coverage and tangent predictions outside the reconstructed hair support. The resulting 5-channel tensor $\YY_i = [\CC^t_i, \TT^t_i, \LL^t_i]$ is then stored as recurrent history for the next frame.

\subsection{Two-Stage Training}
We train the spatial and temporal modules in two stages. We first train the spatial module on independently sampled frames using coverage, tangent, and support-mask supervision. We then freeze the spatial module and train the temporal module on temporally ordered sequences using the same supervision. {Detailed loss definitions and weights are provided in the supplementary material.}

\section{Tangent-guided Reconstruction} \label{sec:position_reconstruction}

The neural stage reconstructs coverage and tangent; however, deferred shading with shadows also requires reliable world-space positions. As discussed above, the undersampled rasterized position map is often sparse and corrupted, and some hair pixels may lack valid position samples altogether. To address this issue, we introduce a tangent-guided reconstruction module that exploits local directional consistency in the reconstructed tangent field to recover world-space positions. Specifically, we design a bidirectional propagation strategy, consisting of backward repair and forward voting, to infer missing world-space positions while preserving anisotropic strand structures.

For each frame, the module takes as input a noisy world-space position map $\PP \in \mathbb{R}^{H \times W \times 3}$, a noisy depth map $\DD \in \mathbb{R}^{H \times W}$, the temporally reconstructed world-space tangent map $\TT^t \in \mathbb{R}^{H \times W \times 3}$, the reconstructed coverage map $\CC^t \in \mathbb{R}^{H \times W}$, and the view-projection matrix. Our goal is to recover a refined position map $\PP^\star$ by combining backward repair for missing samples with forward voting from valid samples.

\subsection{Pixel Classification}
We first determine which pixels belong to the hair support and which of them contain valid position samples. A pixel $\pp$ is considered a hair pixel if its reconstructed coverage is positive. Among hair pixels, we classify $\pp$ as \emph{valid} if its rasterized world-space position is available, and \emph{invalid} otherwise. In practice, following the rasterization convention that missing positions are encoded as zeros, we use $\|\PP(\pp)\| > \epsilon_{\mathrm{pos}}$ as the validity criterion. Valid pixels may cast votes during reconstruction, whereas invalid pixels may only receive them.

\subsection{Depth Inpainting}
To support tangent-guided propagation at all hair pixels, we first fill missing depth values. For each hair pixel $\pp$ whose reconstructed coverage is positive but whose depth is missing, we copy the depth from its nearest valid neighbor in image space:
\begin{align}
\DD(\pp) \leftarrow \DD(\pp^\star),
\quad
\pp^\star = \arg\min_{\qq:\,\DD(\qq) > 0} \|\pp-\qq\|_2.
\end{align}
This simple inpainting step ensures that every hair pixel has a depth estimate for subsequent geometric computations.

\subsection{Bidirectional Position Reconstruction}

\paragraph{Stage I. Per-pixel Step Size.}
We first estimate a local world-space step length for each pixel from the inpainted depth map and the camera intrinsics $(f_x, f_y)$. Specifically, we approximate the horizontal and vertical world-space displacements corresponding to one image-space pixel at depth $\DD(\pp)$ as $\DD(\pp)/f_x$ and $\DD(\pp)/f_y$, respectively, and define
\begin{align}
l_{\pp} = \sqrt{\left({\DD(\pp)}/{f_x}\right)^2 + \left({\DD(\pp)}/{f_y}\right)^2}.
\end{align}
Intuitively, $l_{\pp}$ converts a one-pixel displacement in image space into an approximate world-space propagation step at pixel $\pp$.

\paragraph{Stage II. Screen-space Curvature Center Estimation.}
For each hair pixel $\pp$, we estimate a 2D curvature center $\cc_{\pp}$ in screen space by fitting a local circle to neighboring hair pixels within a $11 \times 11$ window, masked by the reconstructed coverage. The estimated curvature center provides a geometric prior for selecting the most plausible source sample when repairing a missing target pixel. The full implementation is provided in the supplemental material.

\paragraph{Stage III. Screen-space Tangent Projection.}
For each hair pixel $\pp$, we project the reconstructed 3D tangent $\TT^\star(\pp)$ into the image plane using the view-projection matrix, and normalize the result to obtain a 2D screen-space tangent $\TT_{\mathrm{screen}}(\pp)$. This projected tangent captures the local strand direction in image space and is later used to determine whether a valid pixel should vote for a neighboring target pixel.

\paragraph{Stage IV. Backward Repair for Missing Pixels.}
For each invalid target pixel $\pp$, we inspect its $3 \times 3$ neighborhood $\mathcal{N}_{\pp}$ and select the valid neighbor $\ss^\star$ whose curvature center is closest to that of $\pp$:
\begin{align}
\ss^\star
=
\arg\min_{\ss \in \mathcal{N}_{\pp} \cap \mathcal{V}}
\|\cc_{\ss} - \cc_{\pp}\|_2,
\end{align}
where $\mathcal{V}$ denotes the set of valid pixels. We then initialize the missing world-space position by propagating from the selected source sample along the reconstructed tangent at the target pixel:
\begin{align}
\PP^{\star}(\pp) = \PP(\ss^\star) + l_{\pp}\,\TT^\star(\pp).
\end{align}
Since the neural stage densely reconstructs the tangent field, this strategy remains applicable even at pixels whose original position samples are missing.

\paragraph{Stage V. Forward Voting from Valid Pixels.}
Each valid pixel $\ss$ also casts votes toward neighboring pixels by projecting its world-space position forward along its tangent:
\begin{align}
\PP^{\mathrm{vote}}_{\ss \rightarrow \pp}
=
\PP(\ss) + l_{\ss}\,\TT(\ss).
\end{align}
A vote from $\ss$ to a neighboring target pixel $\pp$ is accepted only if the screen-space tangent at $\ss$ is sufficiently aligned with the direction from $\ss$ to $\pp$:
\begin{align}
\left|
\TT_{\mathrm{screen}}(\ss)\cdot
\frac{\pp-\ss}{\|\pp-\ss\|_2}
\right|
\geq
\cos\theta_{\max},
\end{align}
where we set $\theta_{\max}=30^\circ$ in all experiments. Each target pixel $\pp$ maintains a candidate pool $\mathcal{C}_{\pp}$ containing up to $K$ accepted votes. When the pool is full, we discard the candidate with the largest depth. The final forward estimate is then chosen as the frontmost candidate:
\begin{align}
\PP^{\star}(\pp)
=
\arg\min_{\qq \in \mathcal{C}_{\pp}}
\text{\fontfamily{lmss}\selectfont depth}(\qq),
\end{align}
where $\text{\fontfamily{lmss}\selectfont depth}(\qq)$ denotes the screen-space depth of a world-space position $\mathbf{q}$ computed using the view-projection matrix. This visibility-aware selection suppresses implausible votes from deeper layers and favors the frontmost visible hair surface.

\begin{figure}[hbt]
\newcommand{\figcap}[1]{\begin{minipage}{0.495\linewidth}\centering#1\end{minipage}}
\centering
\includegraphics[width=\linewidth]{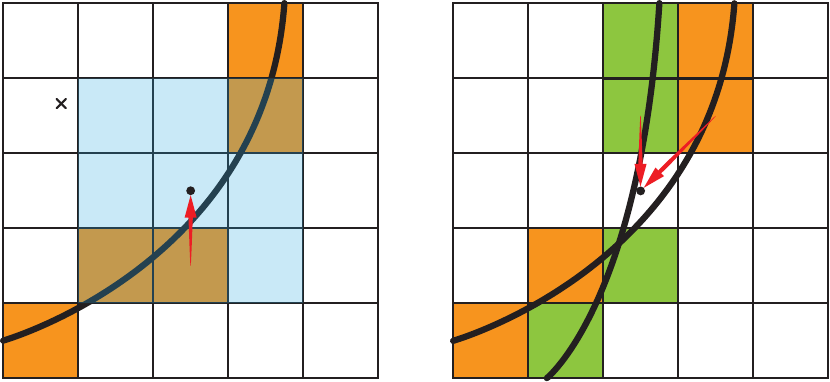}
\put(-194,58){\footnotesize $\pp$}
\put(-235,82){\footnotesize $\cc_\pp$}
\put(-165,28){\footnotesize $\mathcal{N}_\pp$}
\put(-194,28){\footnotesize $\ss_\star$}
\put(-64,58){\footnotesize $\pp$}
\put(-64,80){\footnotesize $\ss_a$}
\put(-33,72){\footnotesize $\ss_b$}
\\
\figcap{\small Backward repair}
\figcap{\small Forward voting} 
\caption{Left: given an empty pixel $\pp$ with reconstructed tangent direction, we estimate its curvature center at $\cc_\pp$. For its $3\times3$ neighborhood $\mathcal{N}_\pp$, we identify the neighbor $\ss_\star$ whose curvature center is closest to $\cc_\pp$ and propagate it to $\pp$. Right: from valid pixels $\ss_a$ and $\ss_b$, which have already been marked as hair, we propagate along the tangent direction. For each reached missing pixel, we assign the value with the minimum depth. }
\label{fig:backward_forward}
\vspace{-15pt}
\Description{}
\end{figure}
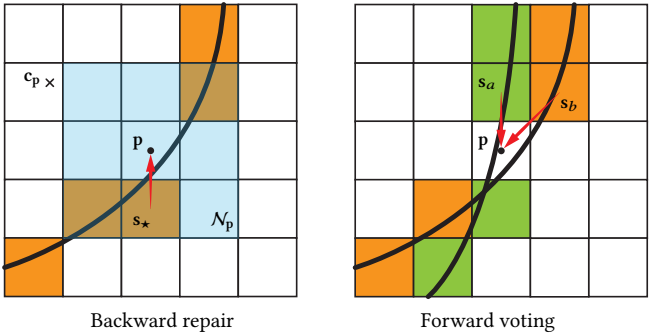

\paragraph{Remark}
The key observation behind our reconstruction strategy is that hair exhibits strong local directional coherence: neighboring pixels that belong to the same visible strand or strand bundle tend to follow a consistent tangent direction in screen space. Once the neural stage reliably reconstructs coverage and tangent, this directional structure provides a strong cue for recovering missing geometric attributes. In particular, a missing position sample can often be inferred either by tracing backward from a nearby valid sample or by receiving a forward proposal from neighboring valid pixels along the local tangent direction. This motivates our bidirectional reconstruction strategy, which combines backward repair and forward voting to improve robustness under sparse rasterization.

For full details of our algorithm, we provide step-by-step pseudocode in the supplementary material.

\section{Results}
\label{sec:results}

We train our models on 1,000 samples from four hairstyles rendered in Unreal Engine under varying viewpoints, distances, hair widths, and dynamics. Unless otherwise specified, experiments are performed at 1920×1080 on a GPU with 14{,}592 cores and 96\,GB of VRAM with an Intel i9-14900KF CPU. Detailed training settings and hyperparameters are provided in the supplementary material.

\begin{figure*}[htb]
\centering

\newcommand{\statichead}[1]{%
  \begin{minipage}{0.12\linewidth}
    \centering
    \small #1
  \end{minipage}%
}

%
\newcommand{\staticresult}[5]{%
  \begingroup
  \setlength{\unitlength}{0.12\linewidth}%
  \begin{picture}(1,#5)

    \put(0,0){%
      \includegraphics[
        trim={#3},
        clip,
        width=\unitlength
      ]{#1}%
    }%

    \put(0,0){%
      \includegraphics[
        trim={#4},
        clip,
        width=0.40\unitlength
      ]{#2}%
    }%

  \end{picture}%
  \endgroup
}

%
\newcommand{\staticref}[2]{%
  \includegraphics[
    trim={#2},
    clip,
    width=0.12\linewidth
  ]{#1}%
}

\statichead{Input}%
\hspace{0.004\linewidth}%
\statichead{MSAA}%
\hspace{0.004\linewidth}%
\statichead{SSAA}%
\hspace{0.004\linewidth}%
\statichead{TAA}%
\hspace{0.004\linewidth}%
\statichead{DLSS}%
\hspace{0.004\linewidth}%
\statichead{FSR}%
\hspace{0.004\linewidth}%
\statichead{Ours}%
\hspace{0.004\linewidth}%
\statichead{Reference}%
\\[2pt]

\staticresult
  {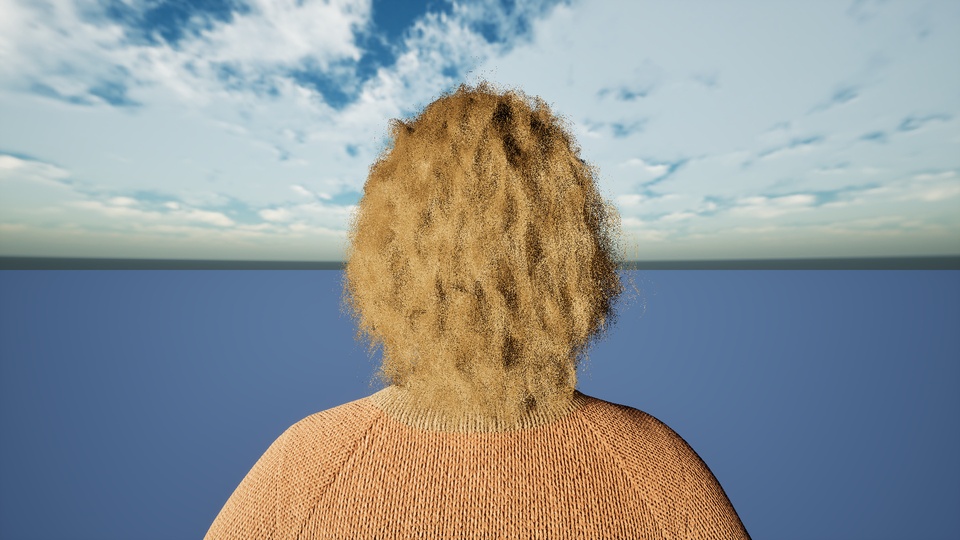}
  {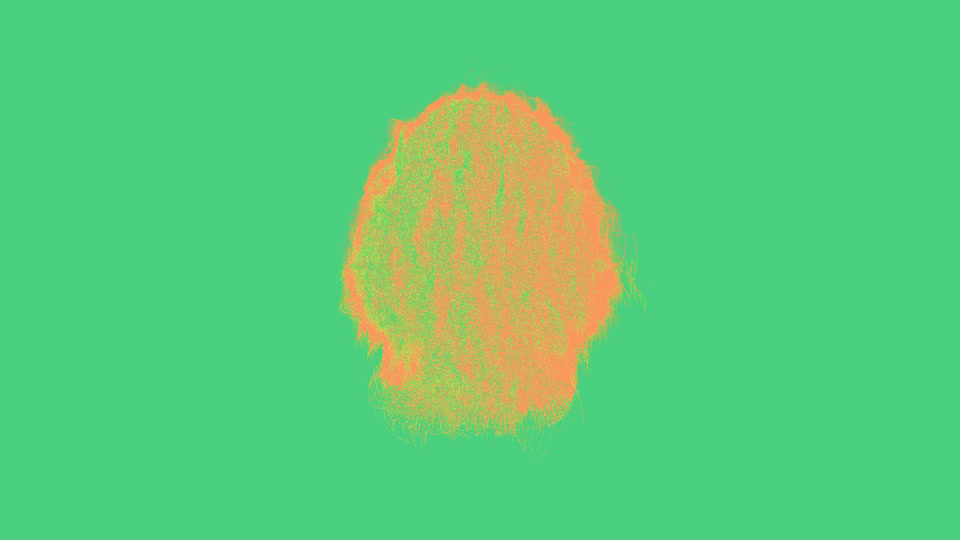}
  {200 75 200 75}
  {200 75 200 75}
  {0.696}
\hspace{0.00\linewidth}%
\staticresult
  {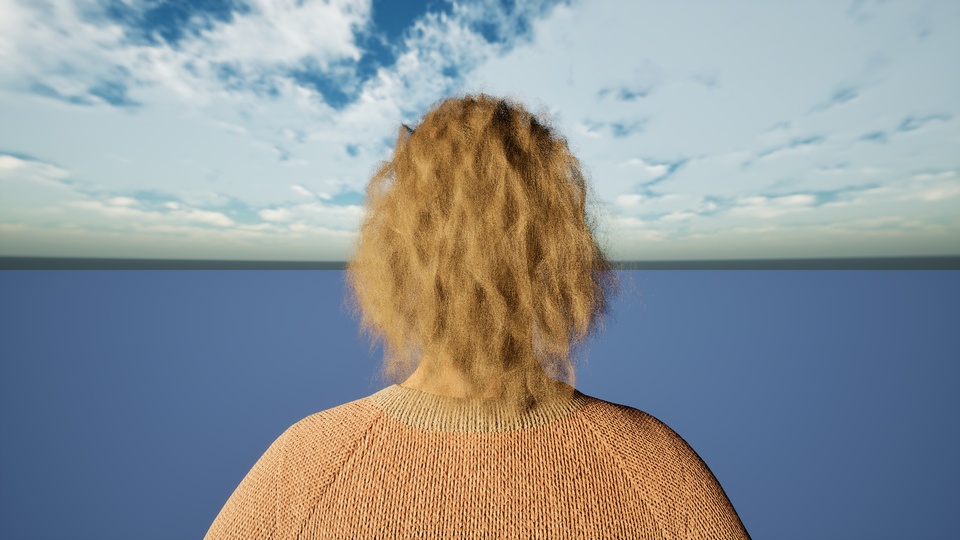}
  {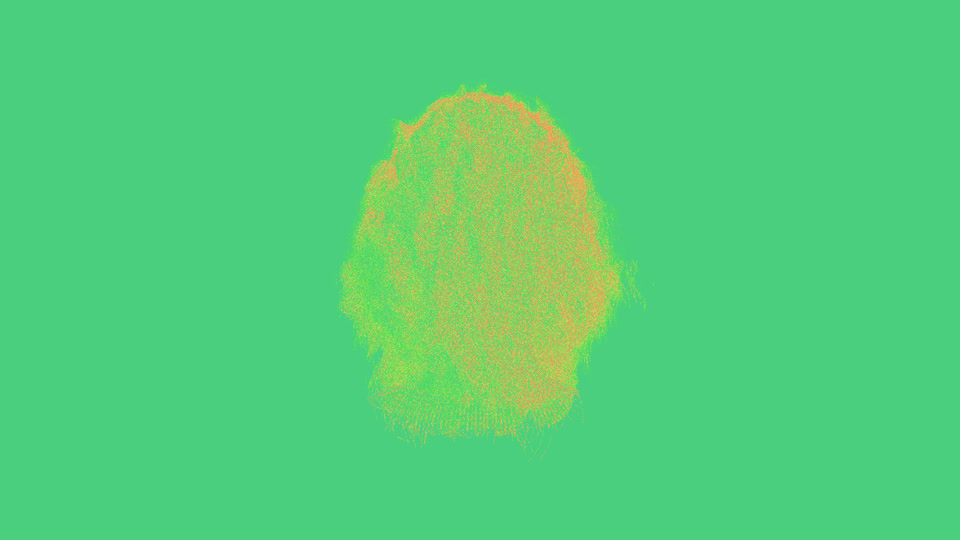}
  {200 75 200 75}
  {200 75 200 75}
  {0.696}%
\hspace{0.004\linewidth}%
\staticresult
  {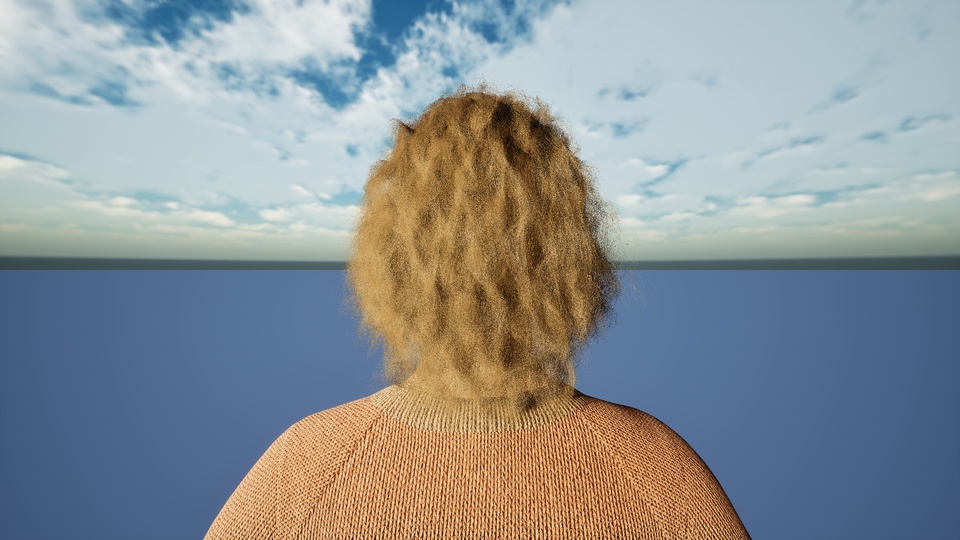}
  {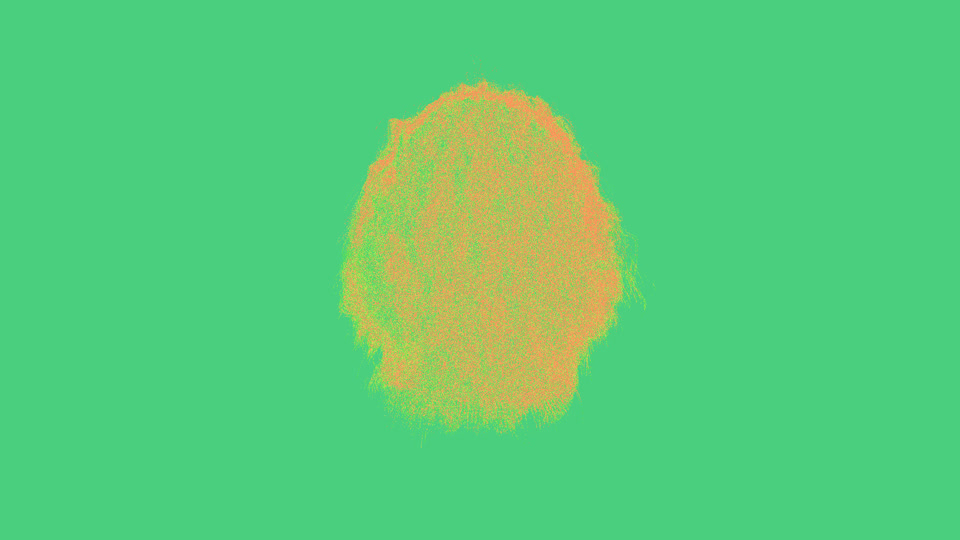}
  {200 75 200 75}
  {200 75 200 75}
  {0.696}%
\hspace{0.004\linewidth}%
\staticresult
  {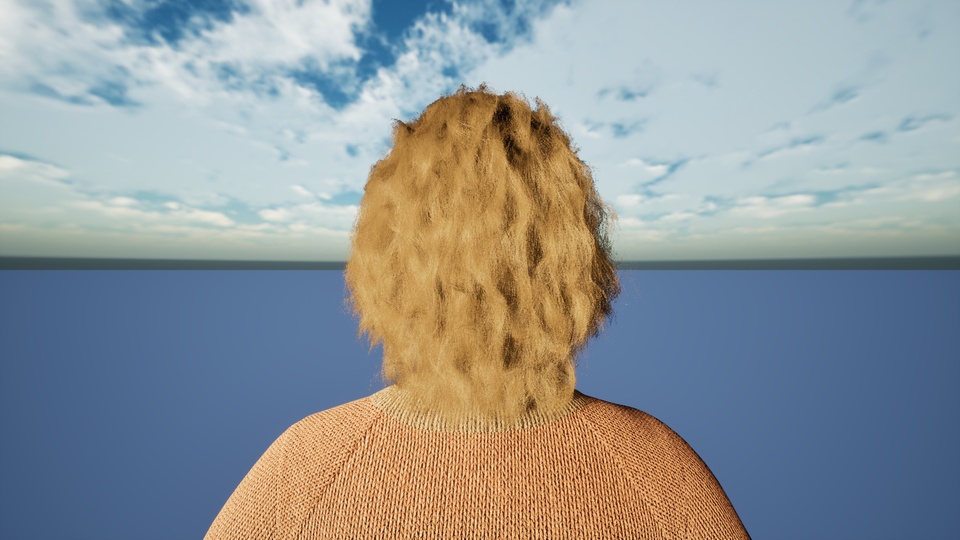}
  {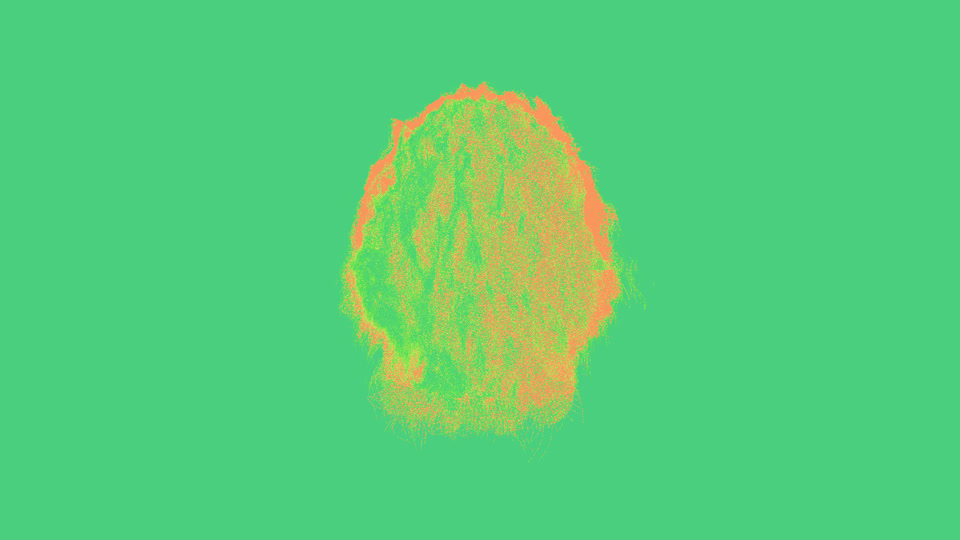}
  {200 75 200 75}
  {200 75 200 75}
  {0.696}%
\hspace{0.004\linewidth}%
\staticresult
  {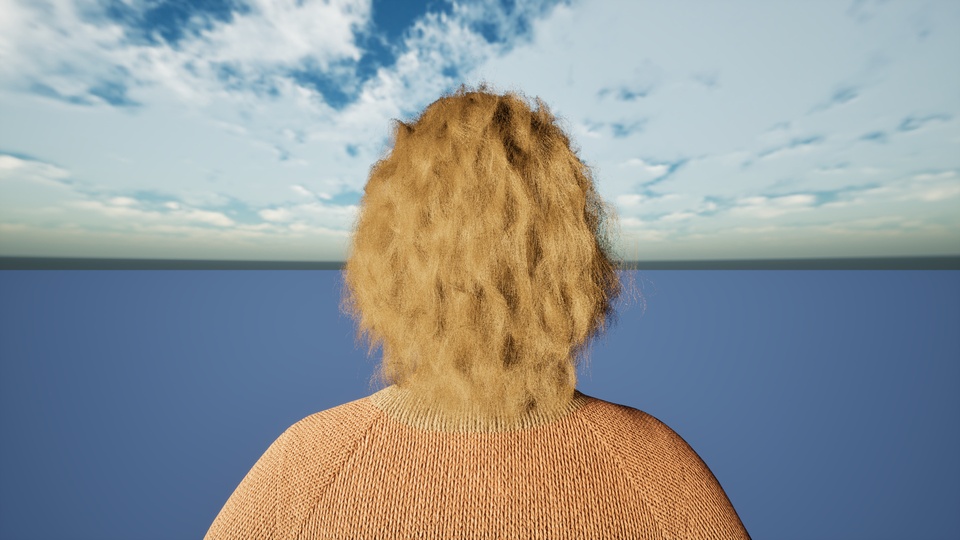}
  {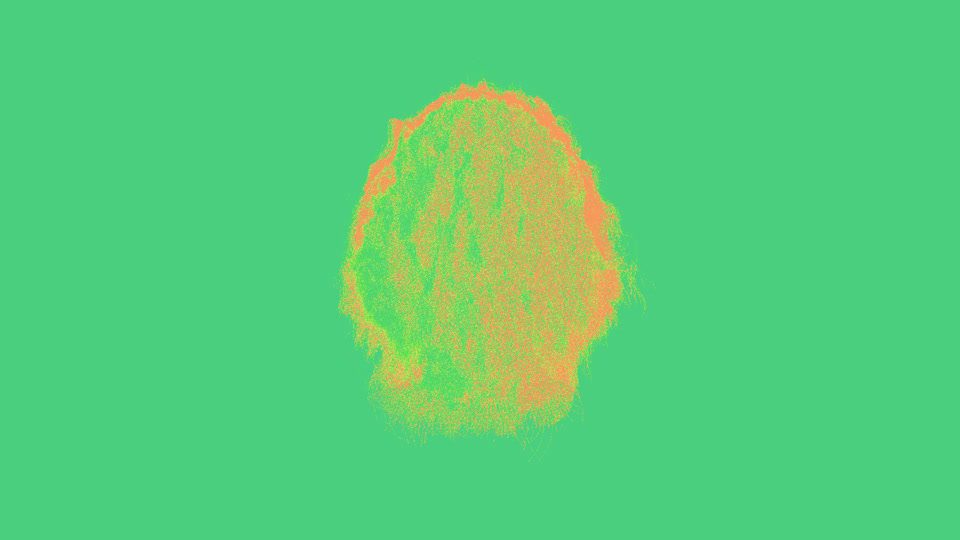}
  {200 75 200 75}
  {200 75 200 75}
  {0.696}%
\hspace{0.004\linewidth}%
\staticresult
  {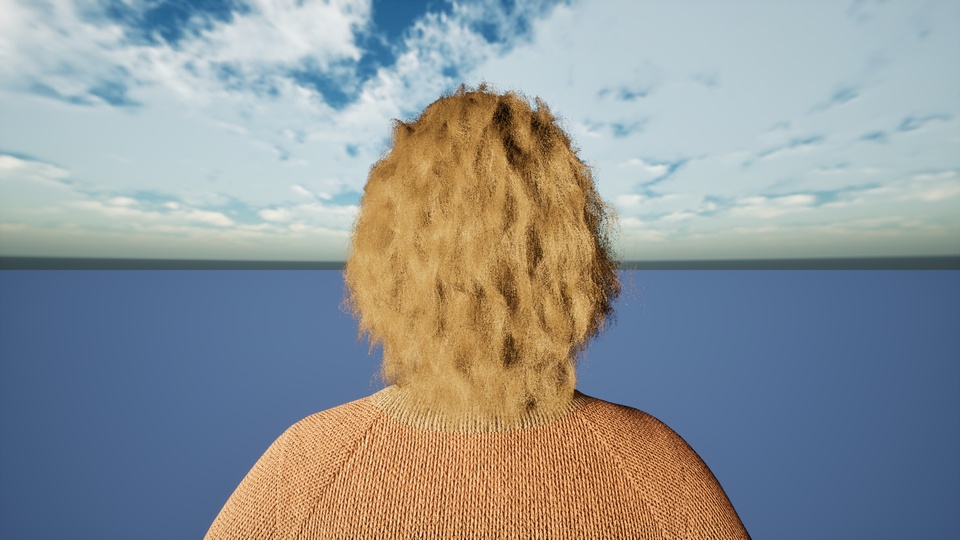}
  {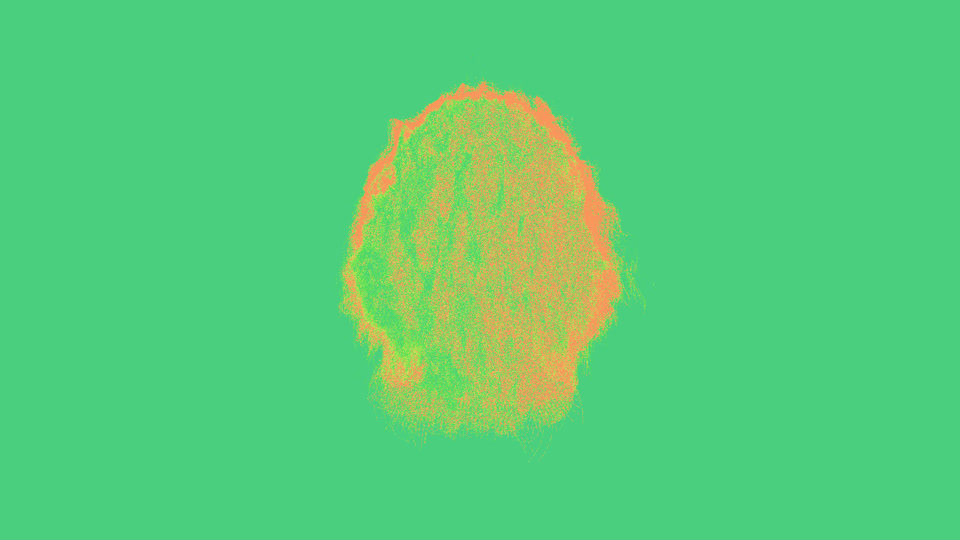}
  {200 75 200 75}
  {200 75 200 75}
  {0.696}%
\hspace{0.004\linewidth}%
\staticresult
  {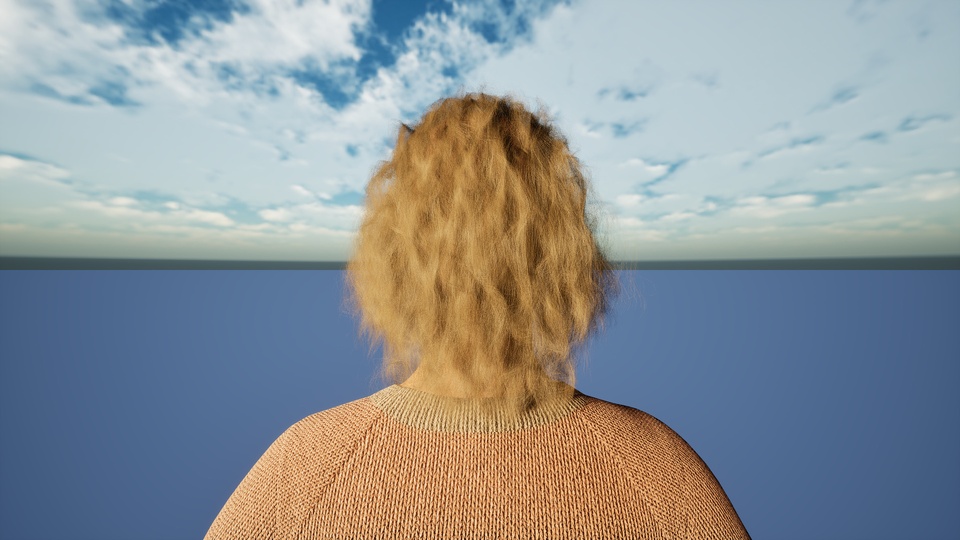}
  {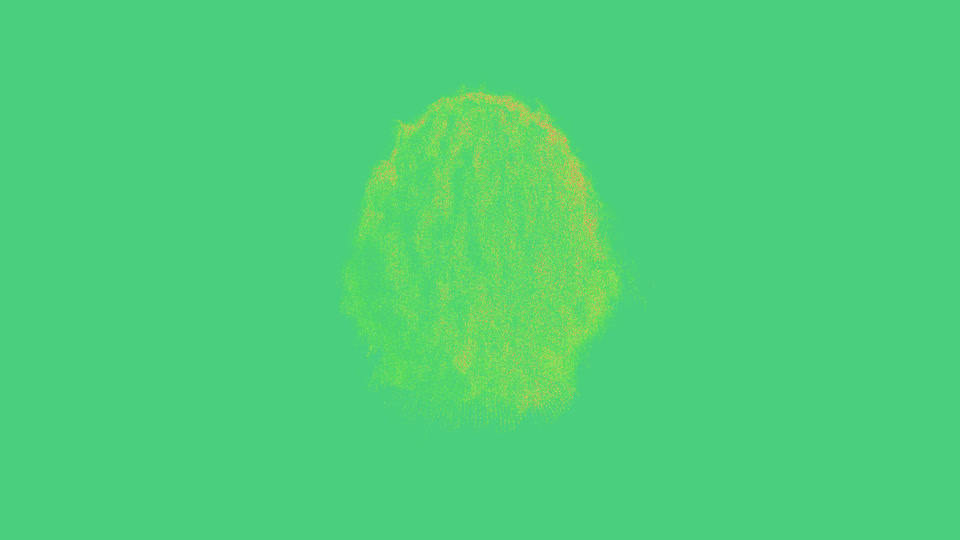}
  {200 75 200 75}
  {200 75 200 75}
  {0.696}
\hspace{0.00\linewidth}%
\staticref
  {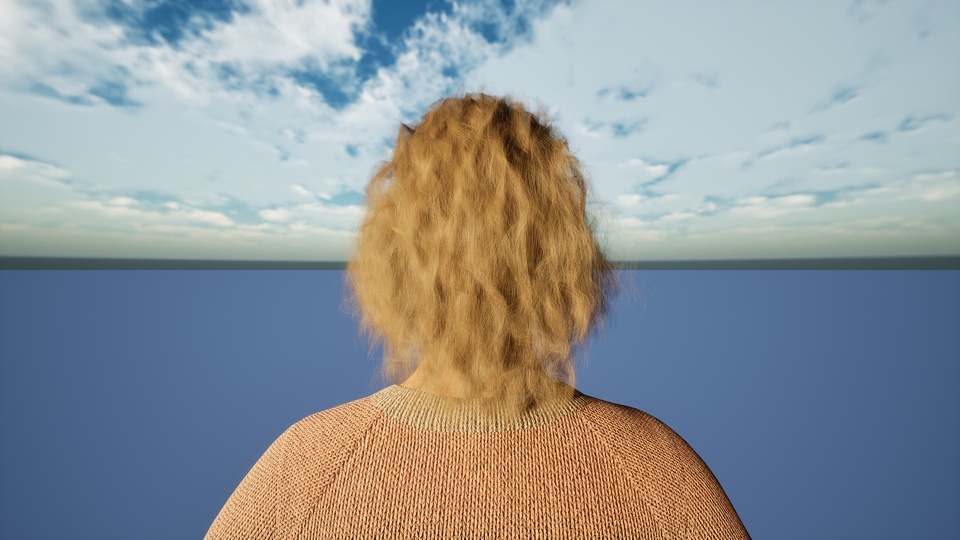}
  {200 75 200 75}
\\[2pt]

\staticresult
  {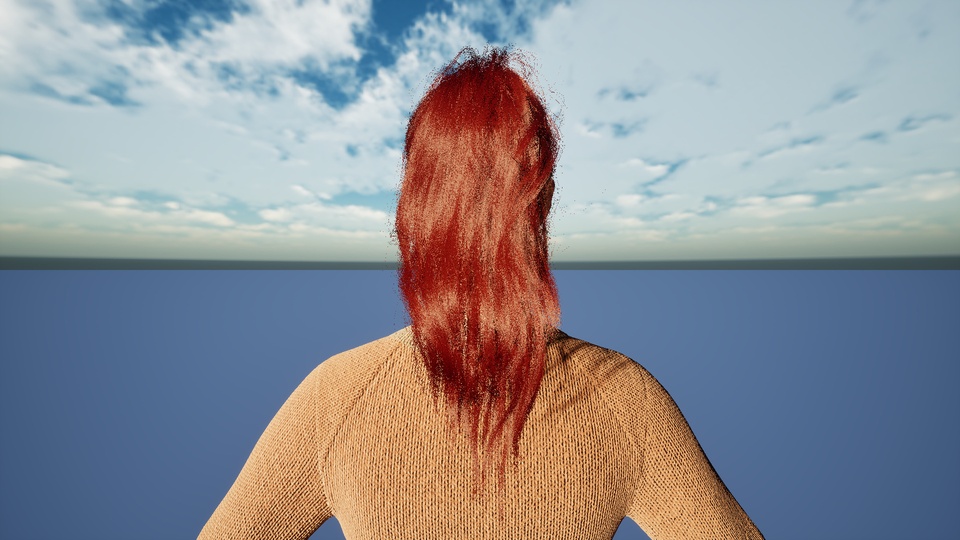}
  {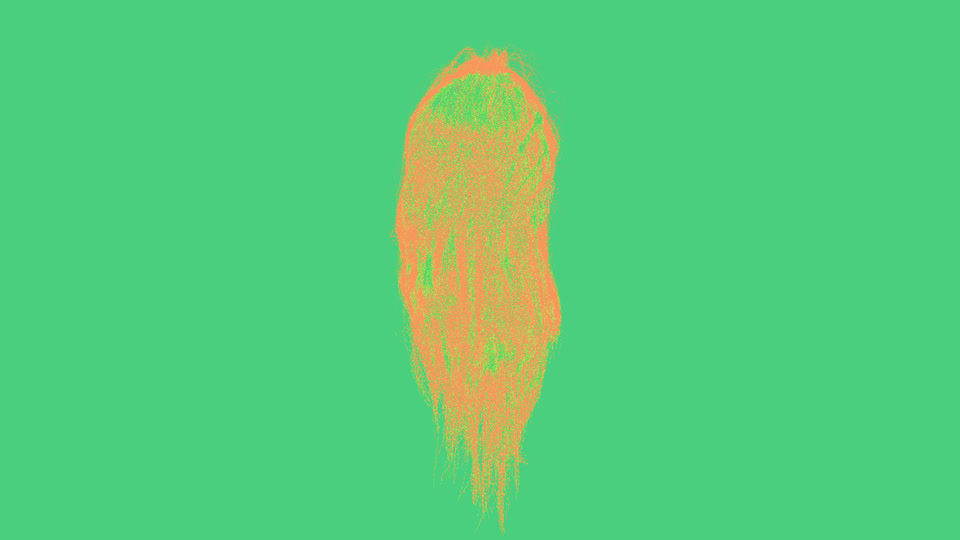}
  {150 25 150 25}
  {150 25 150 25}
  {0.742}%
\hspace{0.004\linewidth}%
\staticresult
  {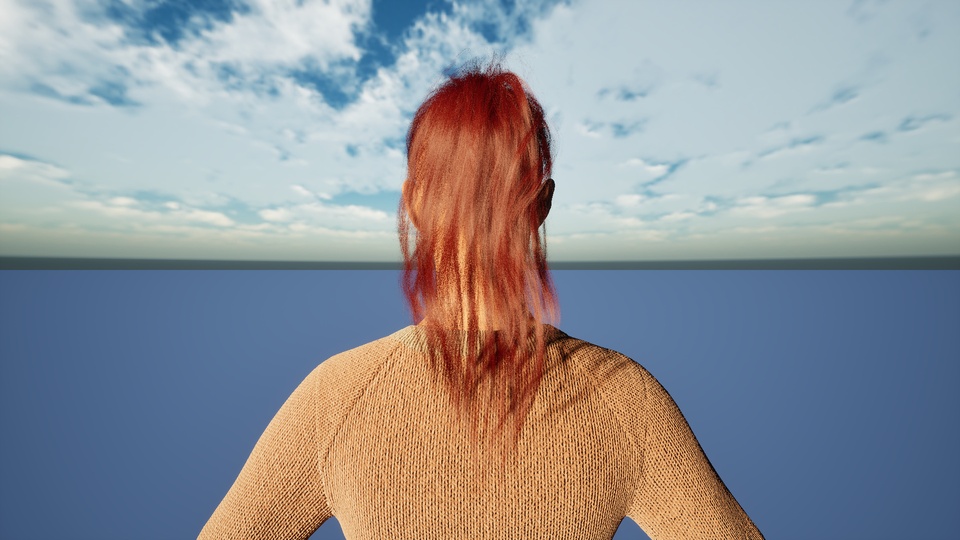}
  {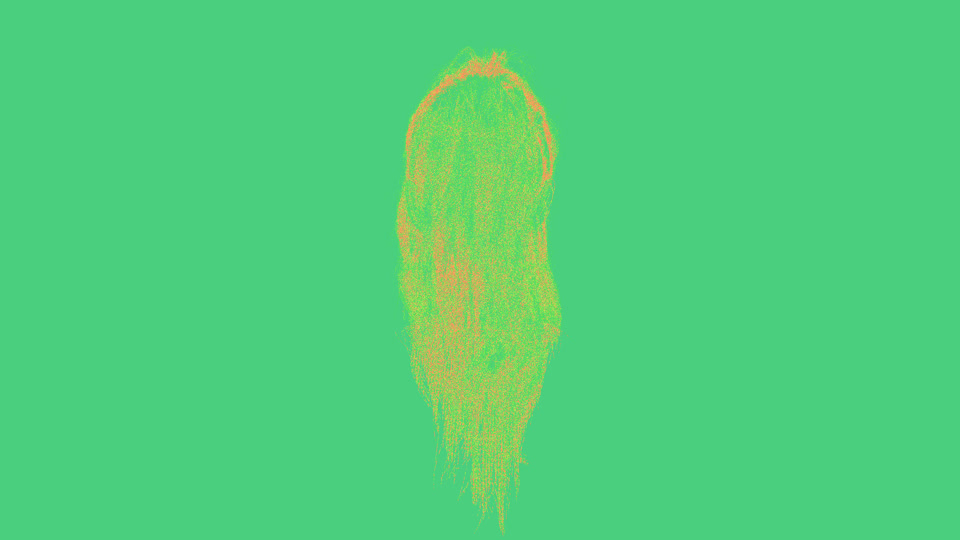}
  {150 25 150 25}
  {150 25 150 25}
  {0.742}%
\hspace{0.004\linewidth}%
\staticresult
  {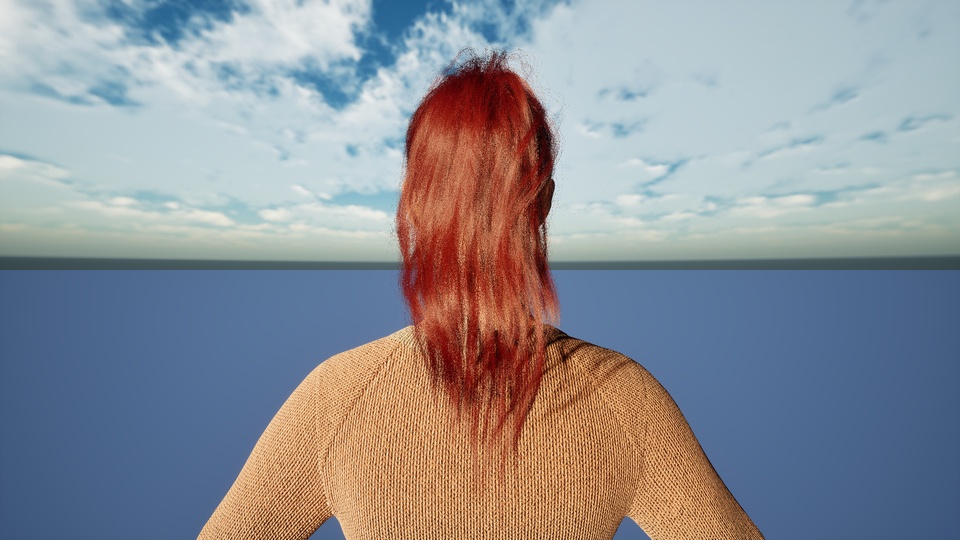}
  {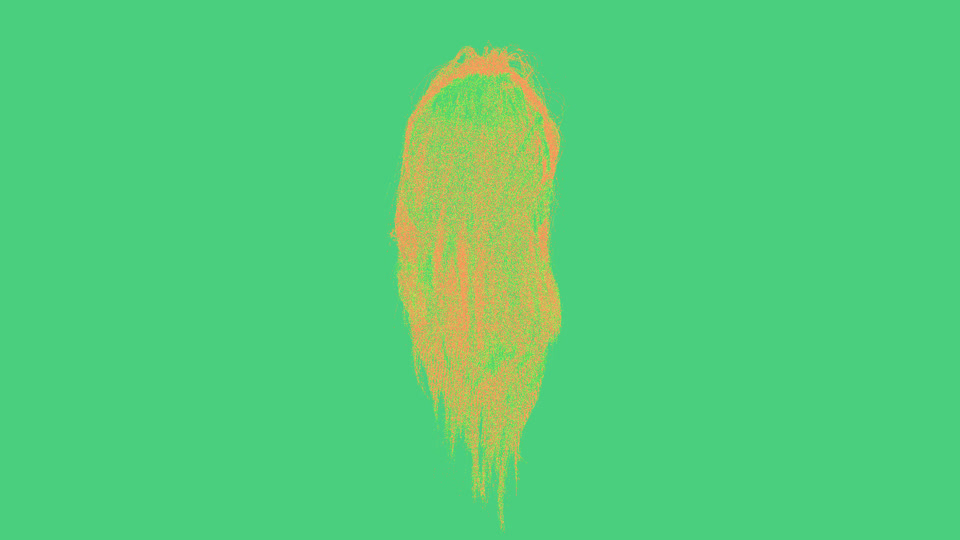}
  {150 25 150 25}
  {150 25 150 25}
  {0.742}%
\hspace{0.004\linewidth}%
\staticresult
  {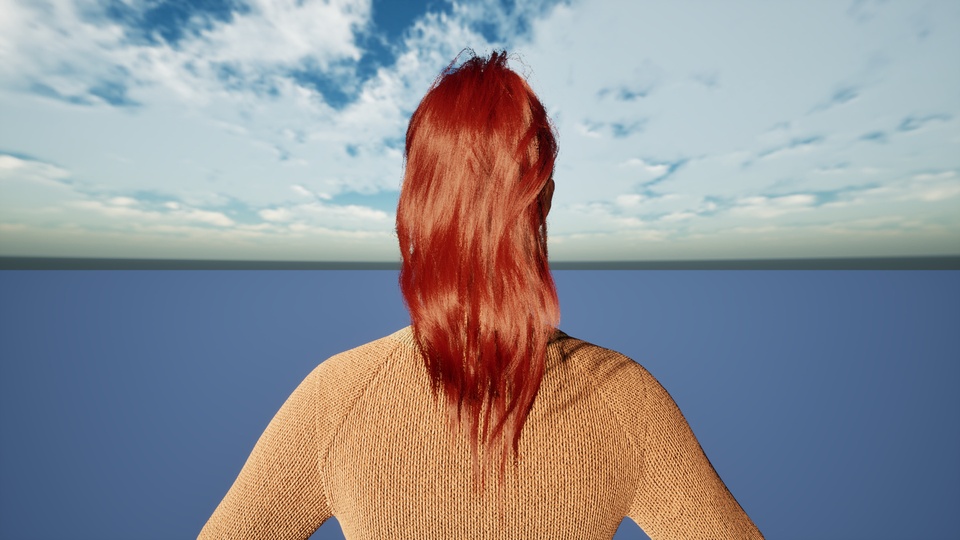}
  {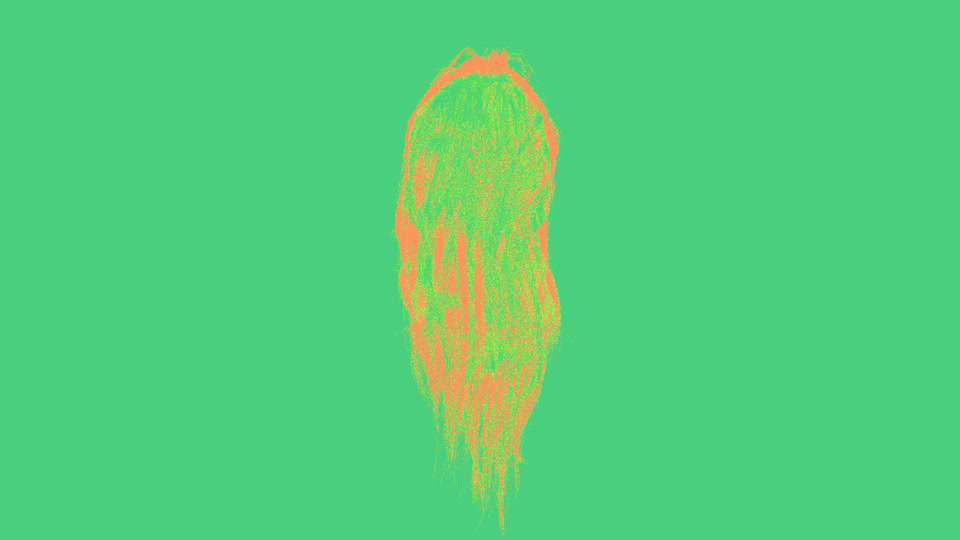}
  {150 25 150 25}
  {150 25 150 25}
  {0.742}%
\hspace{0.004\linewidth}%
\staticresult
  {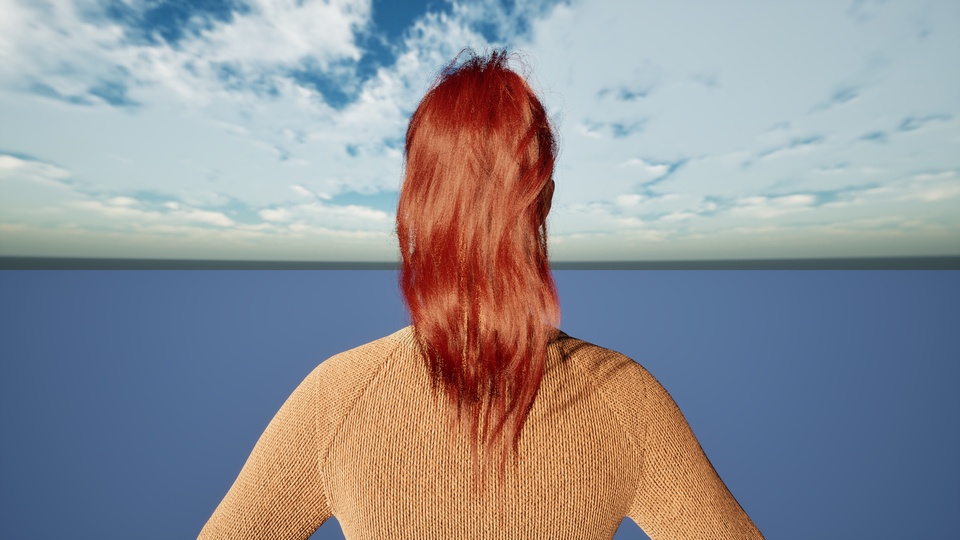}
  {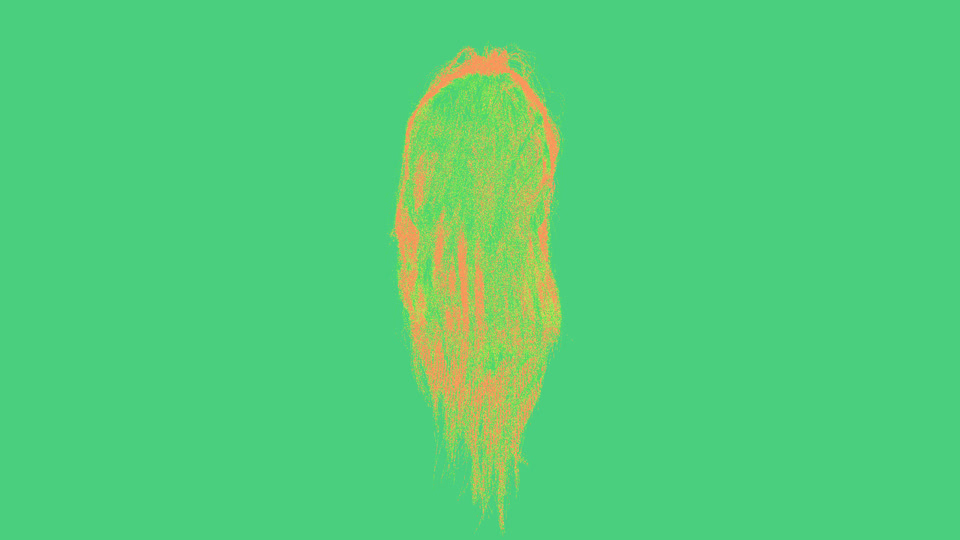}
  {150 25 150 25}
  {150 25 150 25}
  {0.742}%
\hspace{0.004\linewidth}%
\staticresult
  {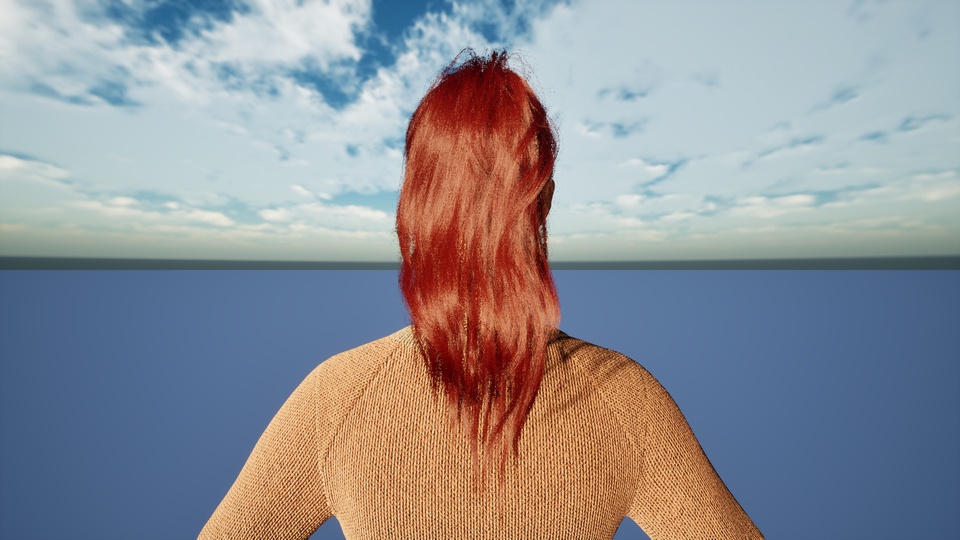}
  {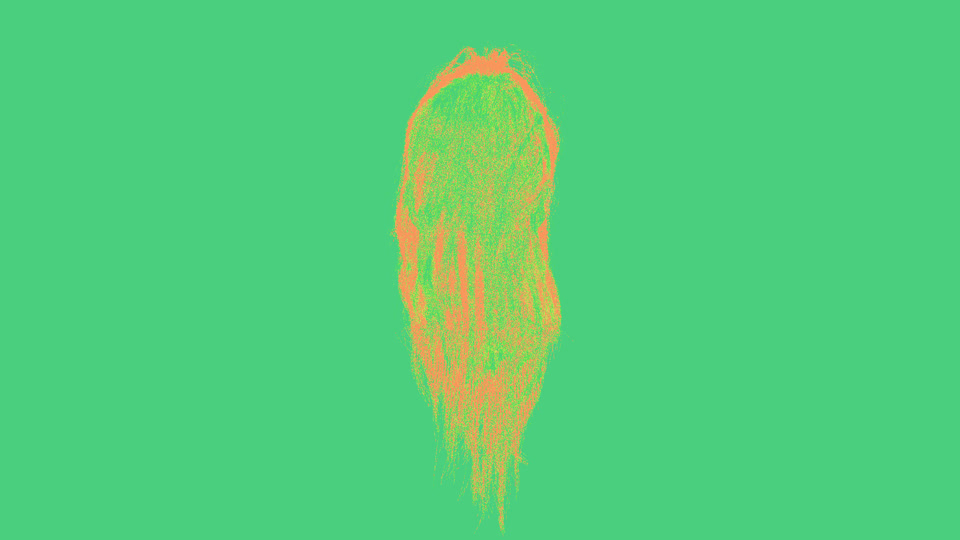}
  {150 25 150 25}
  {150 25 150 25}
  {0.742}%
\hspace{0.004\linewidth}%
\staticresult
  {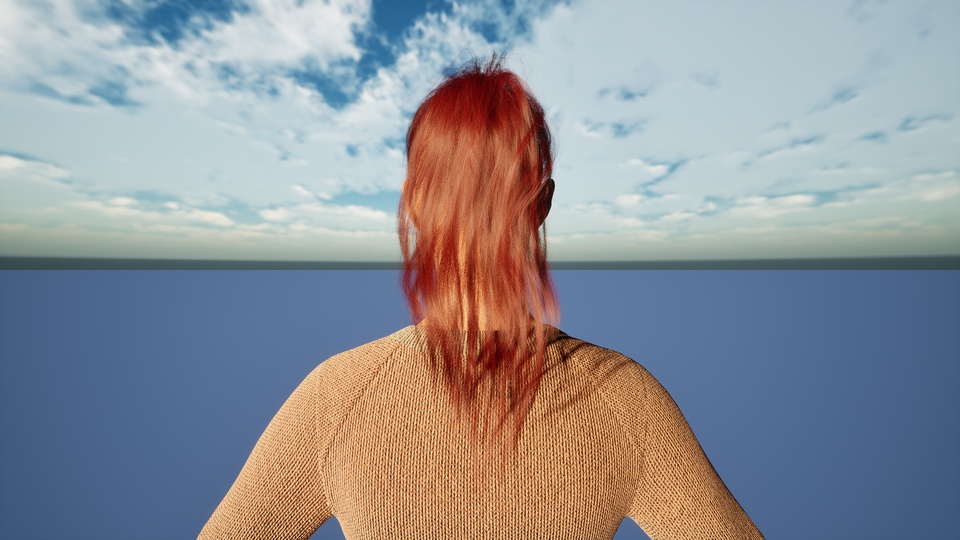}
  {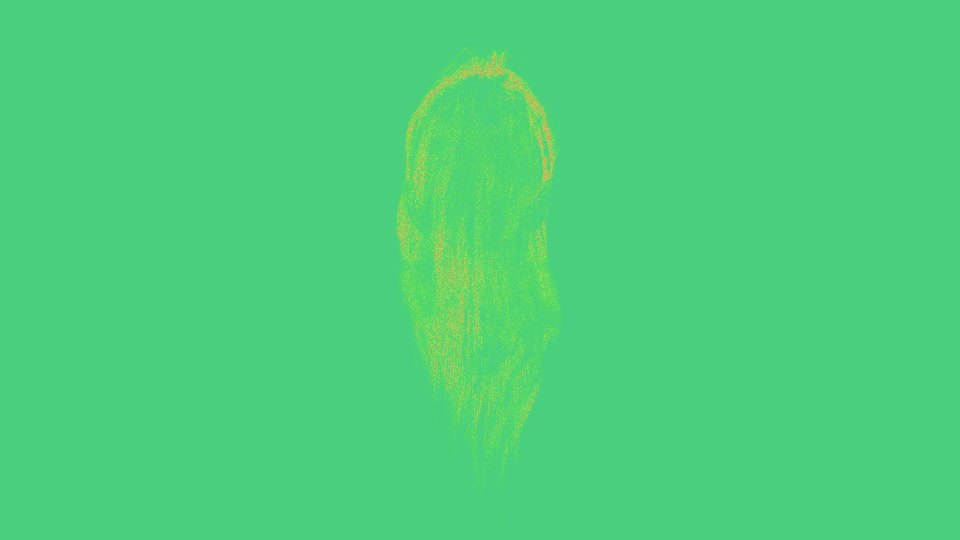}
  {150 25 150 25}
  {150 25 150 25}
  {0.742}%
\hspace{0.004\linewidth}%
\staticref
  {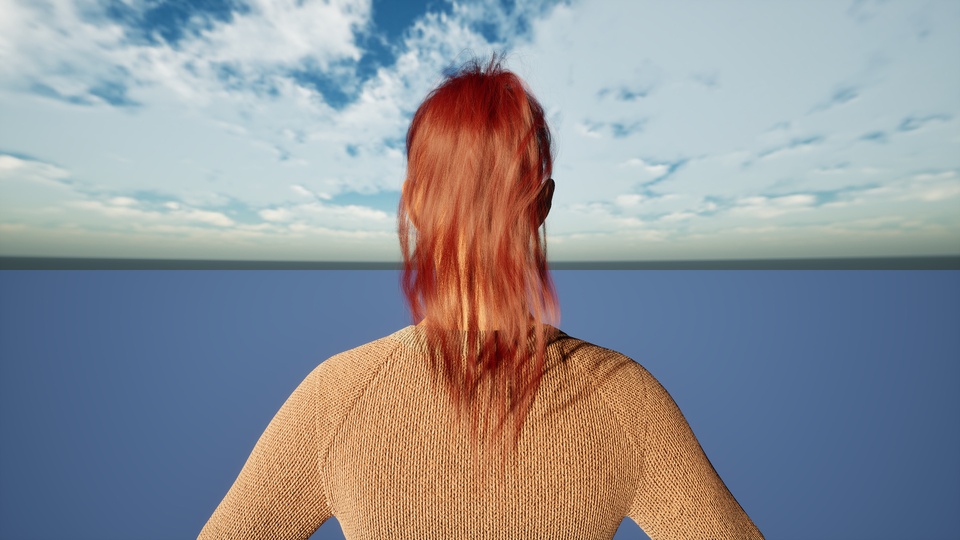}
  {150 25 150 25}
\\[2pt]

\staticresult
  {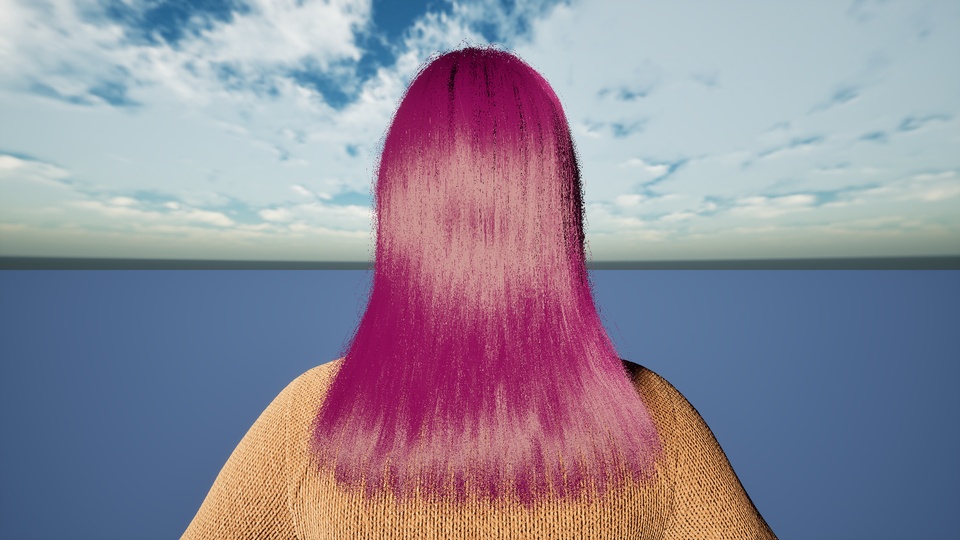}
  {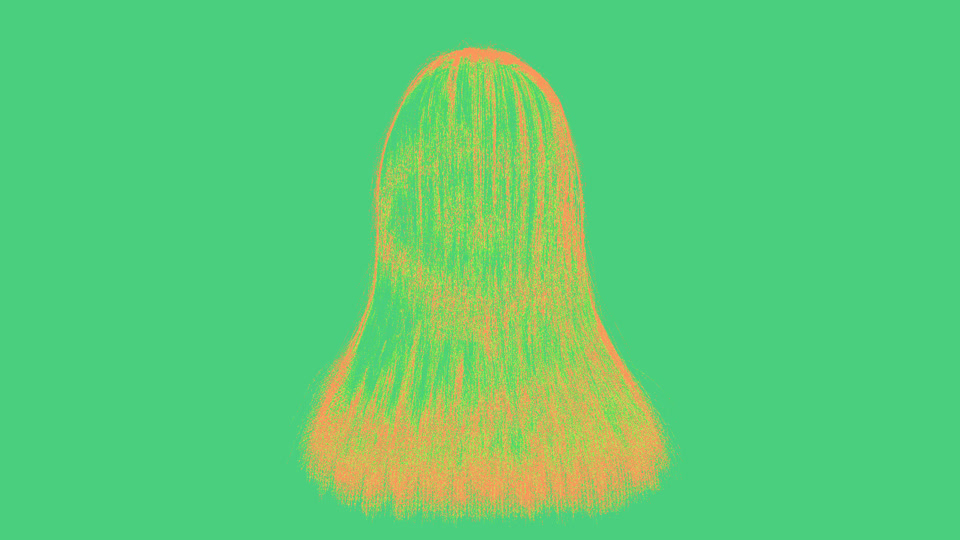}
  {150 25 150 25}
  {150 25 150 25}
  {0.742}%
\hspace{0.004\linewidth}%
\staticresult
  {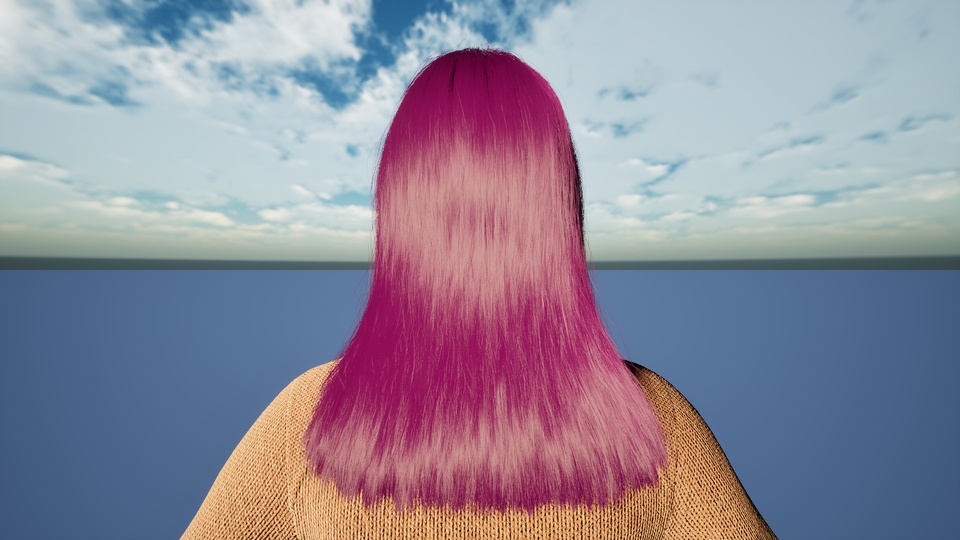}
  {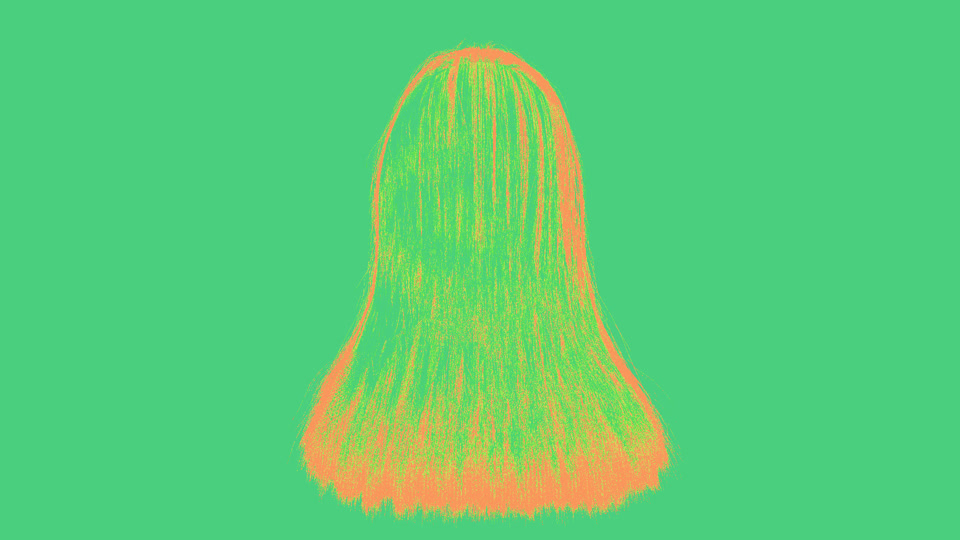}
  {150 25 150 25}
  {150 25 150 25}
  {0.742}%
\hspace{0.004\linewidth}%
\staticresult
  {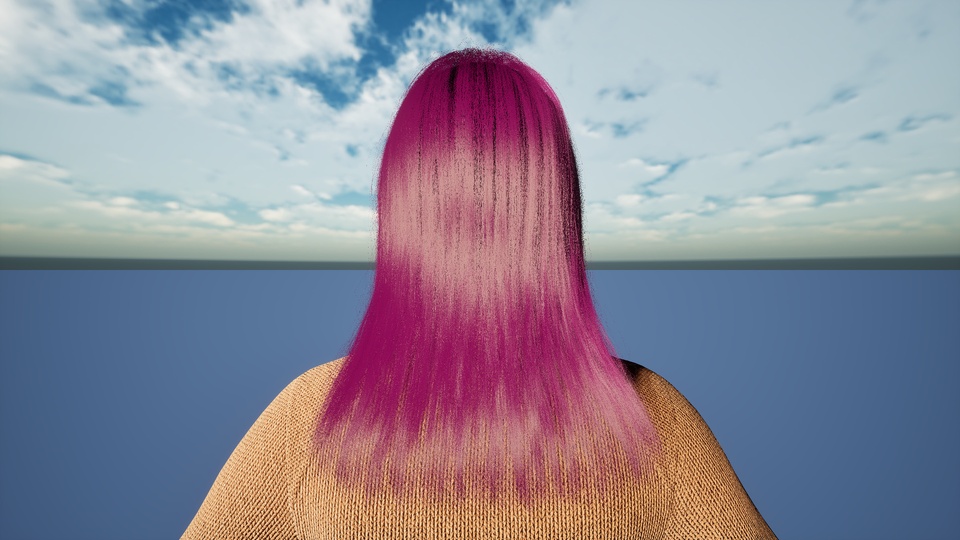}
  {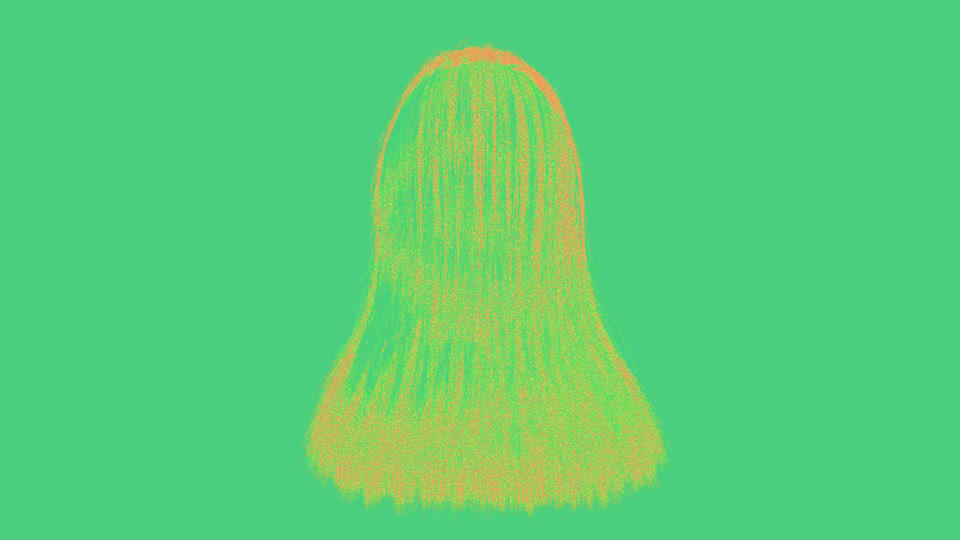}
  {150 25 150 25}
  {150 25 150 25}
  {0.742}%
\hspace{0.004\linewidth}%
\staticresult
  {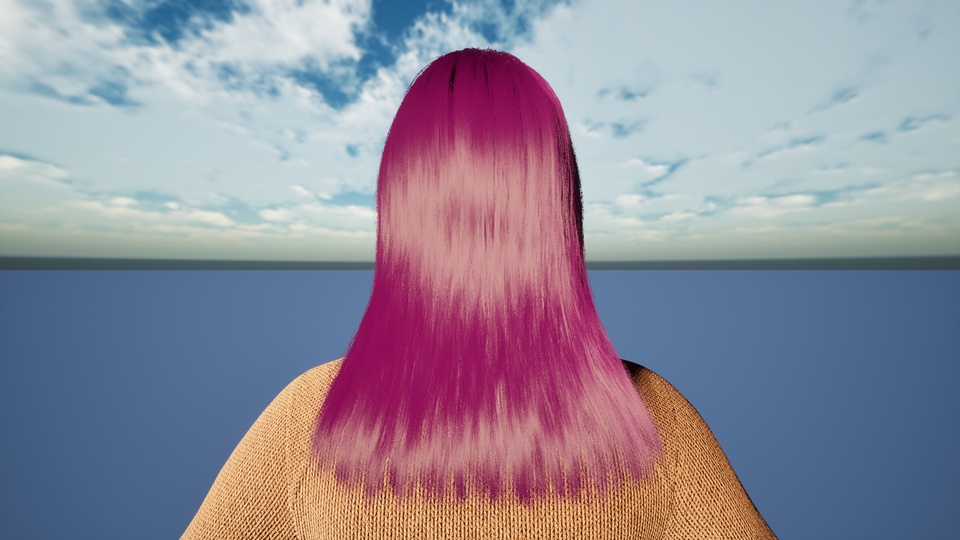}
  {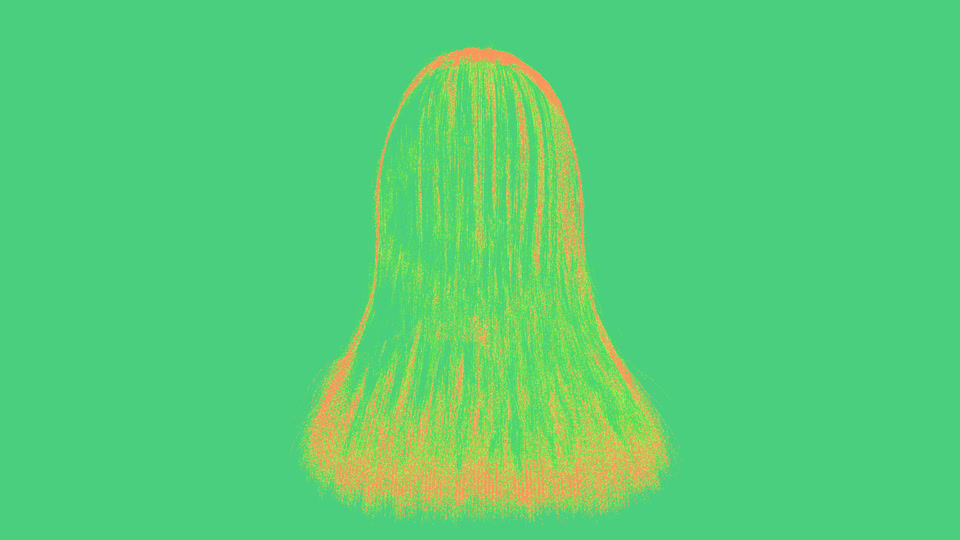}
  {150 25 150 25}
  {150 25 150 25}
  {0.742}%
\hspace{0.004\linewidth}%
\staticresult
  {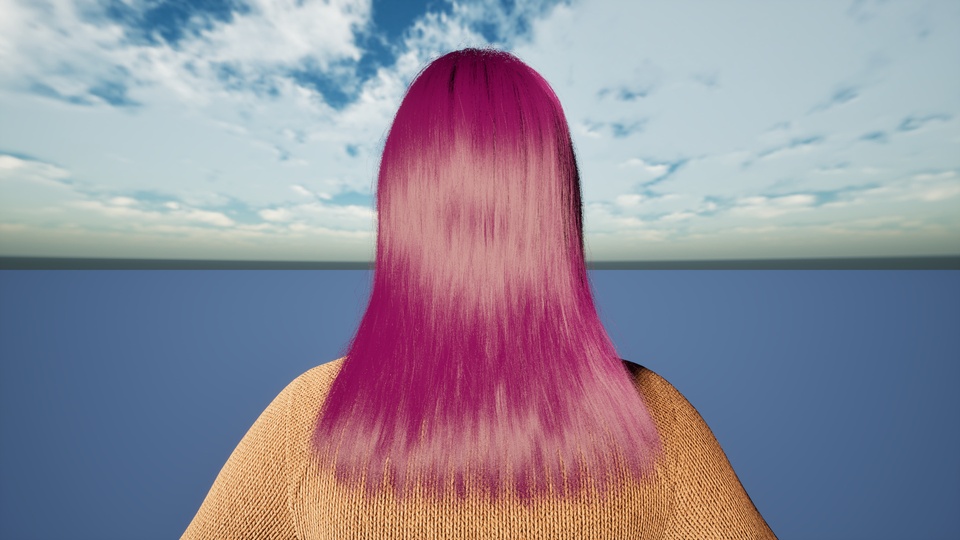}
  {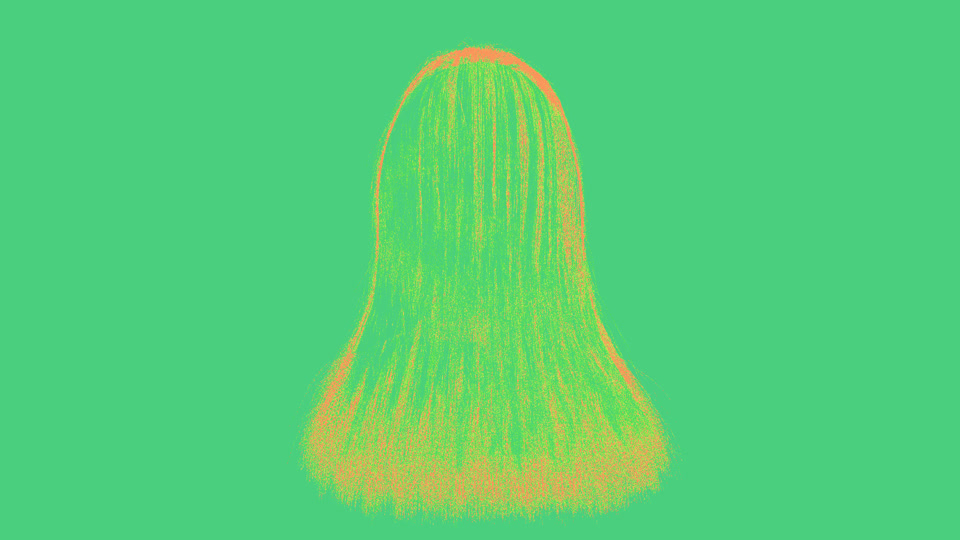}
  {150 25 150 25}
  {150 25 150 25}
  {0.742}%
\hspace{0.004\linewidth}%
\staticresult
  {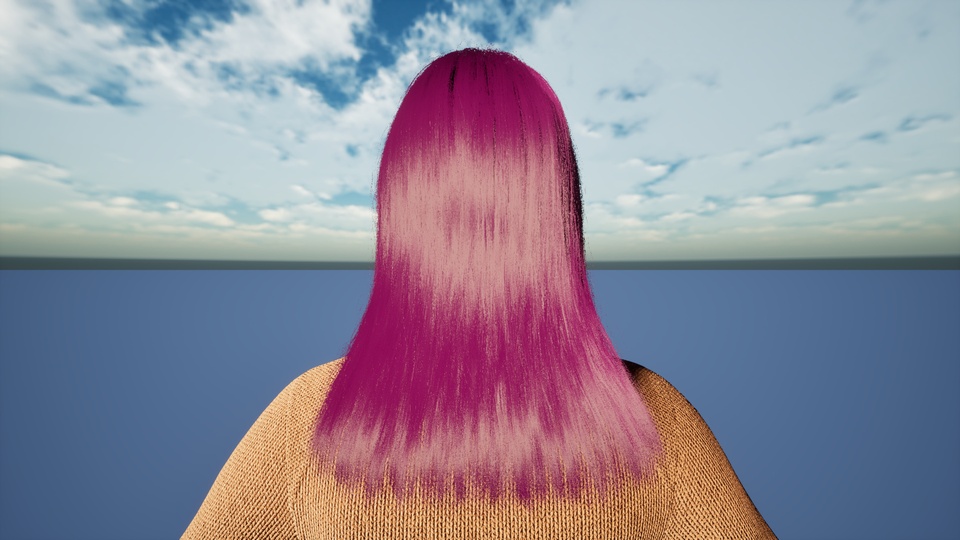}
  {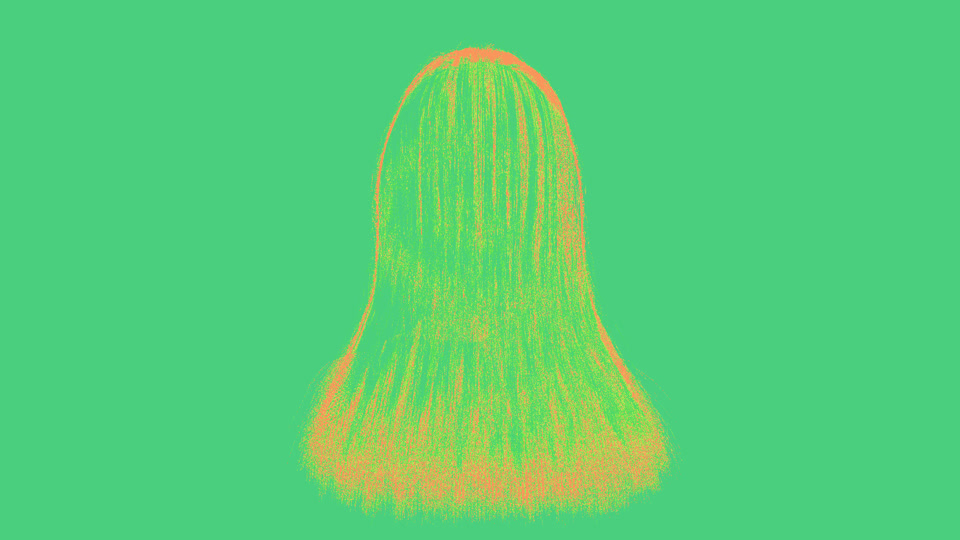}
  {150 25 150 25}
  {150 25 150 25}
  {0.742}%
\hspace{0.004\linewidth}%
\staticresult
  {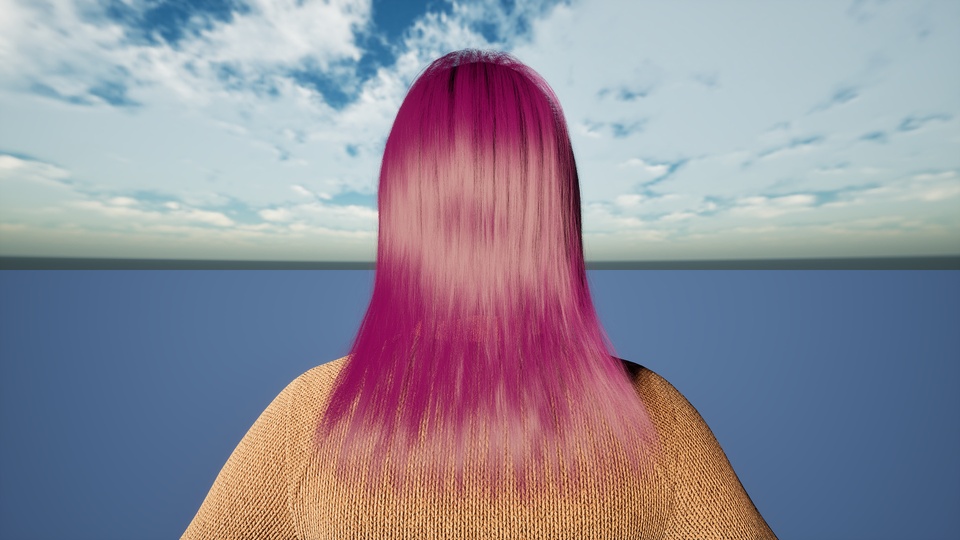}
  {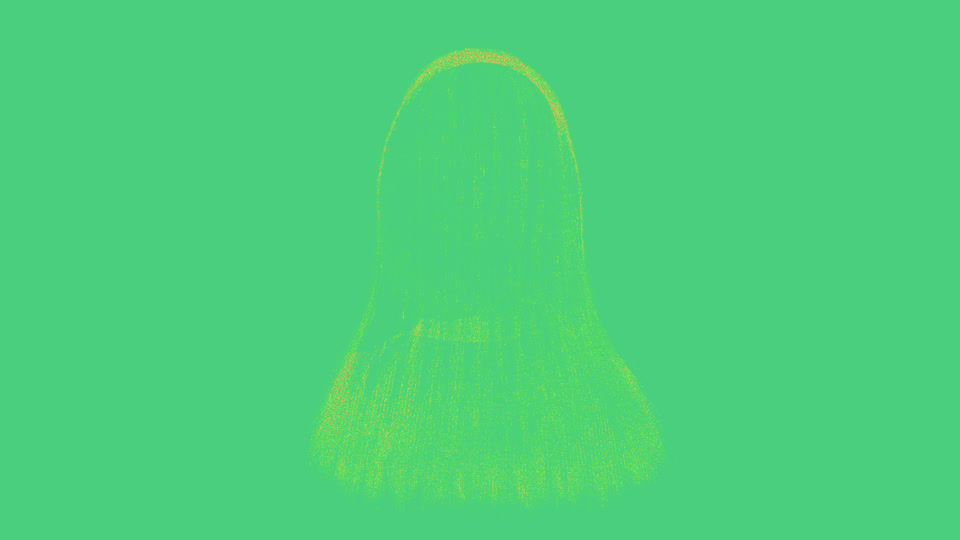}
  {150 25 150 25}
  {150 25 150 25}
  {0.742}%
\hspace{0.004\linewidth}%
\staticref
  {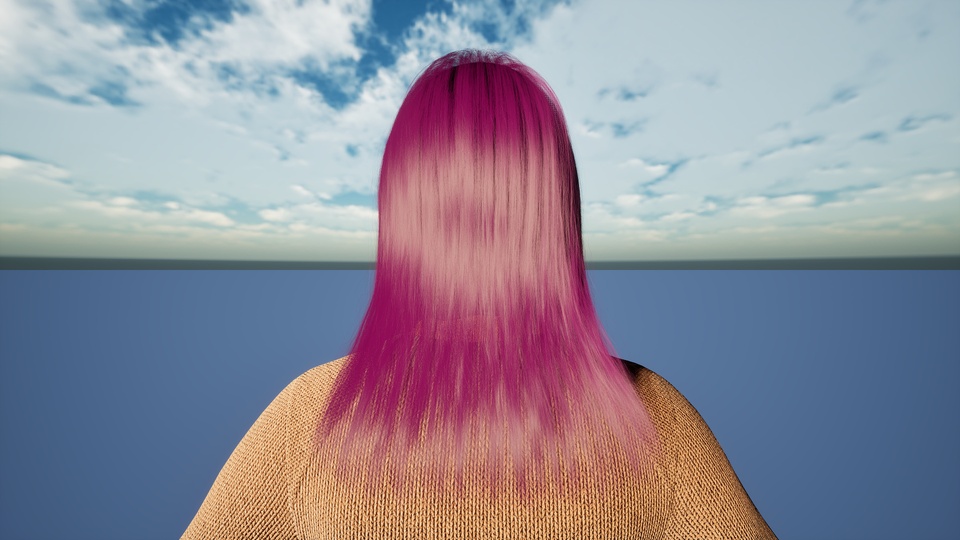}
  {150 25 150 25}
\\[2pt]

\staticresult
  {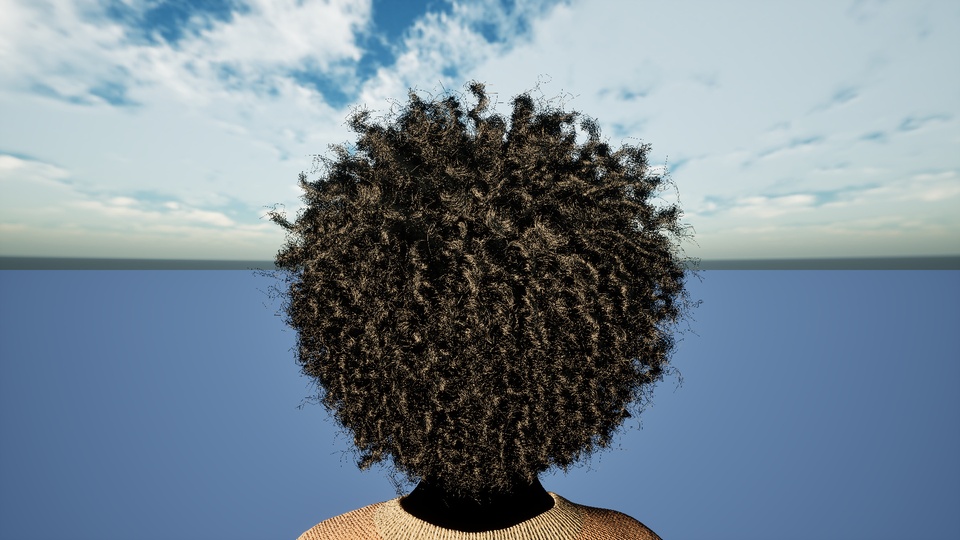}
  {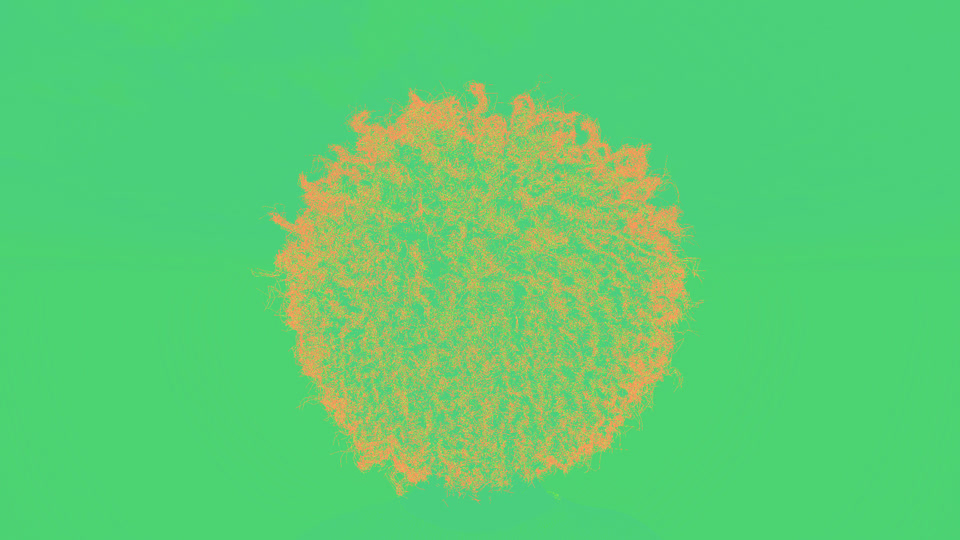}
  {150 25 150 25}
  {150 25 150 25}
  {0.742}%
\hspace{0.004\linewidth}%
\staticresult
  {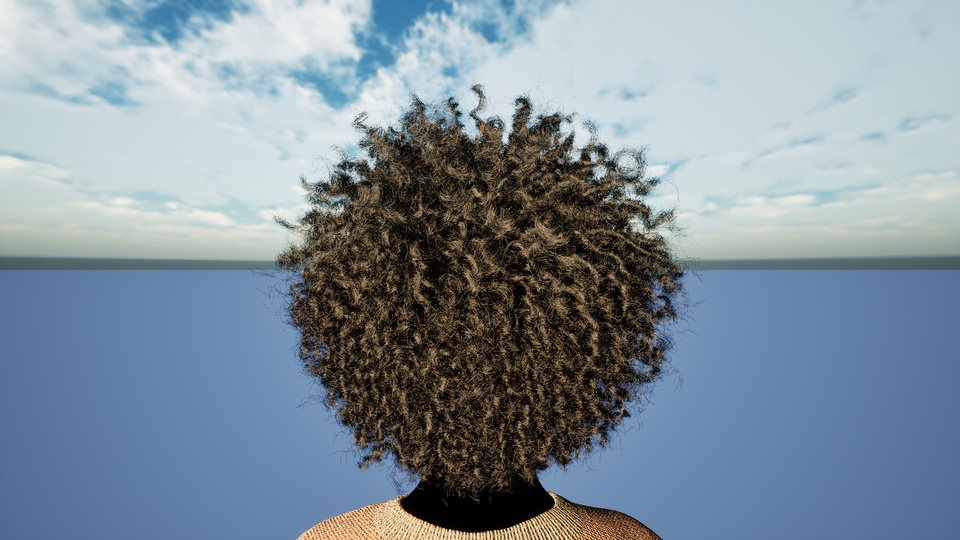}
  {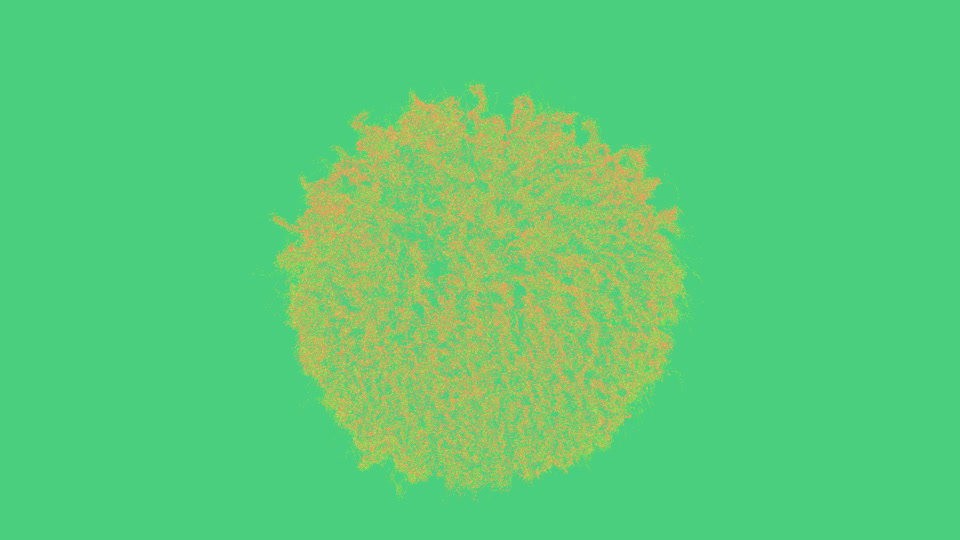}
  {150 25 150 25}
  {150 25 150 25}
  {0.742}%
\hspace{0.004\linewidth}%
\staticresult
  {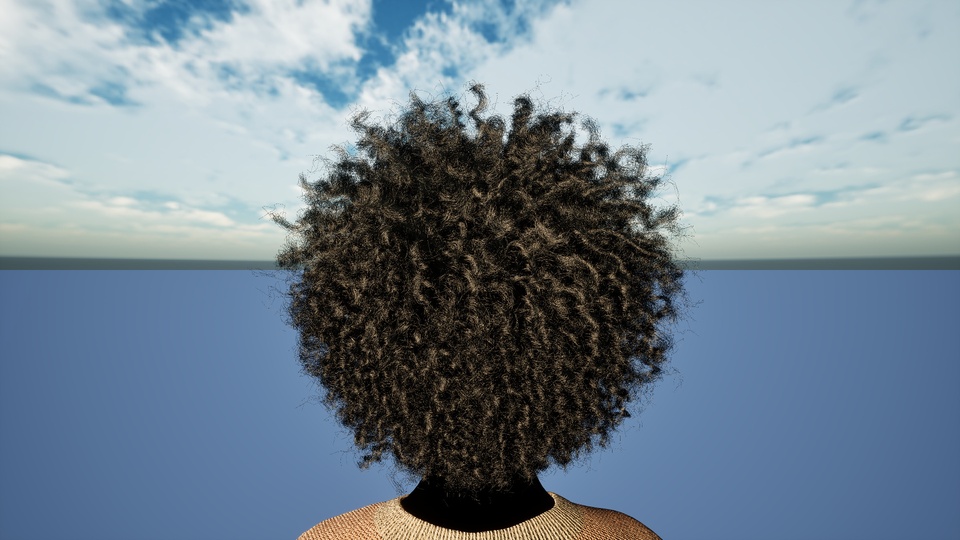}
  {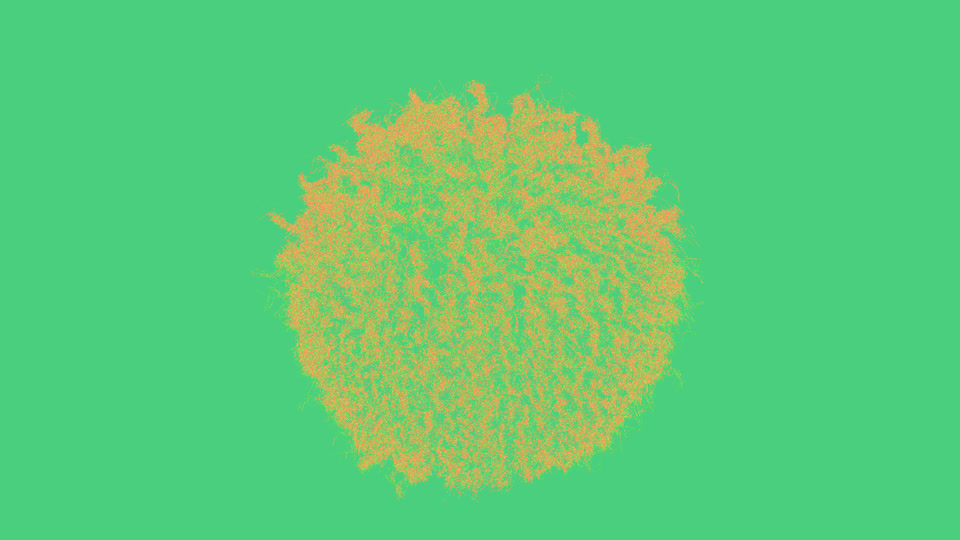}
  {150 25 150 25}
  {150 25 150 25}
  {0.742}%
\hspace{0.004\linewidth}%
\staticresult
  {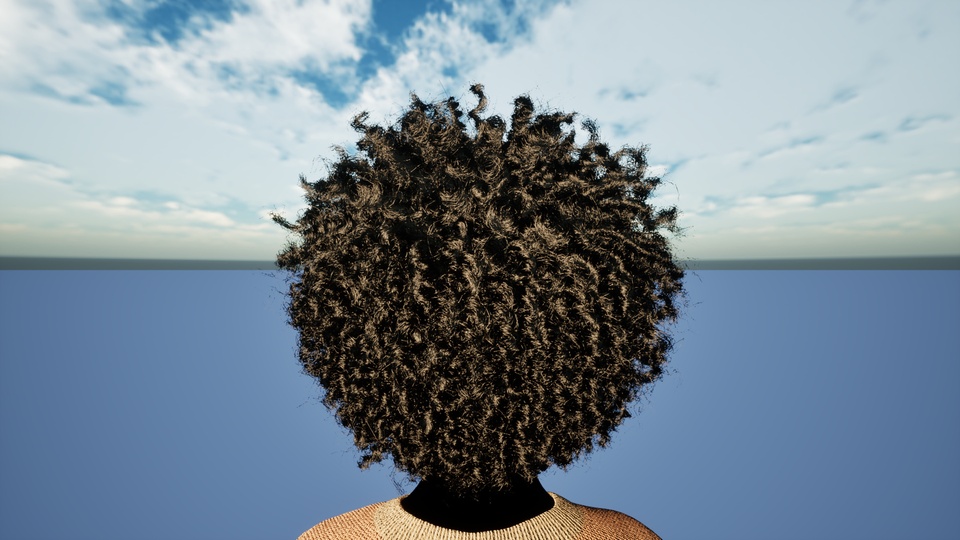}
  {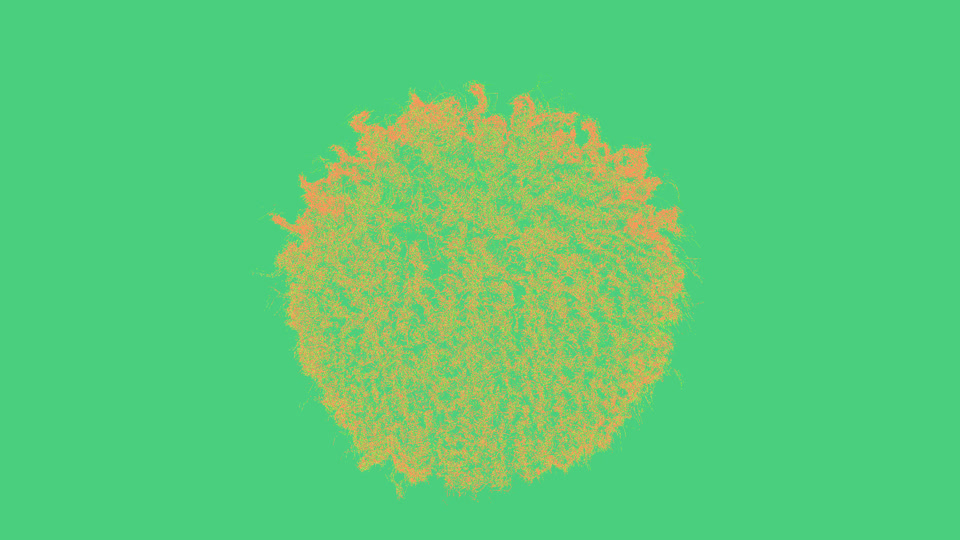}
  {150 25 150 25}
  {150 25 150 25}
  {0.742}%
\hspace{0.004\linewidth}%
\staticresult
  {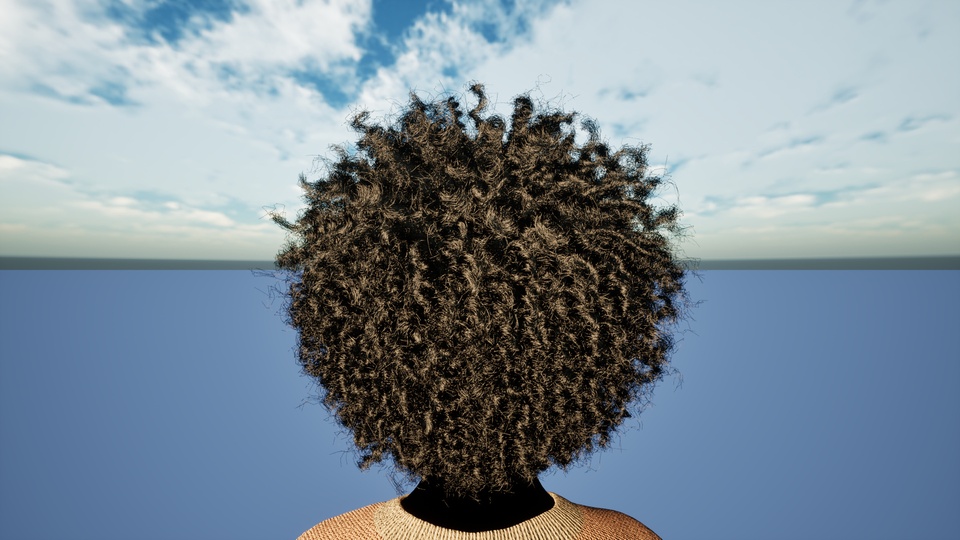}
  {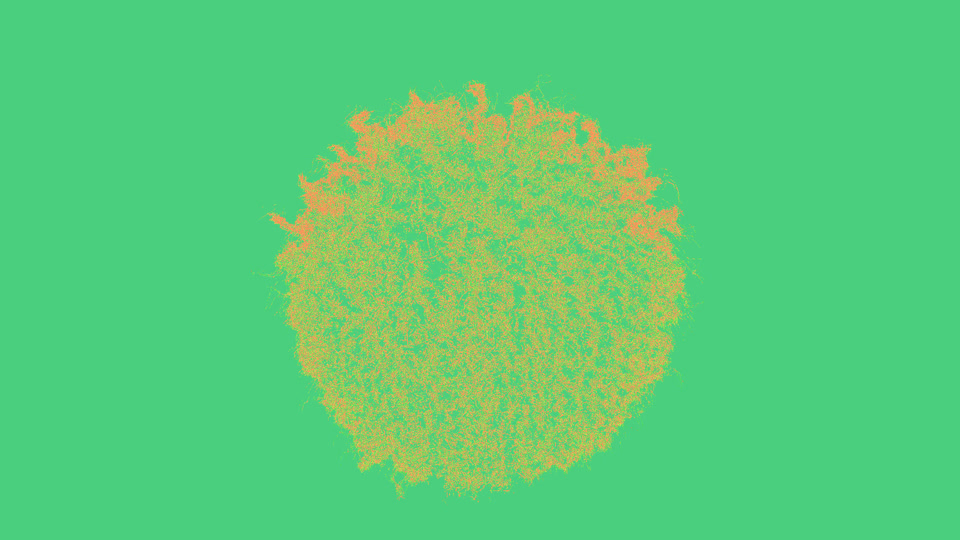}
  {150 25 150 25}
  {150 25 150 25}
  {0.742}%
\hspace{0.004\linewidth}%
\staticresult
  {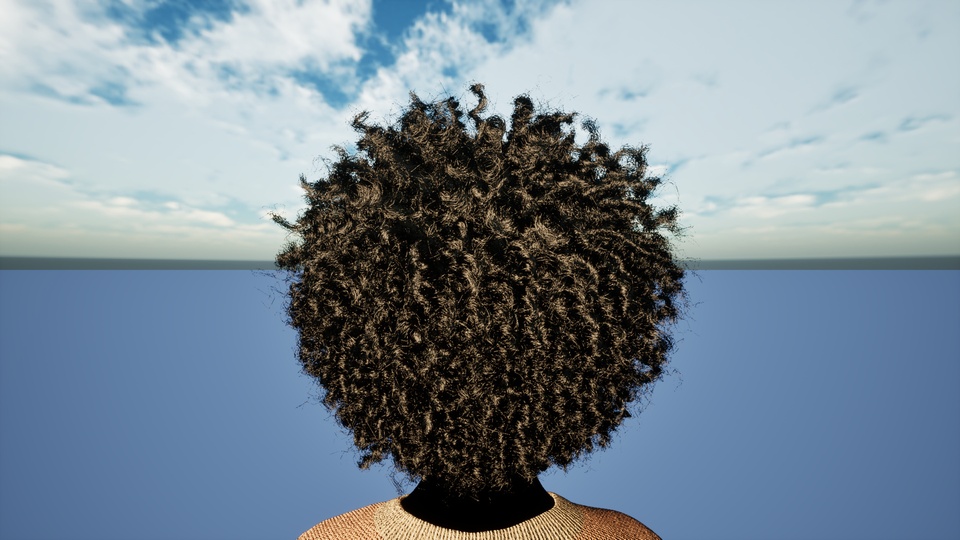}
  {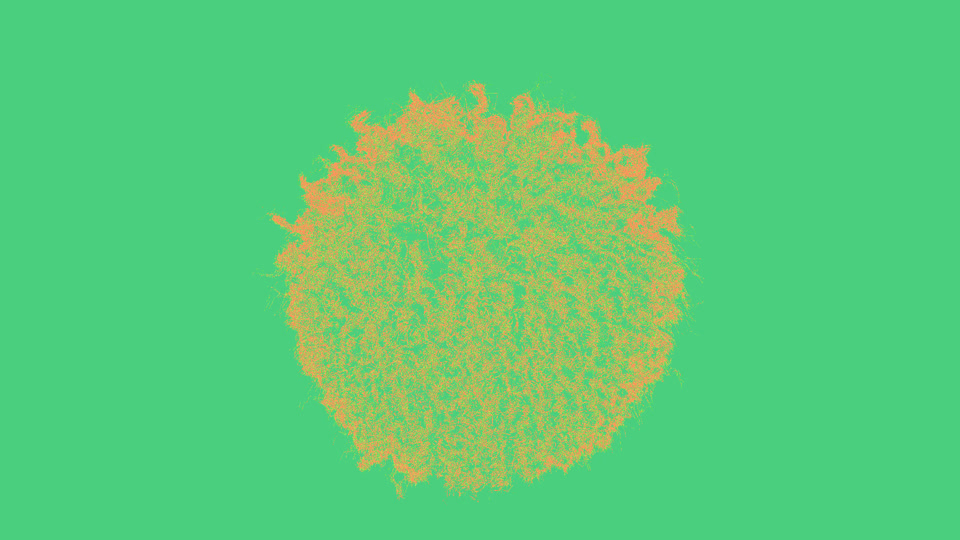}
  {150 25 150 25}
  {150 25 150 25}
  {0.742}%
\hspace{0.004\linewidth}%
\staticresult
  {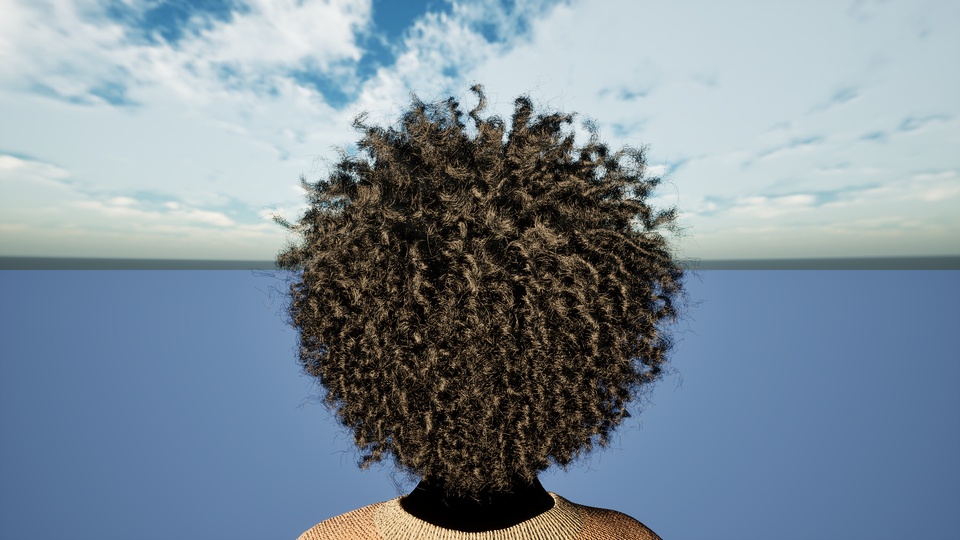}
  {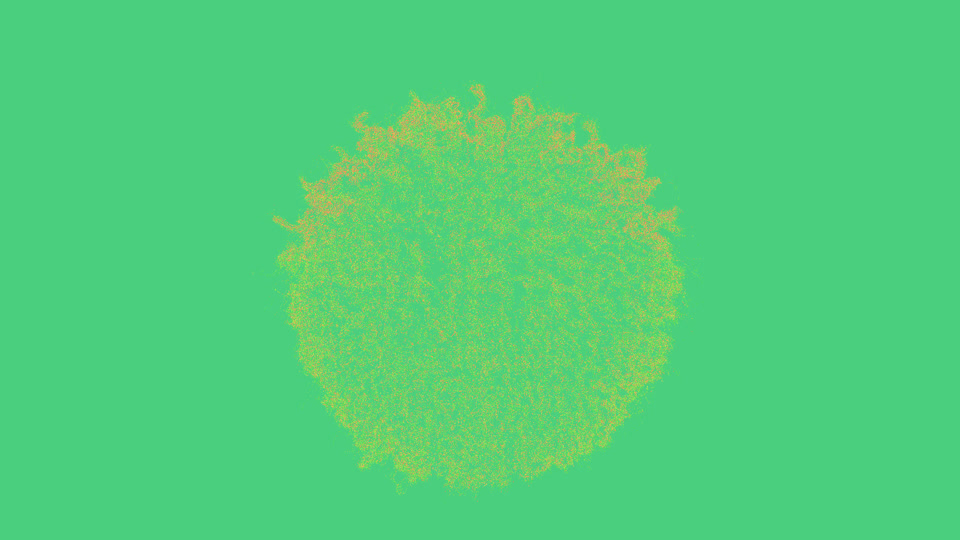}
  {150 25 150 25}
  {150 25 150 25}
  {0.742}%
\hspace{0.004\linewidth}%
\staticref
  {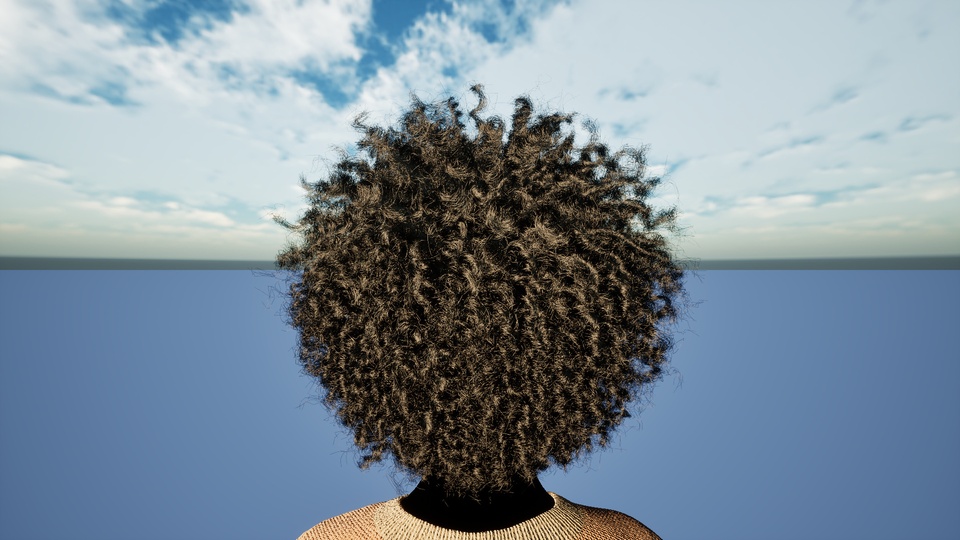}
  {150 25 150 25}

\caption{Comparisons of different static hairstyles. The bottom-left corner shows the error map.}
\Description{}
\label{fig:comparison_static_scene}

\end{figure*}

\paragraph{Comparison with Conventional and Industrial Learning-based Methods} 
We compare our method against conventional anti-aliasing baselines provided by Unreal Engine~5, including 8-spp MSAA, $4\times$ SSAA, and TAA. MSAA uses 8 samples per pixel. SSAA renders at $2\times$ resolution in both width and height and then downsamples the result to the target resolution. TAA accumulates 8 jittered frames using motion-vector-based history reprojection and default history clamping. We also compare our method against industrial learning-based methods, including DLSS~4~\cite{nvidia2025dlss4} and FSR~\cite{amd_fsr}, integrated as plugins in Unreal Engine~5. For DLSS, we use DLAA mode without Ray Reconstruction or Frame Generation. For FSR, we use Native AA mode without sharpening or Frame Generation. We report MSE, PSNR, SSIM, LPIPS, and FLIP in~\autoref{tab:static} for the static hairstyle shown in~\autoref{fig:comparison_static_scene}. All metrics are computed only on pixels corresponding to hair in the reference image. Our method consistently outperforms all competing methods across all quantitative metrics. In terms of performance, our method is comparable to heavily engineered industrial solutions while requiring only 0.8\,ms per frame, making it well-suited for practical use.

\begin{table}[hbt]
\centering
\small
\caption{Quantitative comparison on static hairstyles shown in~\autoref{fig:comparison_static_scene}. 
} 
\label{tab:static}
\renewcommand{\tabcolsep}{0.15cm}
\begin{tabular}{lcccccc}
\toprule
Method 
& MSE$\downarrow$ 
& PSNR$\uparrow$ 
& SSIM$\uparrow$ 
& LPIPS$\downarrow$ 
& FLIP$\downarrow$
& Time$\downarrow$ (ms) \\
\midrule
Input & 0.0318 & 14.97 & 0.8845 & 0.1200 & 0.0338 & -- \\
MSAA  & 0.0065 & 21.89 & 0.9264 & 0.0578 & 0.0168 & 2.0 \\
SSAA  & 0.0177 & 17.52 & 0.9000 & 0.0932 & 0.0291 & 3.1 \\
TAA   & 0.0102 & 19.92 & 0.9347 & 0.0591 & 0.0283 & 0.1 \\
FSR   & 0.0124 & 19.07 & 0.9218 & 0.0772 & 0.0289 & 0.5 \\
DLSS  & 0.0098 & 20.08 & 0.9338 & 0.0583 & 0.0289 & 0.5 \\
Ours  & \textbf{0.0018} & \textbf{27.43} & \textbf{0.9659} & \textbf{0.0206} & \textbf{0.0073} & 0.8 \\
\bottomrule
\end{tabular}
\end{table}

\begin{figure}[htb]
\centering

\begin{minipage}[t]{0.34\linewidth}
  \vspace{0pt}
  \centering

  \small Input (whole view)\\[2pt]

  \includegraphics[
    trim={430 0 430 0},
    clip,
    width=\linewidth
  ]{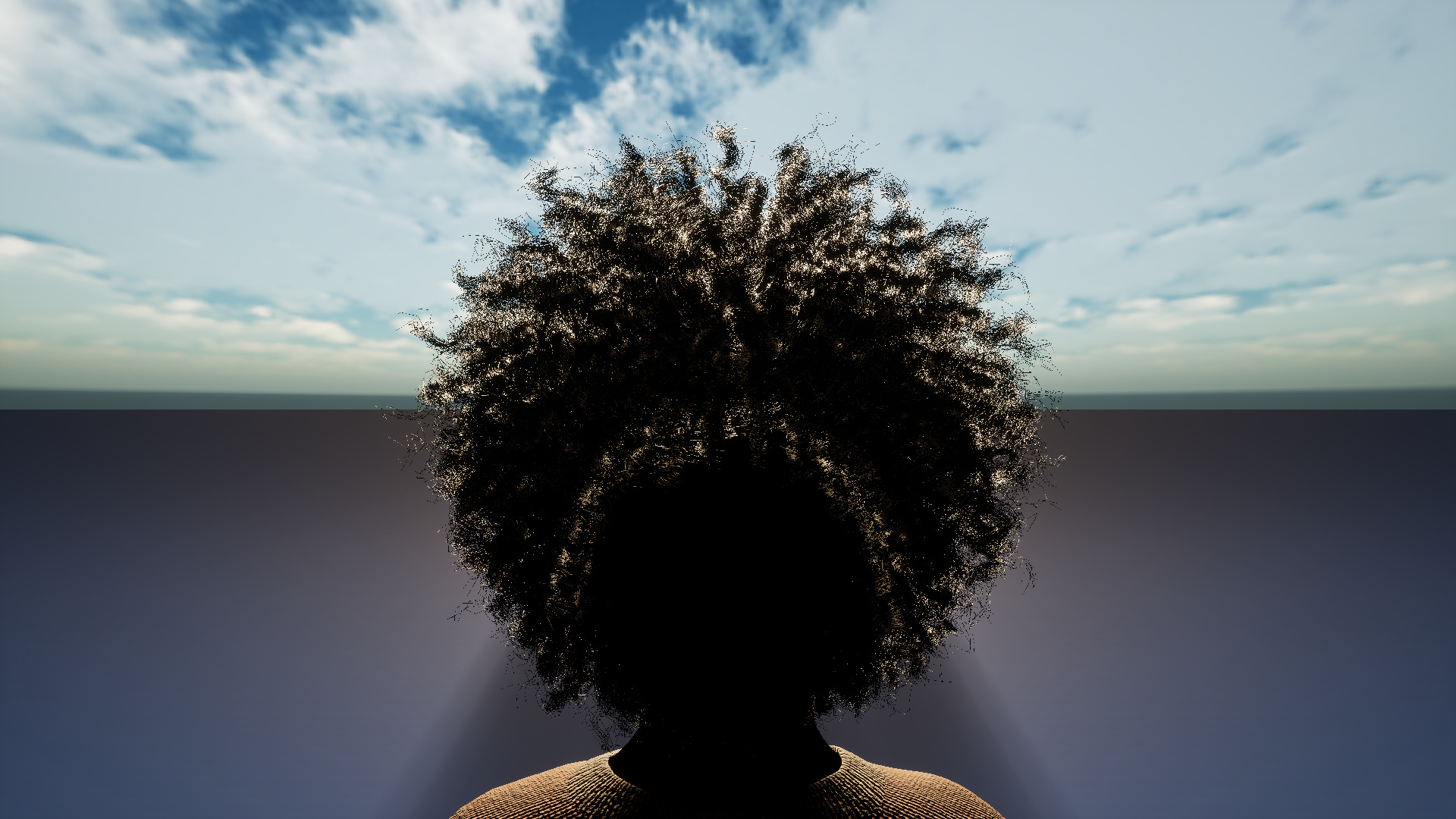}

\end{minipage}%
\hspace{0.015\linewidth}%
\begin{minipage}[t]{0.63\linewidth}
  \vspace{0pt}
  \centering

  \begin{minipage}{0.31\linewidth}
    \centering\small Input
  \end{minipage}%
  \hspace{0.015\linewidth}%
  \begin{minipage}{0.31\linewidth}
    \centering\small TAA
  \end{minipage}%
  \hspace{0.015\linewidth}%
  \begin{minipage}{0.31\linewidth}
    \centering\small DLSS
  \end{minipage}

  \vspace{2pt}

  \includegraphics[
    trim={1025 664 675 263},
    clip,
    width=0.31\linewidth
  ]{Fig_extremely_curly/spp1/Human_afro.0180.jpg}%
  \hspace{0.015\linewidth}%
  \includegraphics[
    trim={1025 664 675 263},
    clip,
    width=0.31\linewidth
  ]{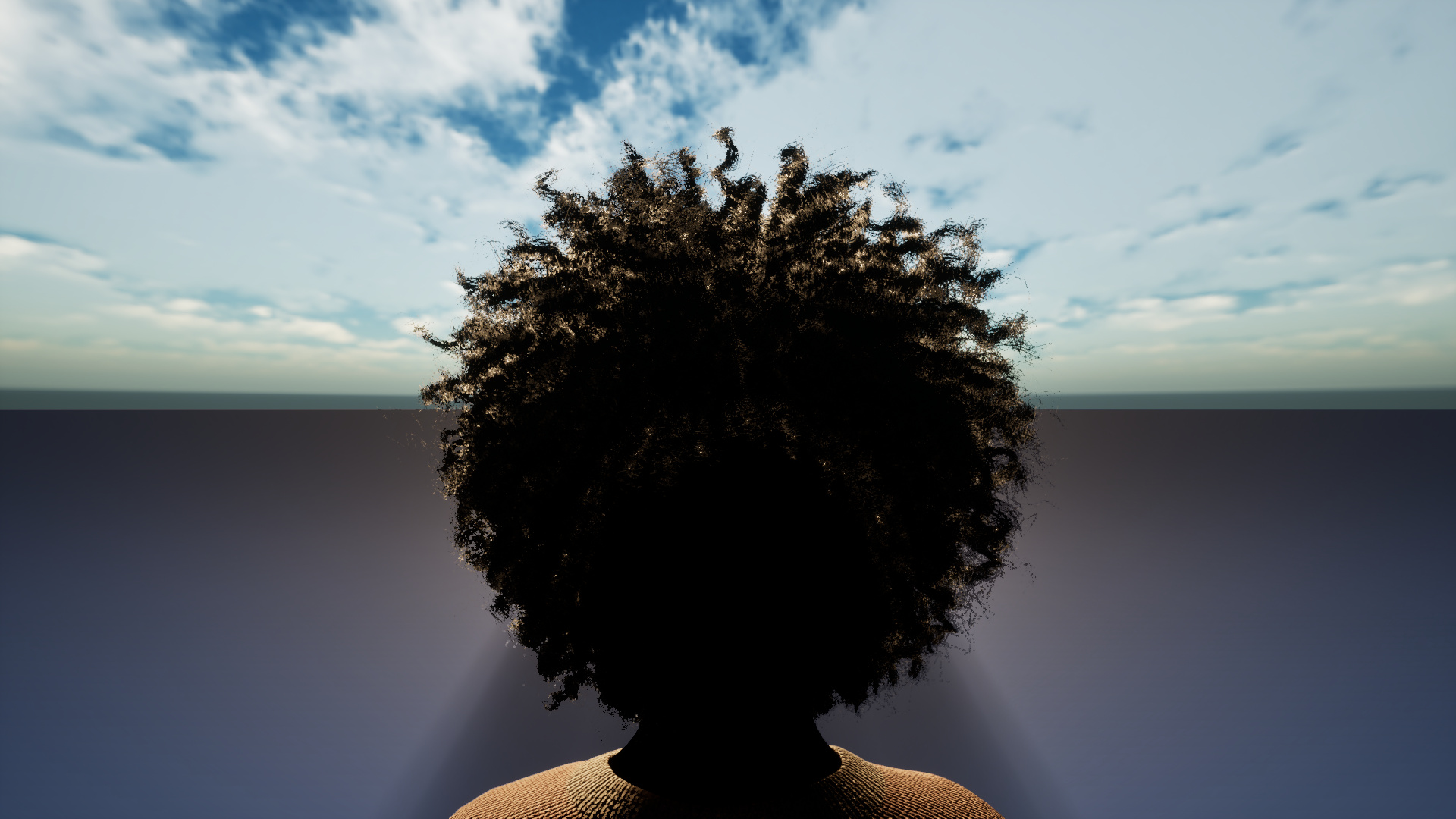}%
  \hspace{0.015\linewidth}%
  \includegraphics[
    trim={1025 664 675 263},
    clip,
    width=0.31\linewidth
  ]{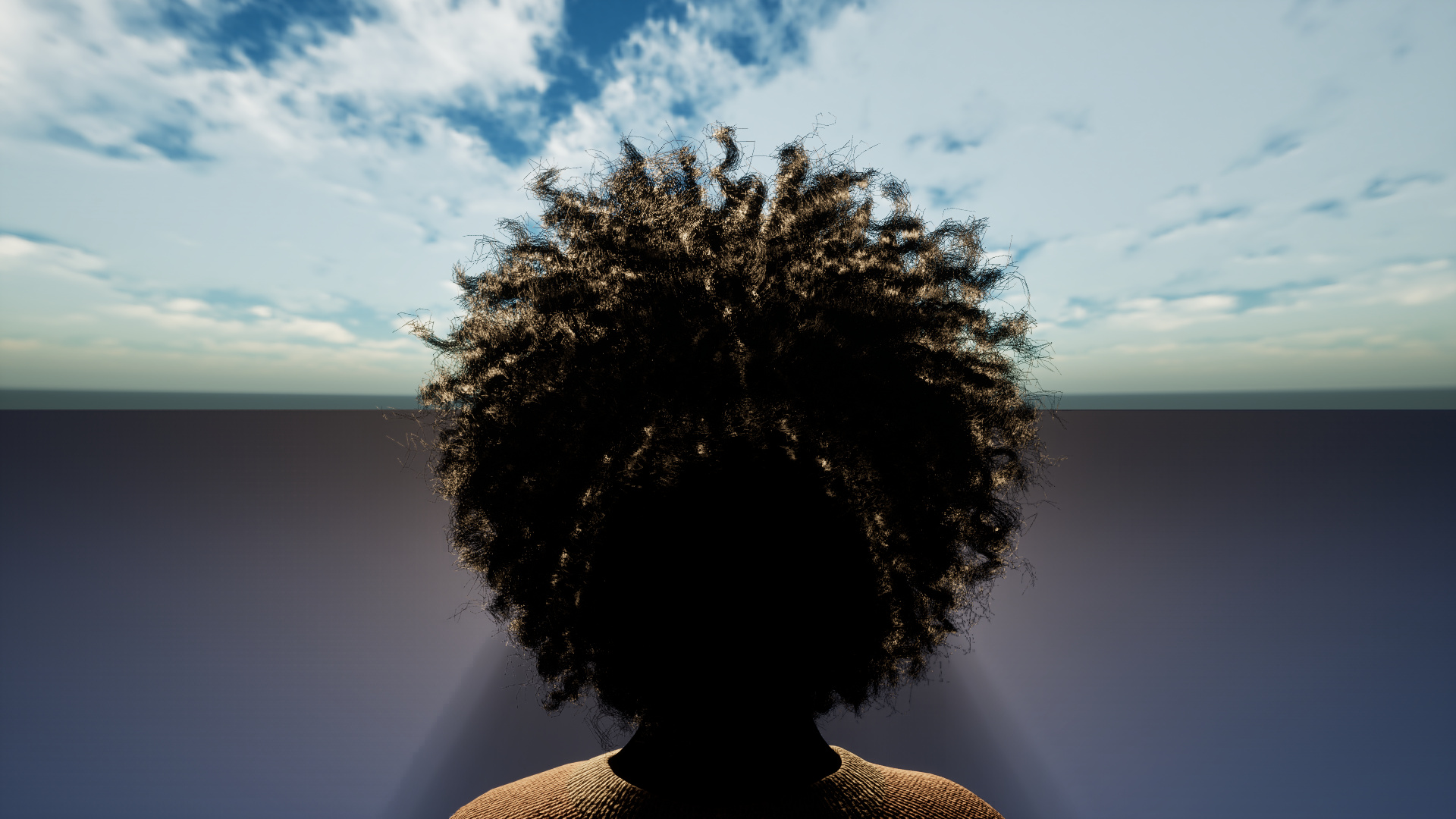}

  \vspace{4pt}

  \begin{minipage}{0.31\linewidth}
    \centering\small FSR
  \end{minipage}%
  \hspace{0.015\linewidth}%
  \begin{minipage}{0.31\linewidth}
    \centering\small Ours
  \end{minipage}%
  \hspace{0.015\linewidth}%
  \begin{minipage}{0.31\linewidth}
    \centering\small Ref
  \end{minipage}

  \vspace{2pt}

  \includegraphics[
    trim={1025 664 675 263},
    clip,
    width=0.31\linewidth
  ]{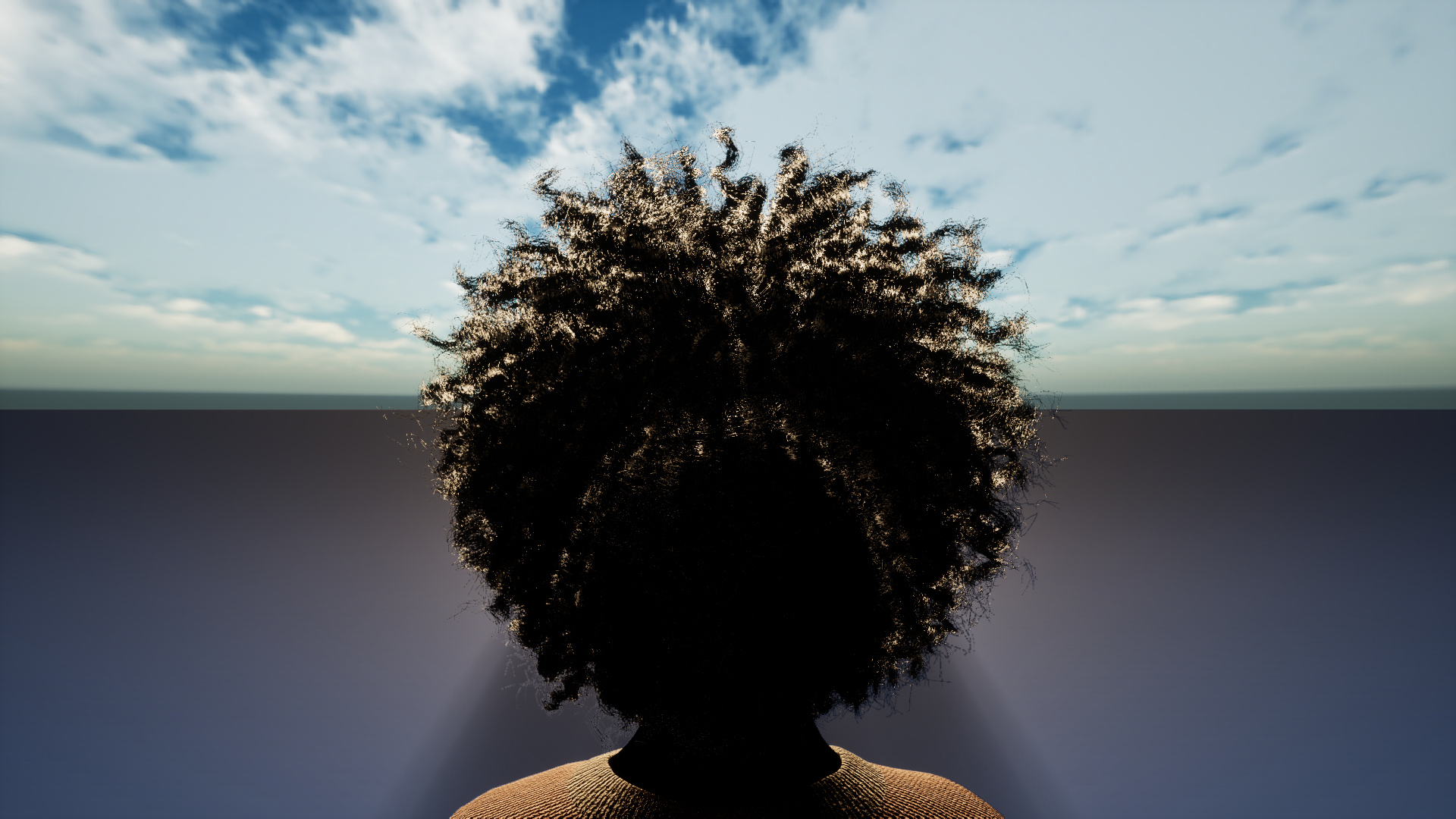}%
  \hspace{0.015\linewidth}%
  \includegraphics[
    trim={1025 664 675 263},
    clip,
    width=0.31\linewidth
  ]{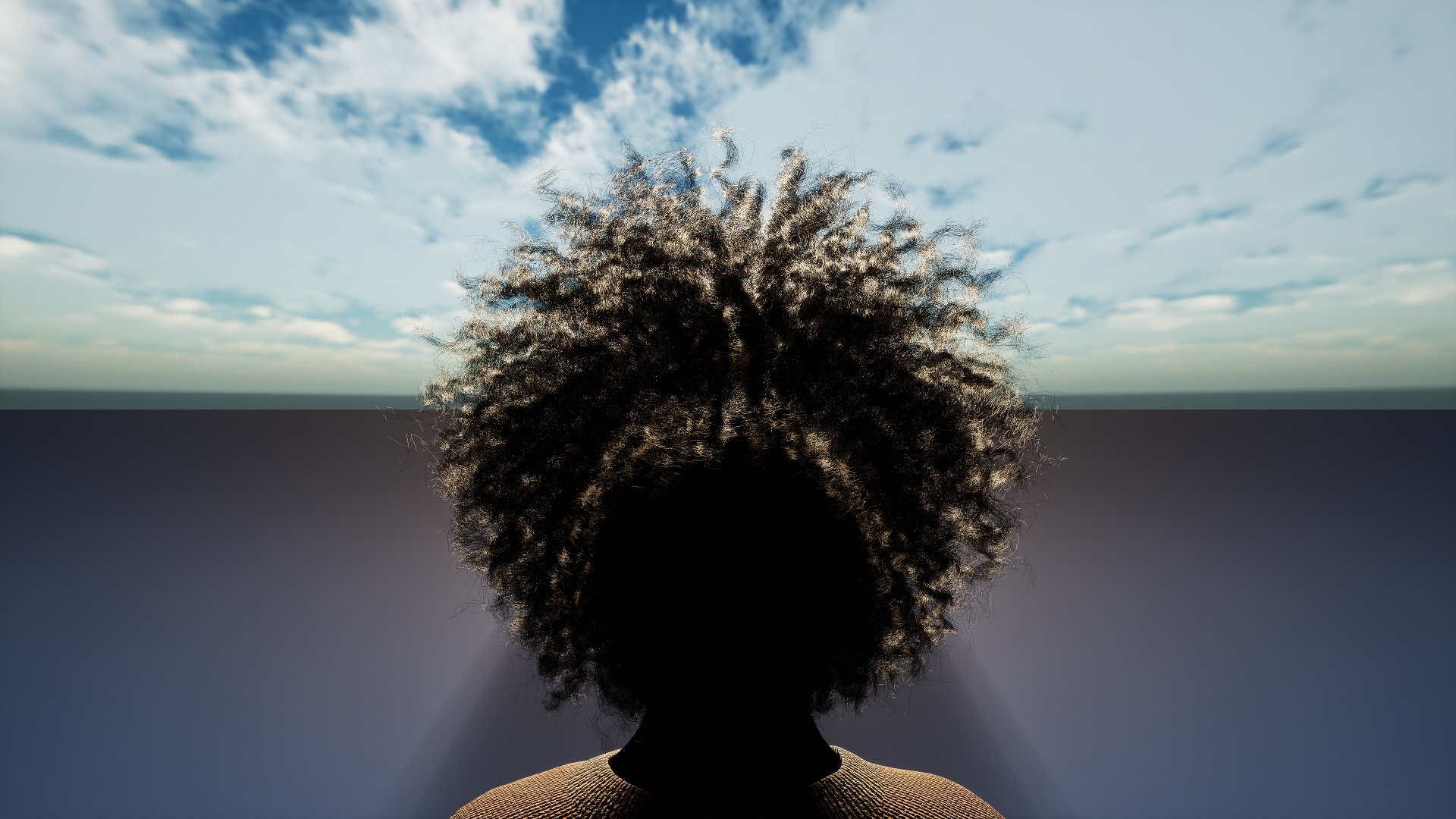}%
  \hspace{0.015\linewidth}%
  \includegraphics[
    trim={1025 664 675 263},
    clip,
    width=0.31\linewidth
  ]{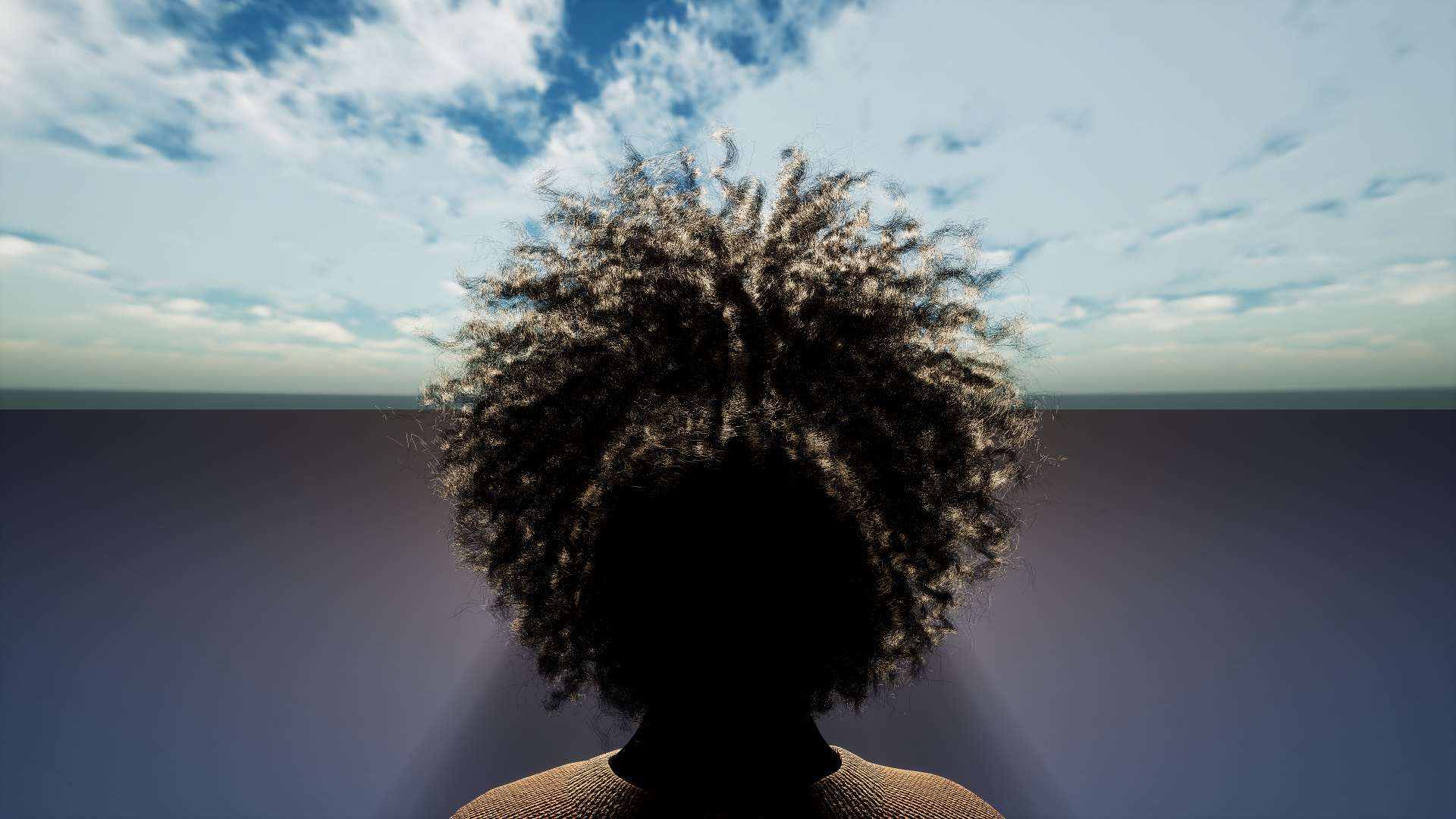}

\end{minipage}

\caption{Zoom-in comparison of extremely curly and highly tangled hairstyle under backlighting. These results demonstrate the benefit of reconstructing geometry-related G-buffers before shading, rather than directly filtering the final RGB image.}
\Description{}
\label{fig:extremely_curly}
\end{figure}

\paragraph{Extremely curved hairstyle}
We highlight the advantage of our geometry-aware G-buffer-based formulation on extremely curved or highly tangled hairstyles, such as the afro hair shown in \autoref{fig:extremely_curly} and the supplementary video. 

These hairstyles contain rapidly changing strand directions, dense overlaps, and complex silhouettes, making them particularly challenging for RGB-space filtering methods. 
Since the final RGB image entangles coverage, visibility, depth ordering, shadows, and multiple scattering, directly filtering RGB values tends to blur fine curved strands or mix shading information from different depth layers. 
In contrast, our method performs reconstruction in G-buffer space before deferred shading. 
By first reconstructing coverage and tangent and then using them to guide position completion, our method preserves local strand directionality and maintains better consistency among coverage, tangent, and position. 
This geometry-aware reconstruction leads to clearer strand boundaries and more stable shading than RGB-based methods in highly curved hair regions.

\paragraph{Comparison with Academic Hair-specialized Methods}
A recent neural filtering method for hair rendering~\cite{Currius2022} addresses a different problem from ours. Their method removes stochastic sampling noise introduced by stochastic transparency, whereas our method addresses artifacts caused by undersampling of hair geometry, where insufficient rasterization samples lead to missing or unreliable G-buffer information. Therefore, the two methods target fundamentally different sources of rendering artifacts and are not directly comparable. We also note a recent real-time hair filtering method by \citet{Huang2025filter}. Unfortunately, its implementation is not publicly available. According to the paper, their method requires 8\,ms per frame for hair containing 50{,}000 strands at a screen resolution of $1920\times1080$. However, the paper does not report the hardware specifications or quantitative evaluation metrics. Under the same setup, their reported runtime is roughly $10\times$ slower than ours, which is reasonable because their analytic hair-linking approach requires explicit pixel--hair intersection tests.

\begin{figure*}[htb]
\centering

\newcommand{\dynamichead}[1]{%
  \begin{minipage}{0.143\linewidth}
    \centering
    \small #1
  \end{minipage}%
}

\newcommand{\dynamicrowlabel}[1]{%
  \begin{minipage}[b][0.101\linewidth][c]{0.03\linewidth}
    \centering
    \rotatebox[origin=c]{90}{\small #1}
  \end{minipage}%
}

%
\newcommand{\dynamicresult}[2]{%
  \begingroup
  \setlength{\unitlength}{0.143\linewidth}%
  \begin{picture}(1,0.705)

    \put(0,0){%
      \includegraphics[
        trim={150 37.5 150 37.5},
        clip,
        width=\unitlength
      ]{#1}%
    }%

    \put(0,0){%
      \includegraphics[
        trim={150 37.5 150 37.5},
        clip,
        width=0.40\unitlength
      ]{#2}%
    }%

  \end{picture}%
  \endgroup
}

\newcommand{\dynamicref}[1]{%
  \includegraphics[
    trim={150 37.5 150 37.5},
    clip,
    width=0.143\linewidth
  ]{#1}%
}

\begin{minipage}{0.03\linewidth}
\mbox{}
\end{minipage}%
\hspace{0.008\linewidth}%
\dynamichead{Input}%
\hspace{0.008\linewidth}%
\dynamichead{TAA}%
\hspace{0.008\linewidth}%
\dynamichead{DLSS}%
\hspace{0.008\linewidth}%
\dynamichead{FSR}%
\hspace{0.008\linewidth}%
\dynamichead{Ours}%
\hspace{0.008\linewidth}%
\dynamichead{Reference}%
\\[1pt]

\dynamicrowlabel{Frame 0}%
\hspace{0.008\linewidth}%
\dynamicresult
  {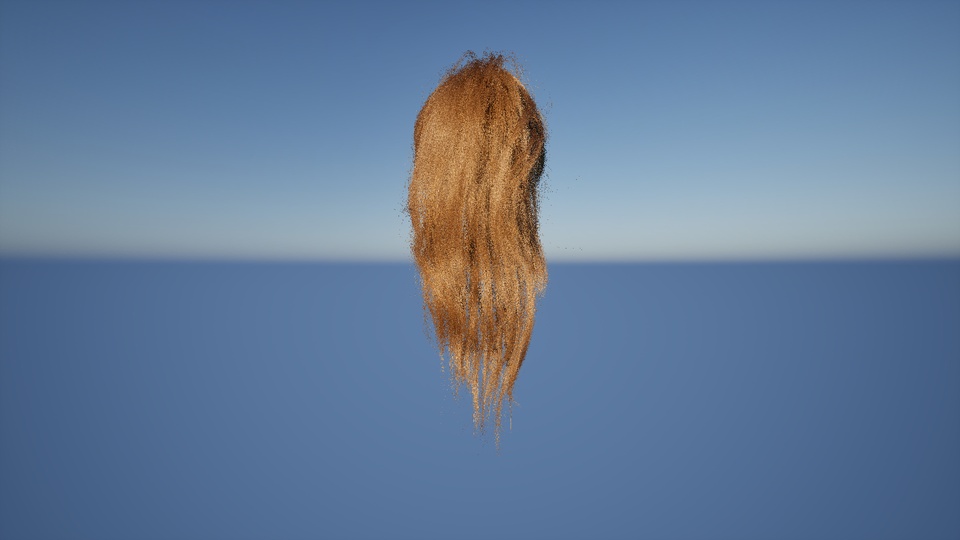}
  {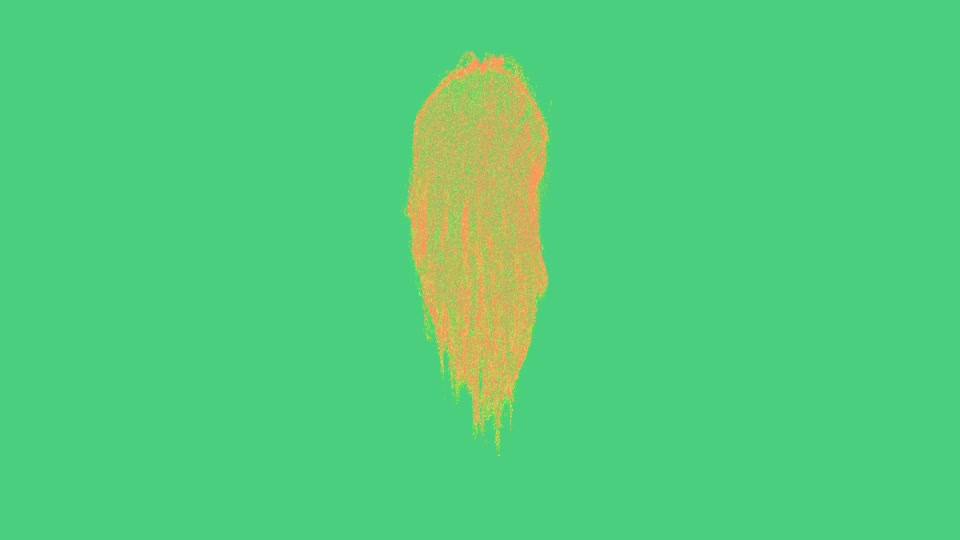}%
\hspace{0.008\linewidth}%
\dynamicresult
  {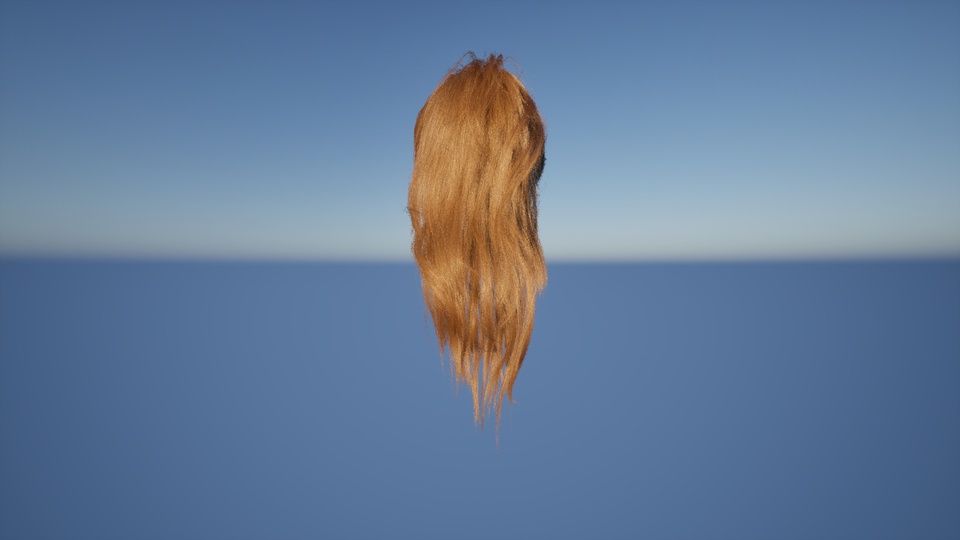}
  {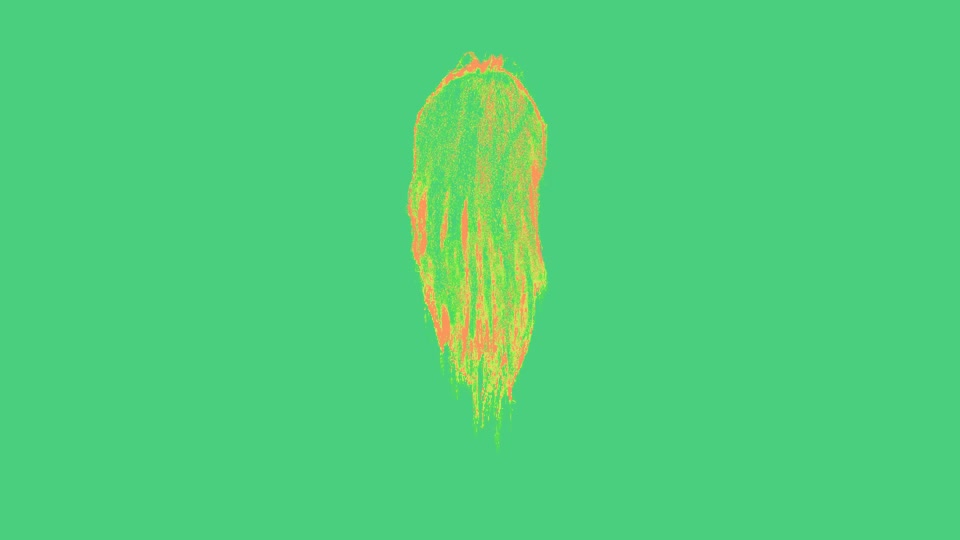}%
\hspace{0.008\linewidth}%
\dynamicresult
  {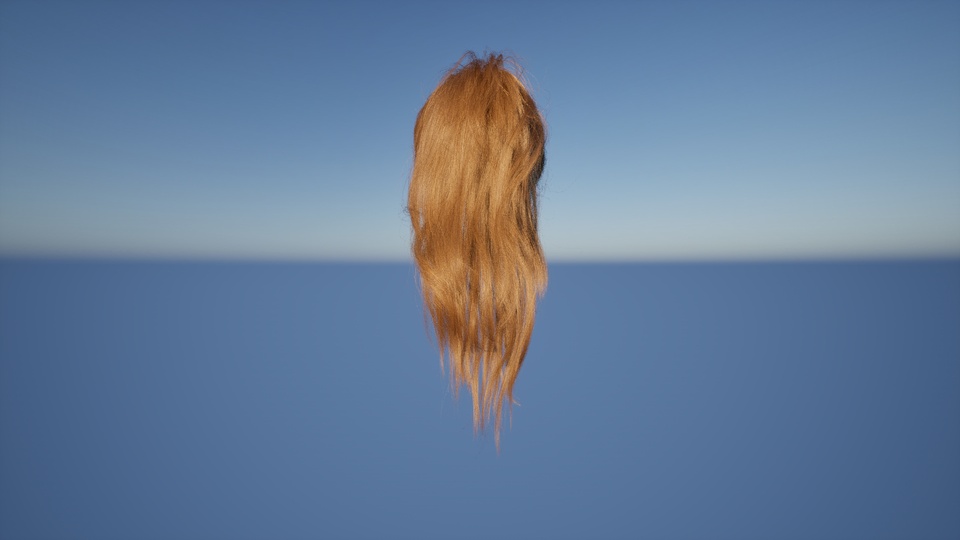}
  {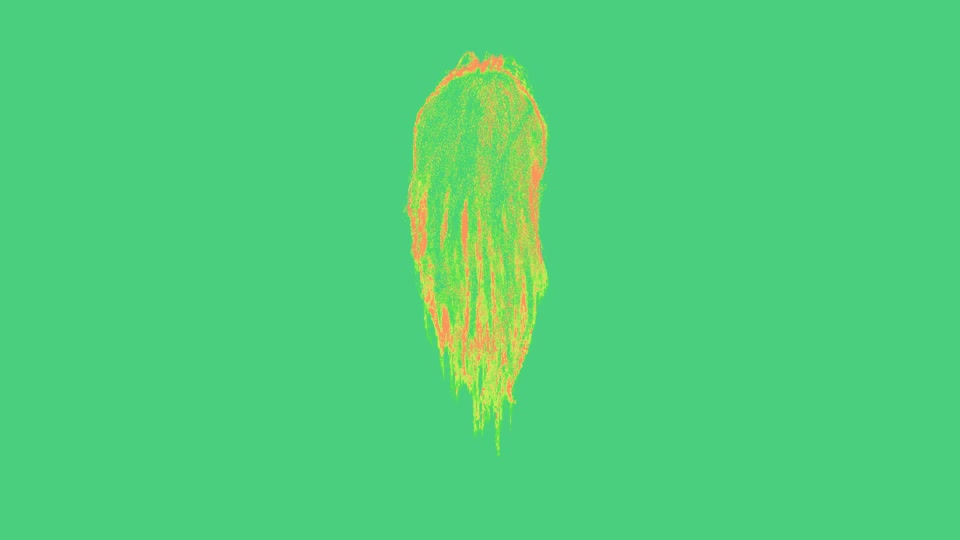}%
\hspace{0.008\linewidth}%
\dynamicresult
  {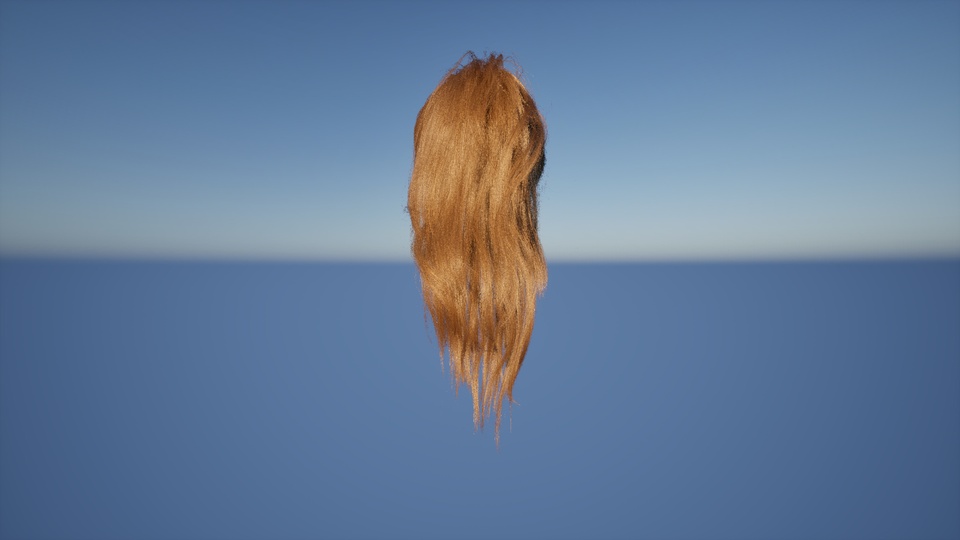}
  {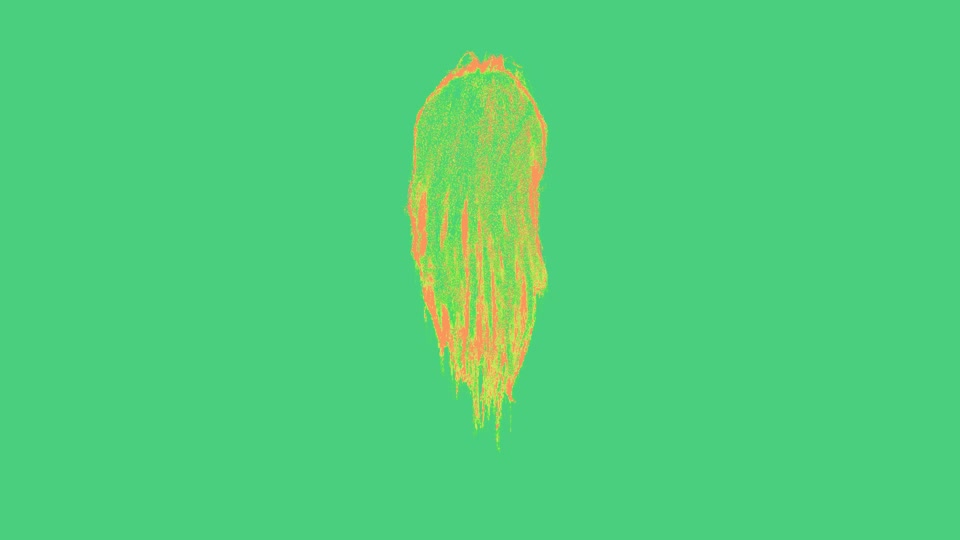}%
\hspace{0.008\linewidth}%
\dynamicresult
  {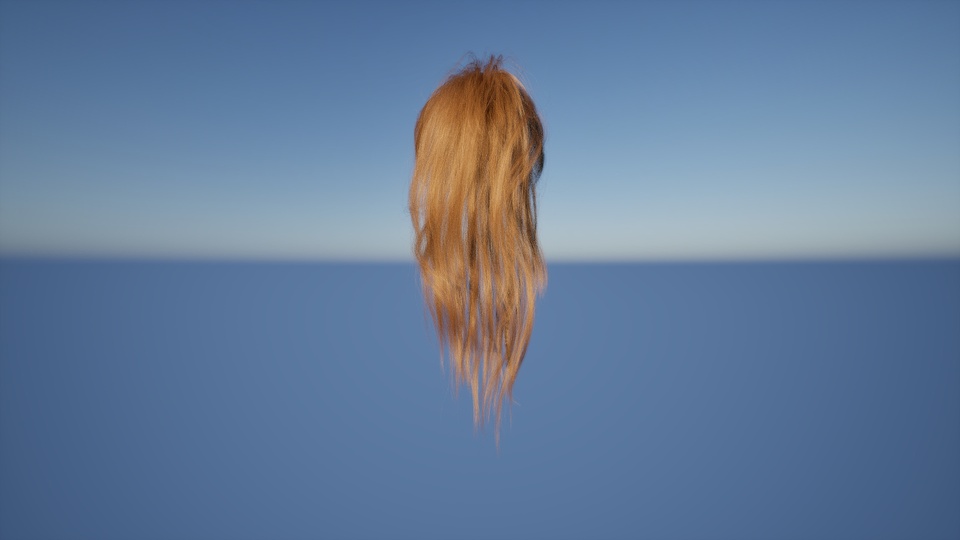}
  {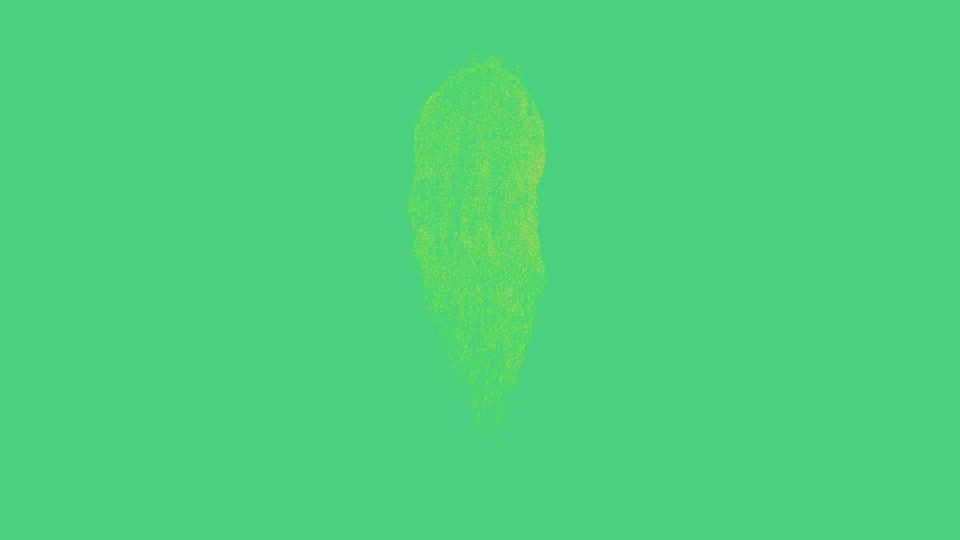}%
\hspace{0.008\linewidth}%
\dynamicref
  {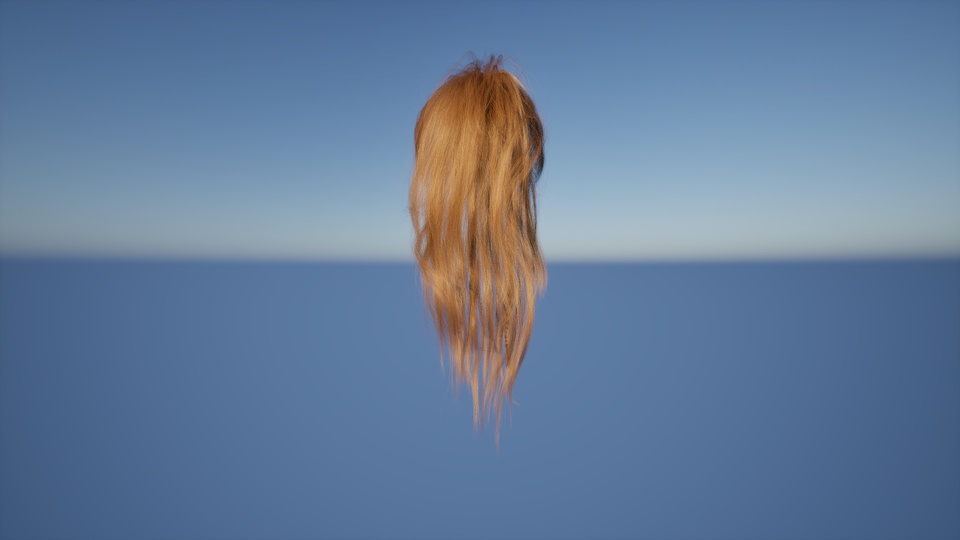}%
\\[2pt]

\dynamicrowlabel{Frame 50}%
\hspace{0.008\linewidth}%
\dynamicresult
  {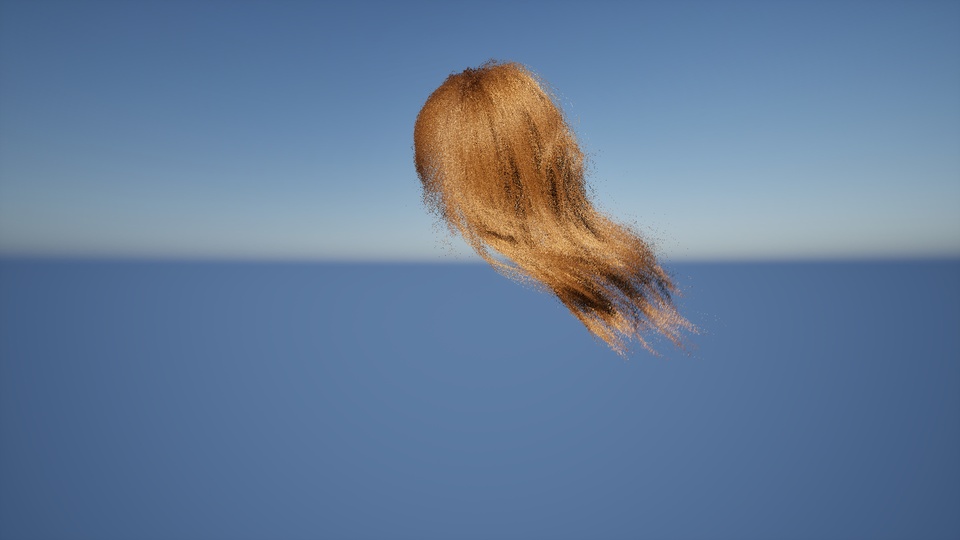}
  {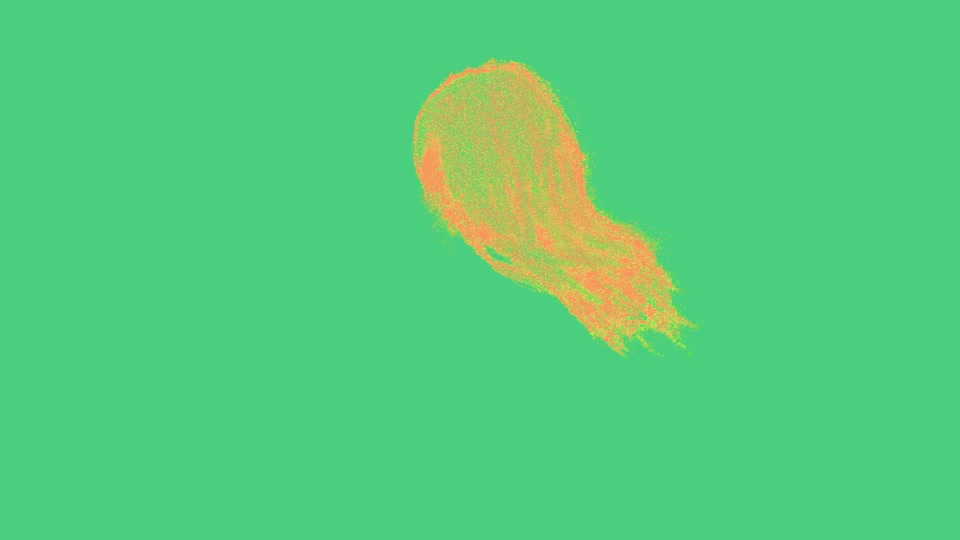}%
\hspace{0.008\linewidth}%
\dynamicresult
  {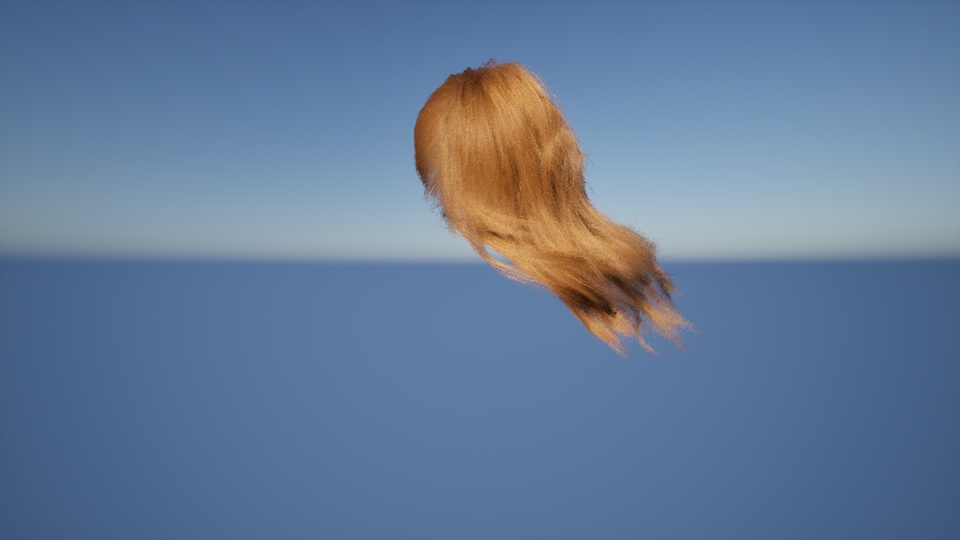}
  {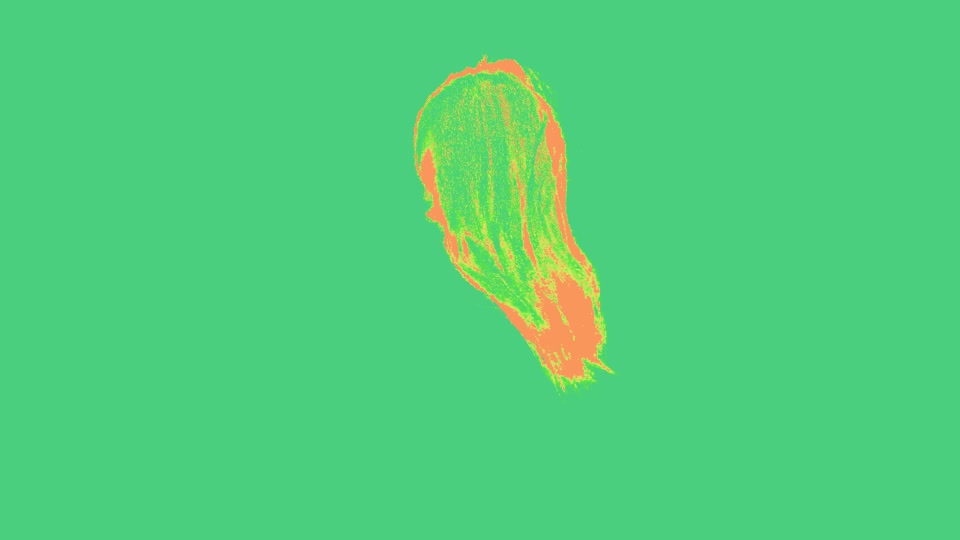}%
\hspace{0.008\linewidth}%
\dynamicresult
  {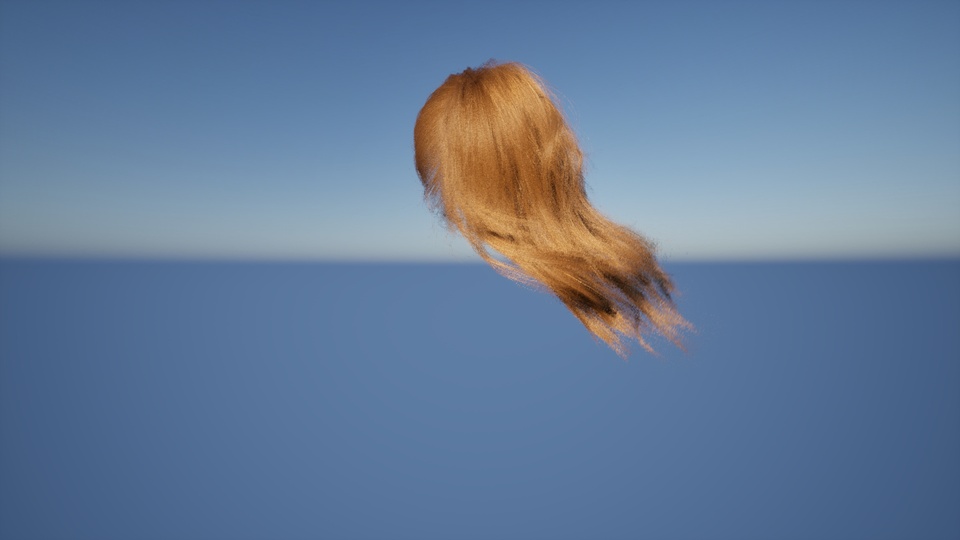}
  {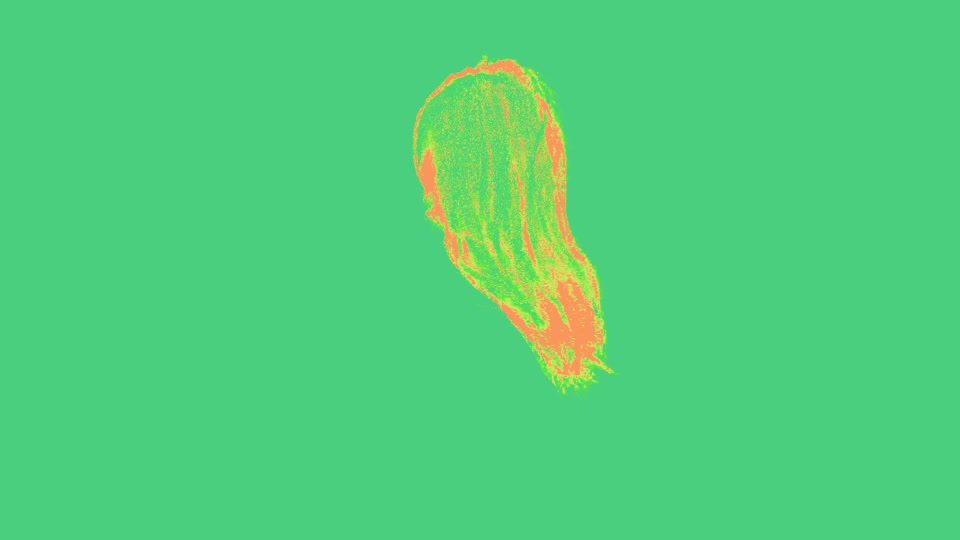}%
\hspace{0.008\linewidth}%
\dynamicresult
  {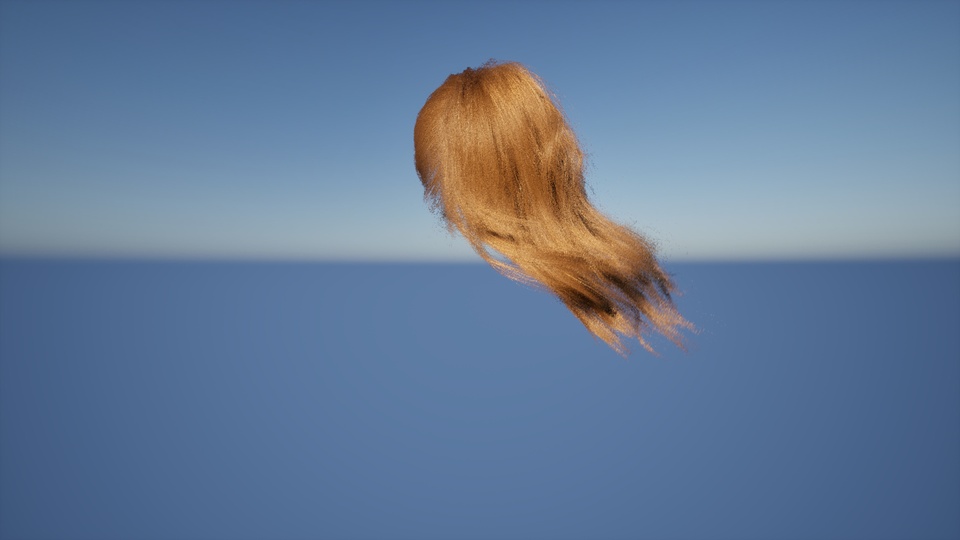}
  {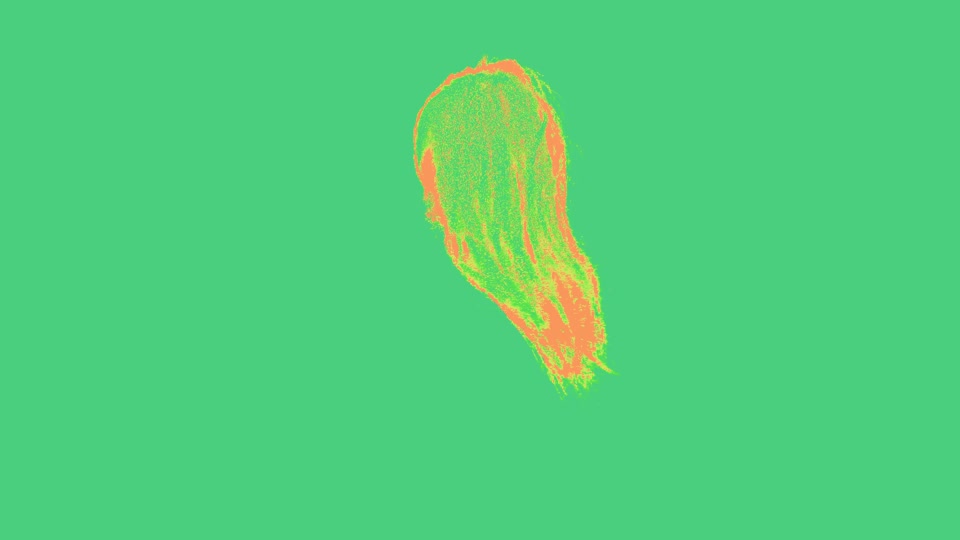}%
\hspace{0.008\linewidth}%
\dynamicresult
  {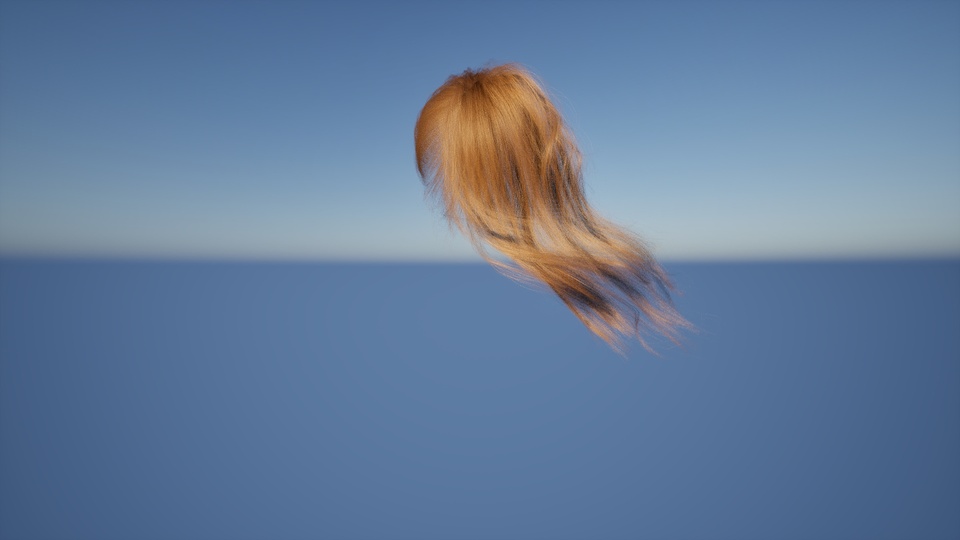}
  {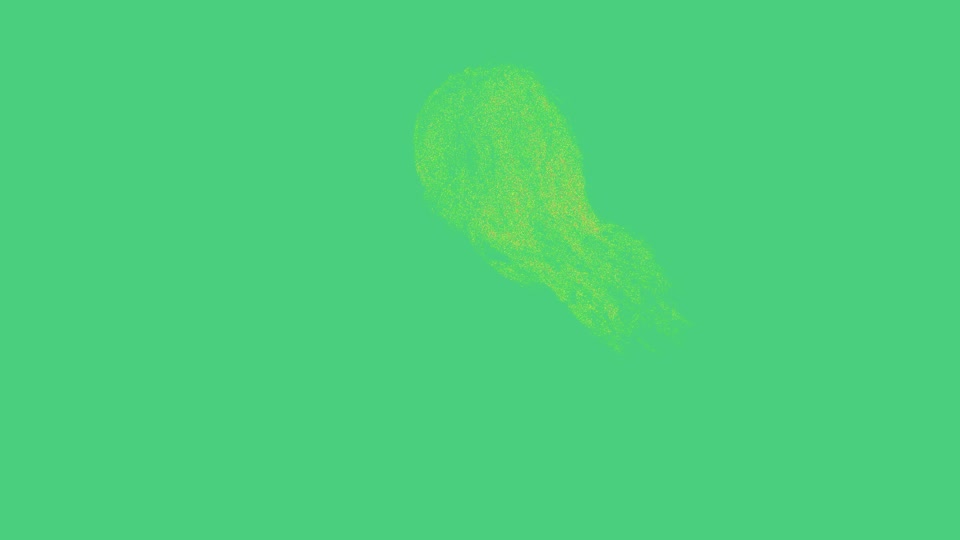}%
\hspace{0.008\linewidth}%
\dynamicref
  {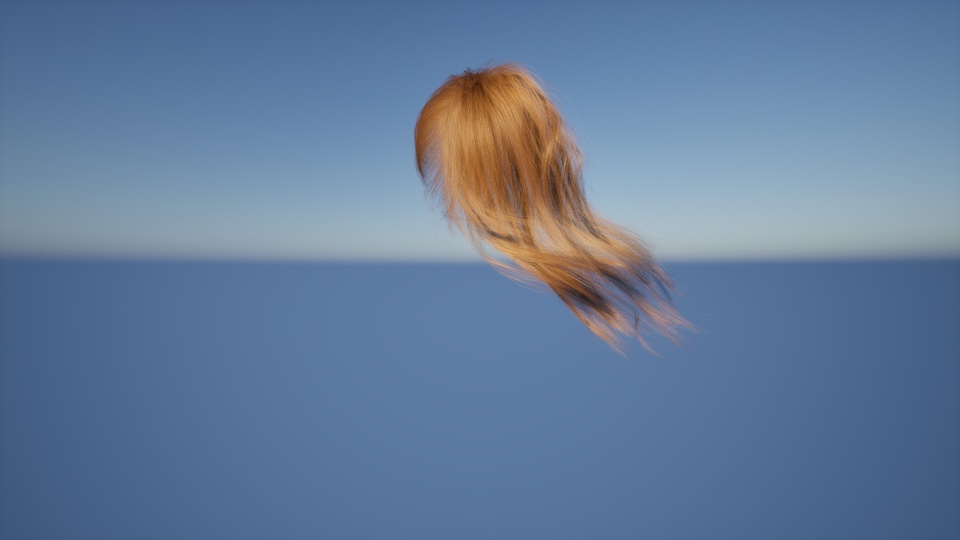}%
\\[2pt]

\dynamicrowlabel{Frame 150}%
\hspace{0.008\linewidth}%
\dynamicresult
  {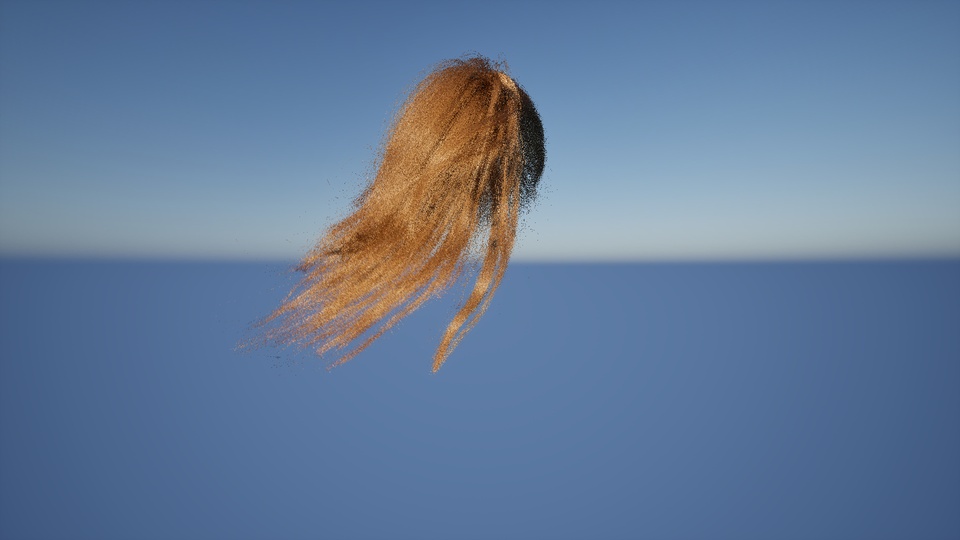}
  {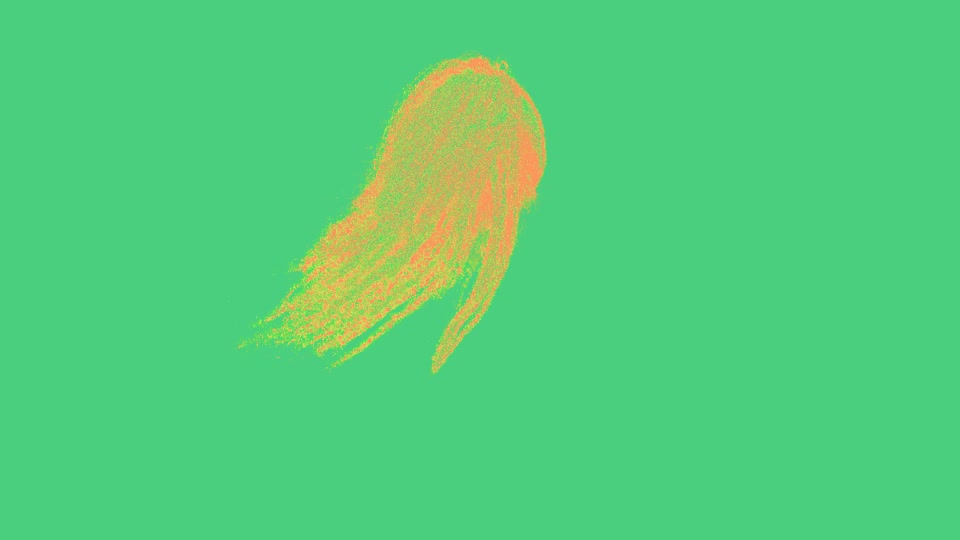}%
\hspace{0.008\linewidth}%
\dynamicresult
  {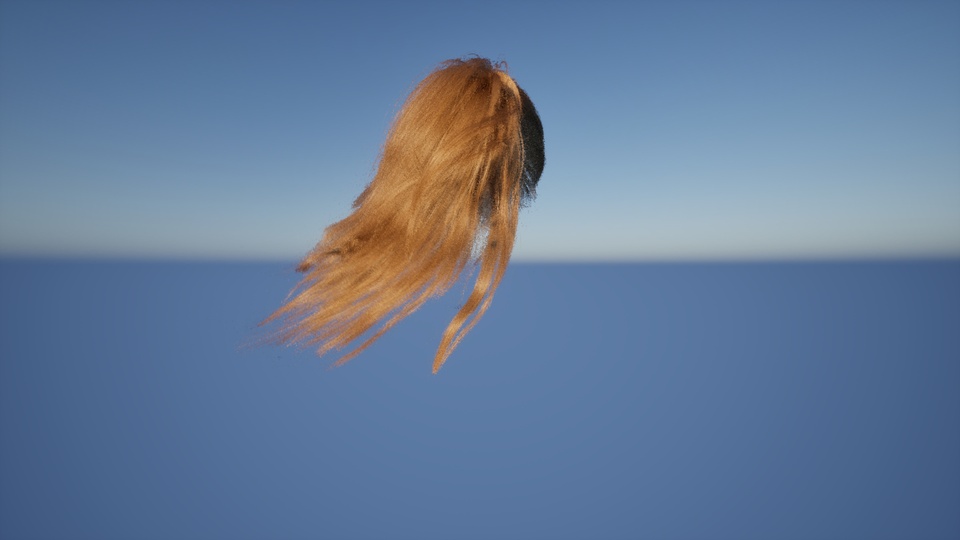}
  {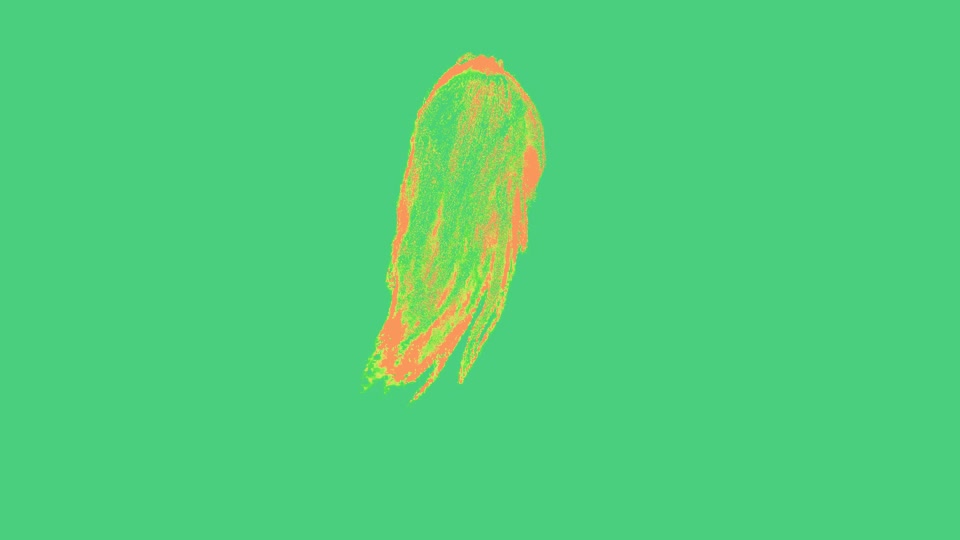}%
\hspace{0.008\linewidth}%
\dynamicresult
  {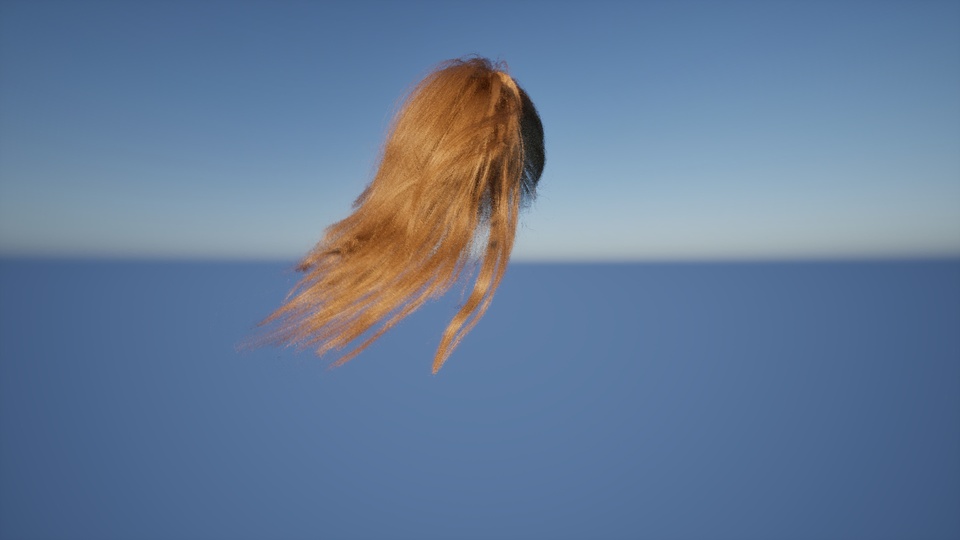}
  {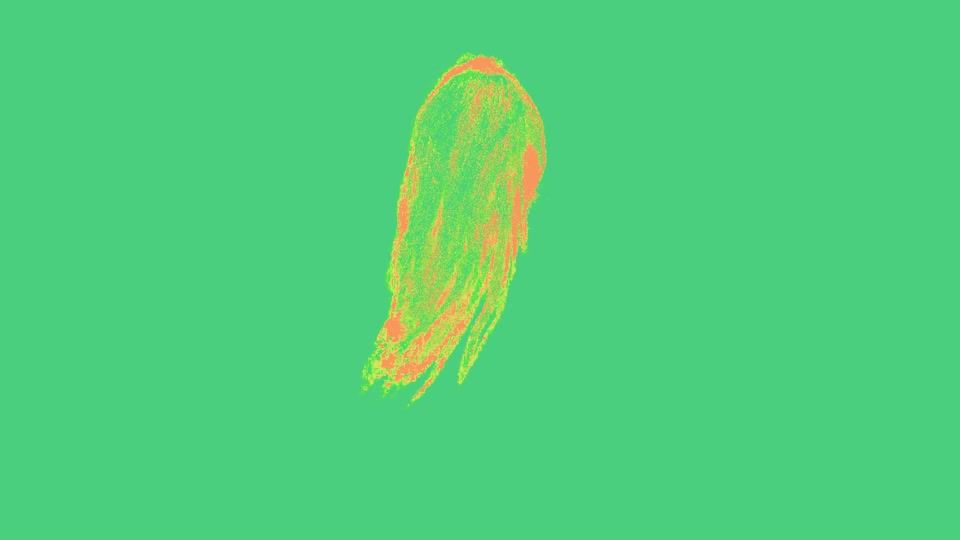}%
\hspace{0.008\linewidth}%
\dynamicresult
  {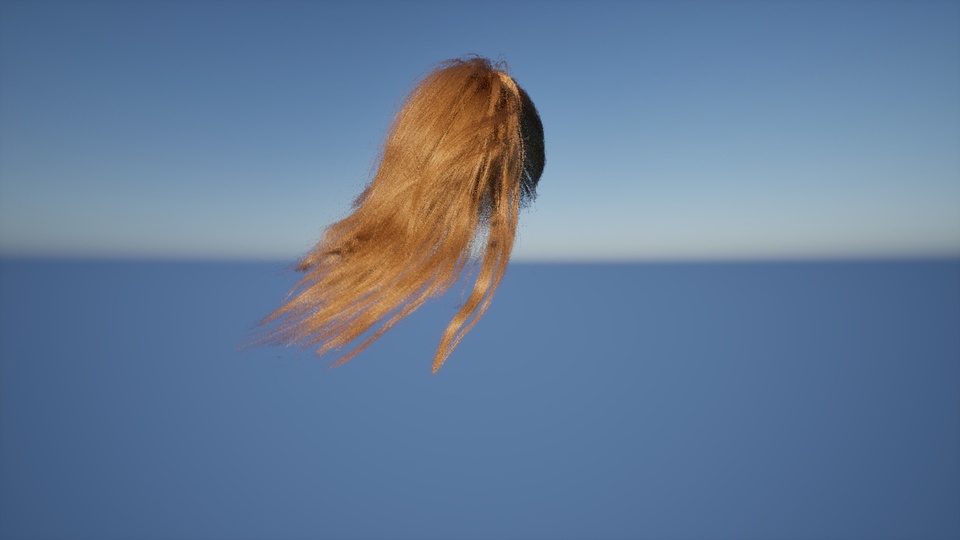}
  {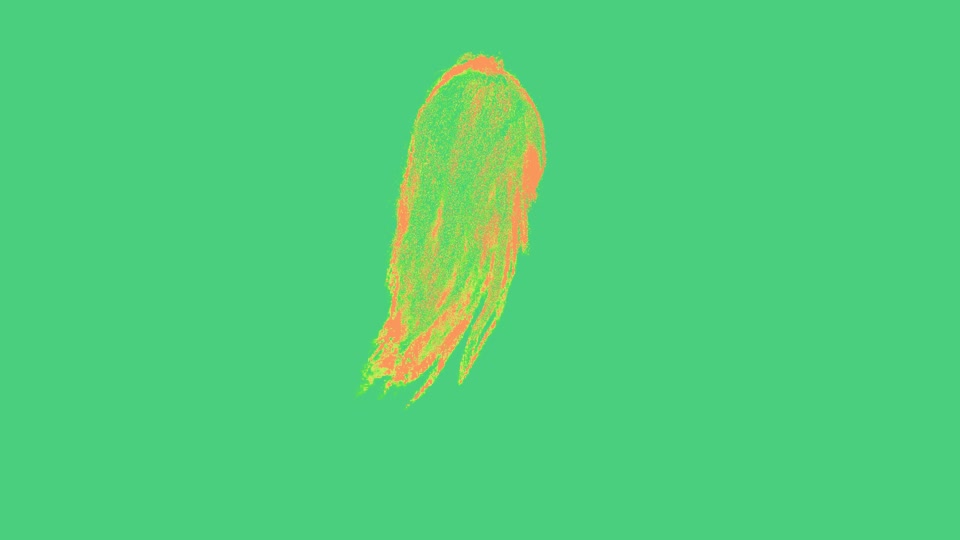}%
\hspace{0.008\linewidth}%
\dynamicresult
  {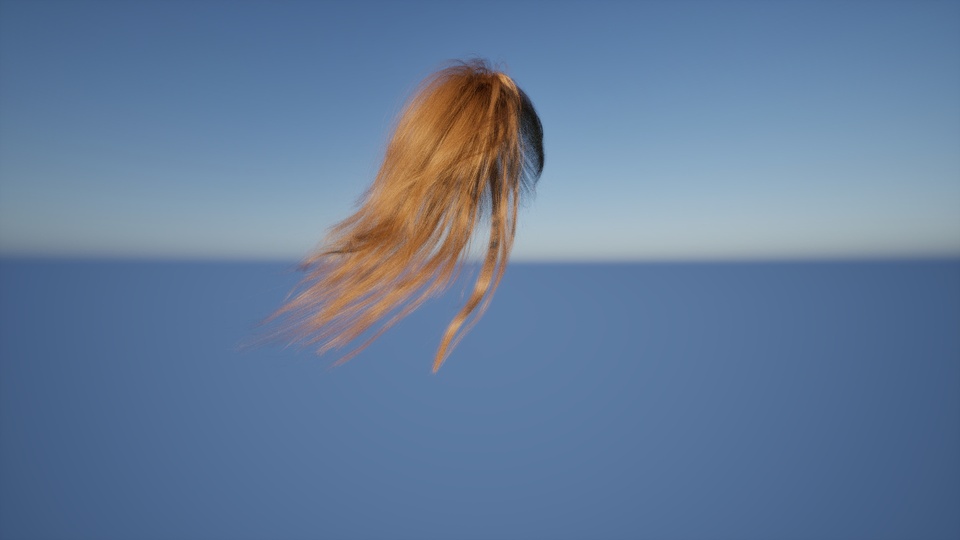}
  {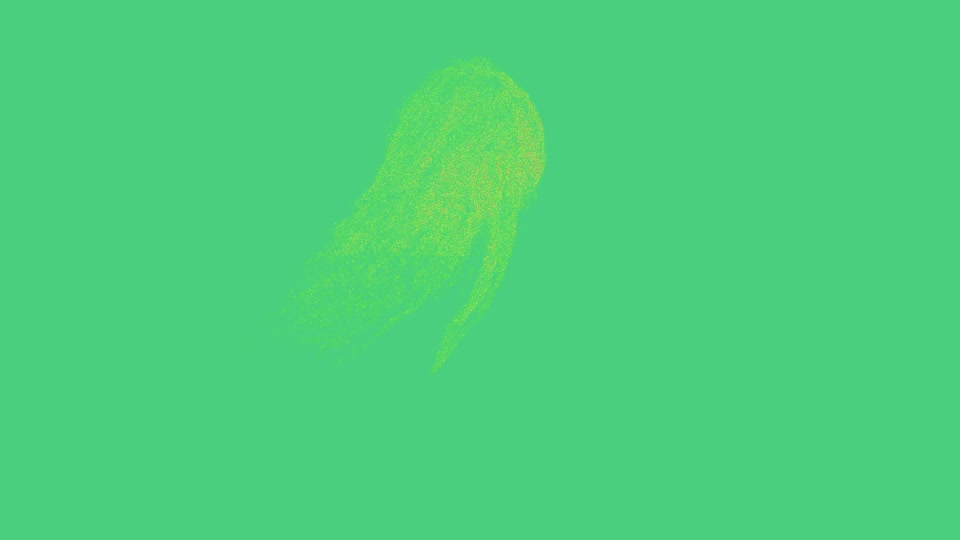}%
\hspace{0.008\linewidth}%
\dynamicref
  {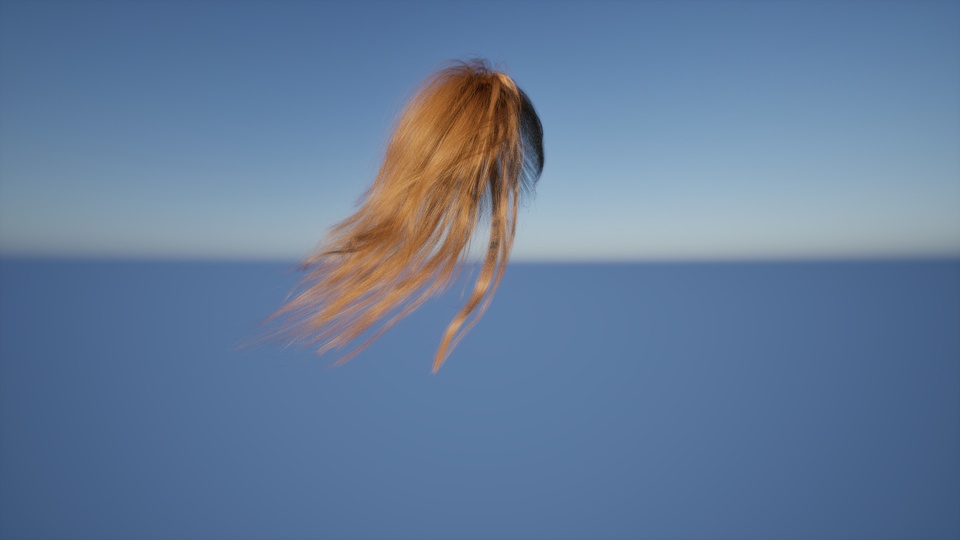}%

\caption{Comparison of dynamic hair sequence.}
\Description{}
\label{fig:comparison_dynamic_scene}

\end{figure*}

\paragraph{Dynamic Hairs}
To evaluate temporal stability, we further test our method on dynamic hair simulation sequences (\autoref{fig:comparison_dynamic_scene}). Unlike static scenes, dynamic scenes introduce temporal changes, disocclusions, and frame-to-frame sampling variation, making flickering and ghosting more noticeable. In addition to spatial metrics, we include the perceptual video metric FVVDP~\cite{mantiuk2021fvvdp} to assess spatiotemporal artifacts, such as flicker and judder. \autoref{tab:dynamic} shows that our method consistently outperforms all competing methods across all quantitative metrics while maintaining a performance cost comparable to industrial solutions.

\begin{table}[hbt]
\centering
\small
\caption{Quantitative metrics on dynamic hairs  on~\autoref{fig:comparison_dynamic_scene}.} 
\label{tab:dynamic}
\renewcommand{\tabcolsep}{0.1cm}
\begin{tabular}{lccccccc}
\toprule
Method 
& MSE$\downarrow$ 
& PSNR$\uparrow$ 
& SSIM$\uparrow$ 
& LPIPS$\downarrow$
& FLIP$\downarrow$
& FVVDP$\uparrow$
& Time$\downarrow$ (ms)\\
\midrule
Input & 0.0111 & 19.72 & 0.9433 & 0.0513 & 0.0192 & 8.42 & -- \\
TAA   & 0.0067 & 21.83 & 0.9706 & 0.0269 & 0.0181 & 8.11 & 0.2\\
FSR   & 0.0068 & 21.74 & 0.9642 & 0.0291 & 0.0180 & 8.14 & 0.8\\
DLSS  & 0.0059 & 22.37 & 0.9680 & 0.0267 & 0.0175 & 8.21 & 0.7\\
Ours  & \textbf{0.0015} & \textbf{28.57} & \textbf{0.9756} & \textbf{0.0155} & \textbf{0.0075} & \textbf{9.27} & 1.0\\
\bottomrule
\end{tabular}
\end{table}

\begin{figure*}[htb]
\centering

\newcommand{\colorhead}[1]{%
  \begin{minipage}{0.143\linewidth}
    \centering
    #1
  \end{minipage}%
}

%
\newcommand{\colorresult}[3]{%
  \begingroup
  \setlength{\unitlength}{0.143\linewidth}%
  \begin{picture}(1,0.696)
    \put(0,0){%
      \includegraphics[
        trim={200 75 200 75},
        clip,
        width=\unitlength
      ]{#1}%
    }%

    \put(0,0){%
      \includegraphics[
        trim={200 75 200 75},
        clip,
        width=0.40\unitlength
      ]{#2}%
    }%

    \put(0.015,0.605){%
      \small #3%
    }%
  \end{picture}%
  \endgroup
}

\newcommand{\colorref}[1]{%
  \includegraphics[
    trim={200 75 200 75},
    clip,
    width=0.143\linewidth
  ]{#1}%
}

\colorhead{Input}%
\hspace{0.01\linewidth}%
\colorhead{TAA}%
\hspace{0.01\linewidth}%
\colorhead{DLSS}%
\hspace{0.01\linewidth}%
\colorhead{FSR}%
\hspace{0.01\linewidth}%
\colorhead{Ours}%
\hspace{0.01\linewidth}%
\colorhead{Ref}%
\\[2pt]

\colorresult
  {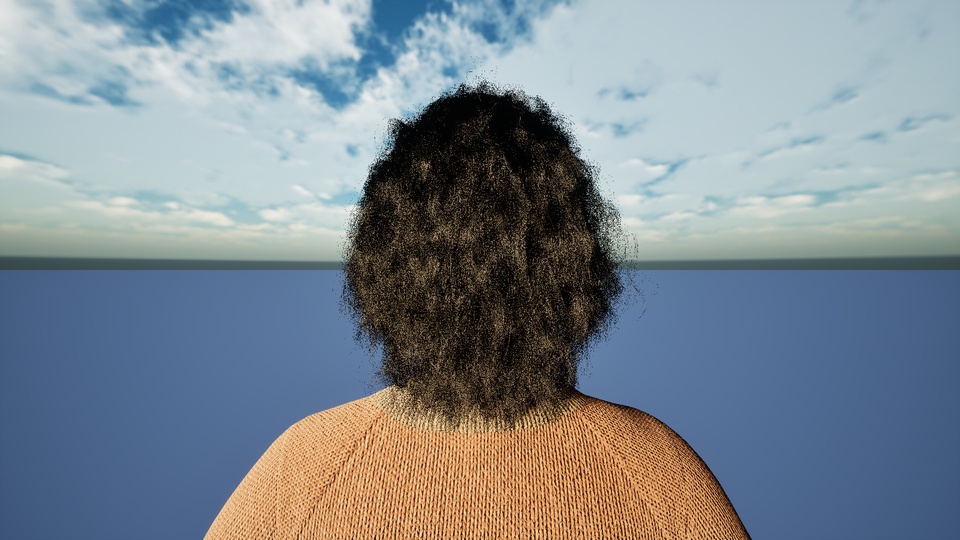}
  {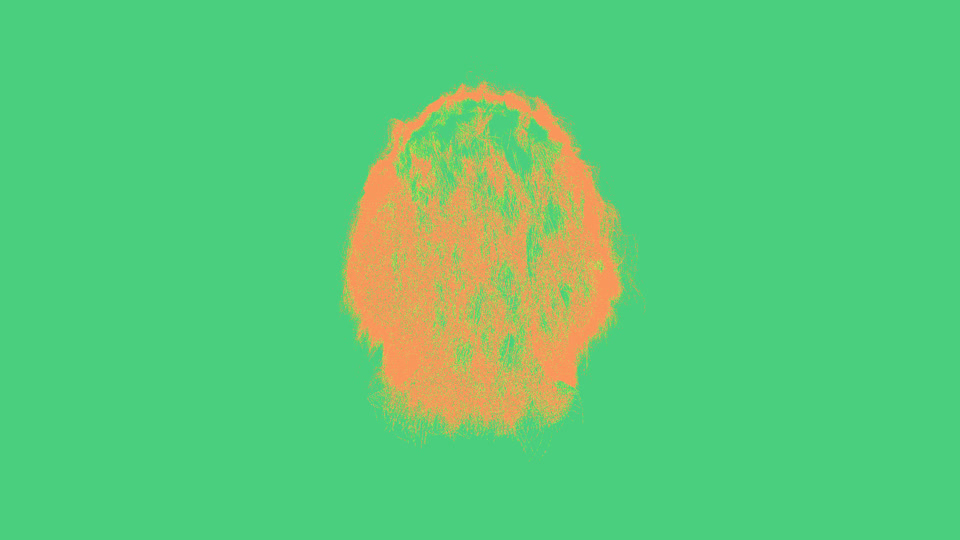}
  {12.80}%
\hspace{0.01\linewidth}%
\colorresult
  {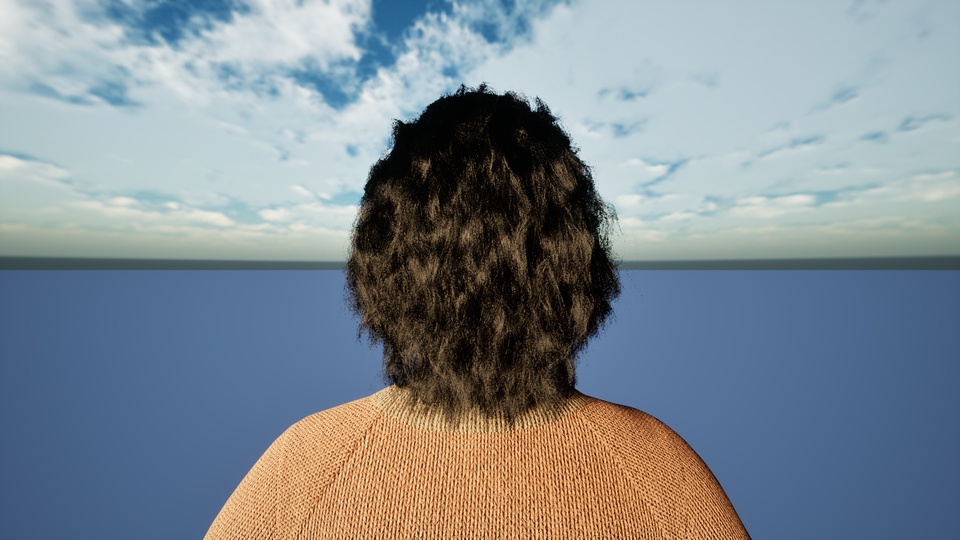}
  {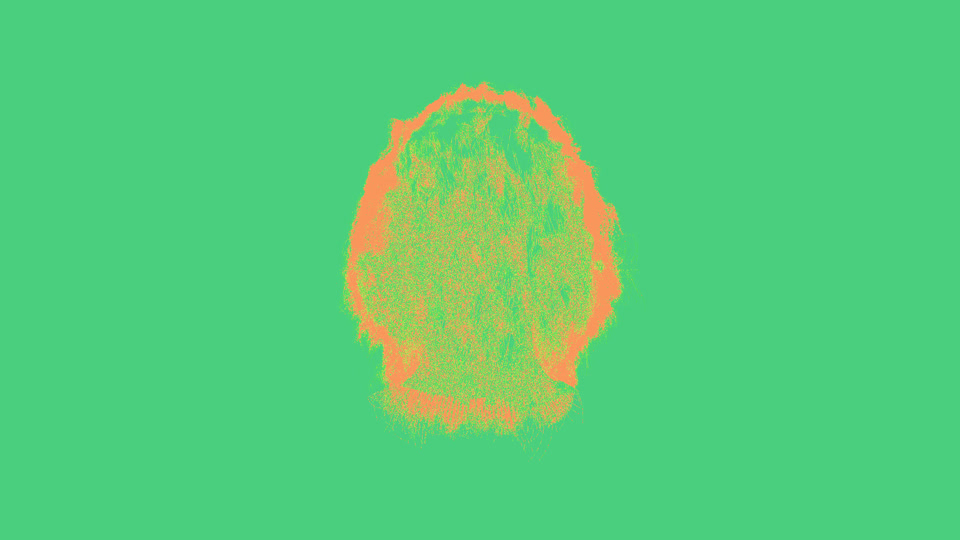}
  {16.35}%
\hspace{0.01\linewidth}%
\colorresult
  {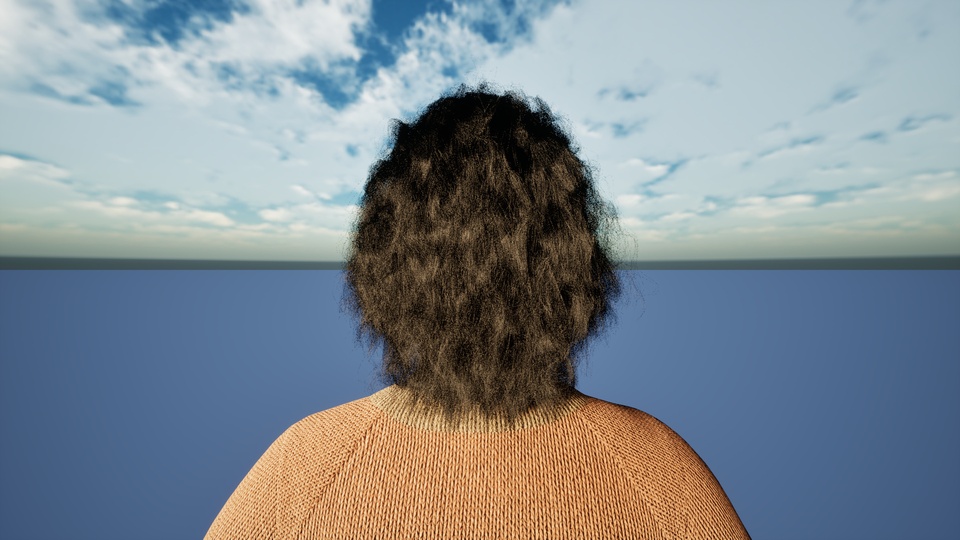}
  {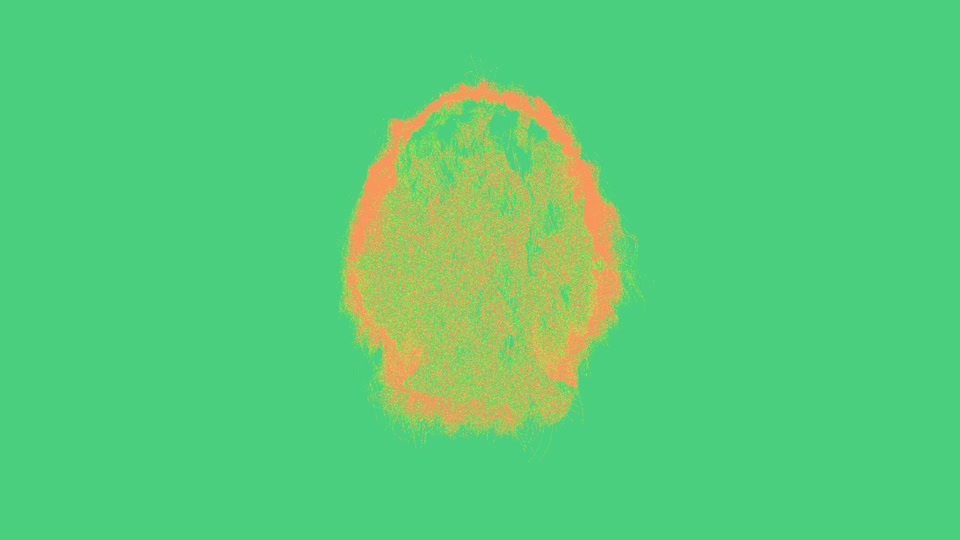}
  {17.00}%
\hspace{0.01\linewidth}%
\colorresult
  {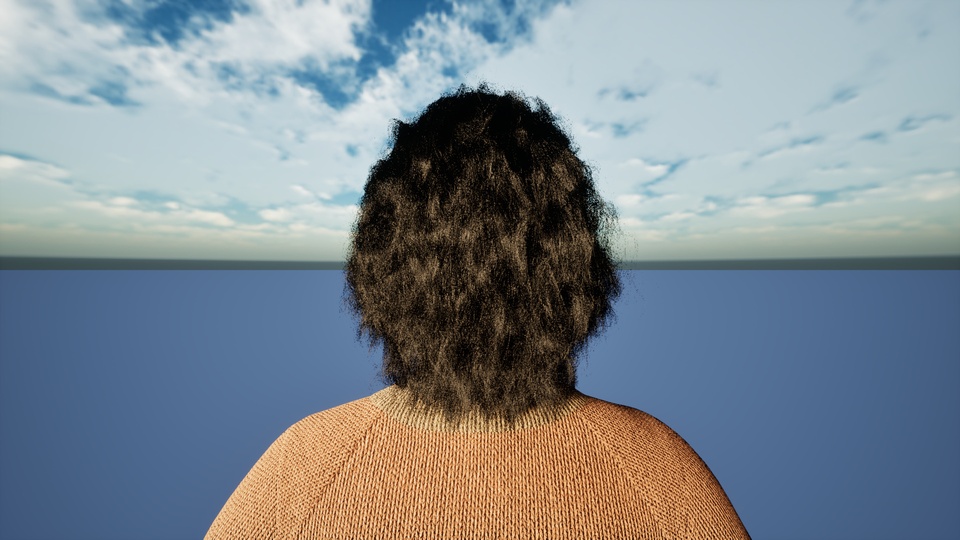}
  {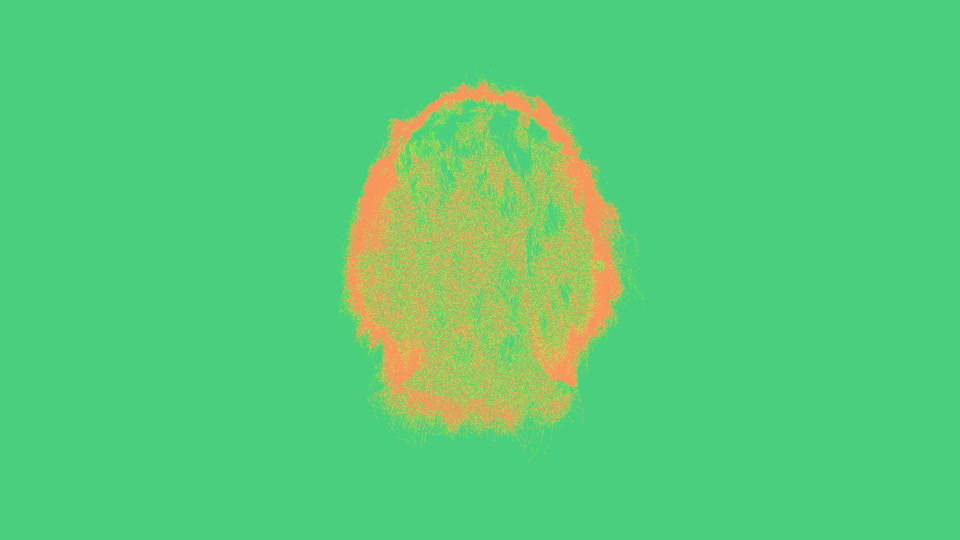}
  {16.15}%
\hspace{0.01\linewidth}%
\colorresult
  {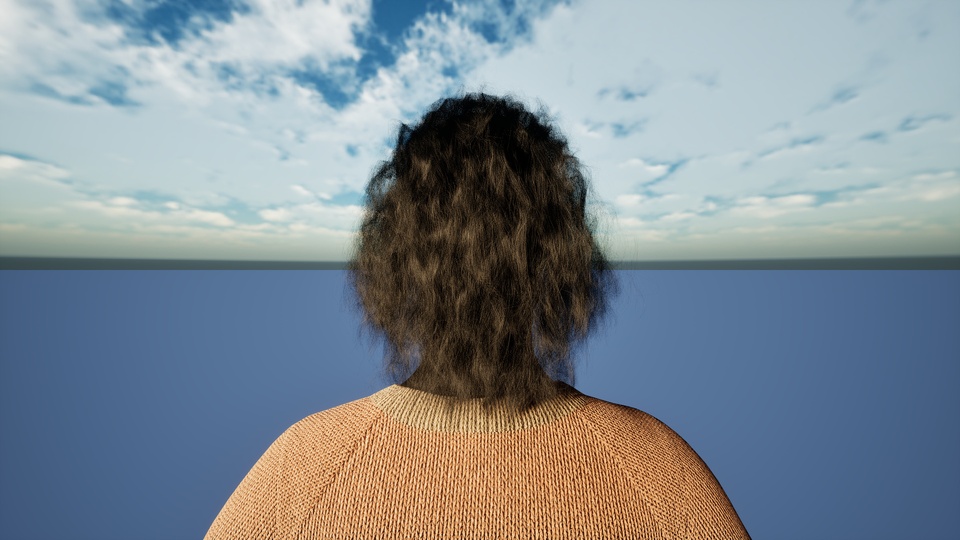}
  {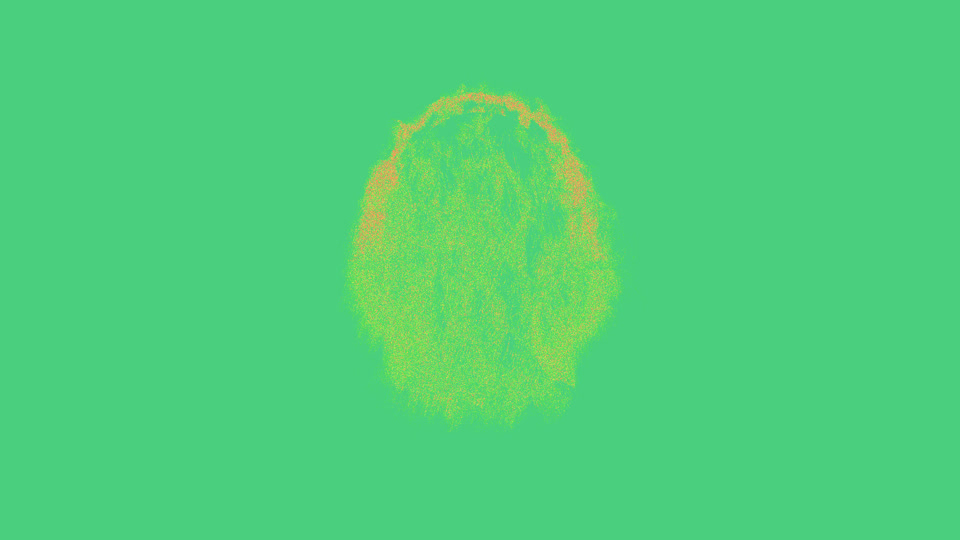}
  {25.72}%
\hspace{0.01\linewidth}%
\colorref{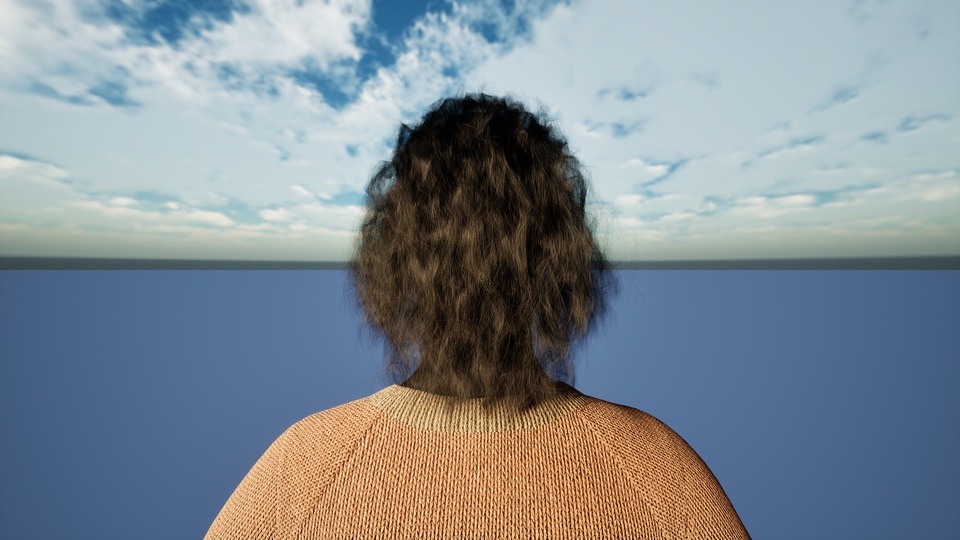}%
\\[2pt]

\colorresult
  {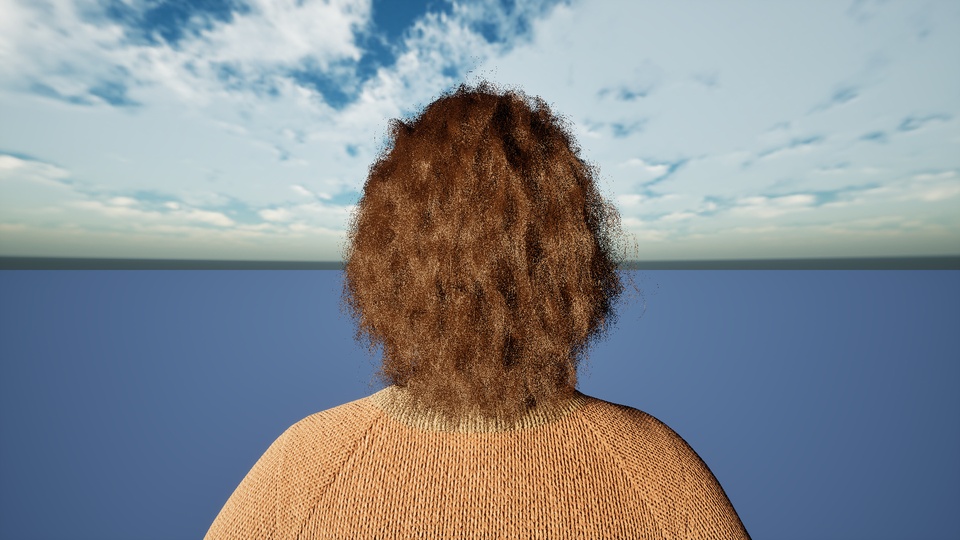}
  {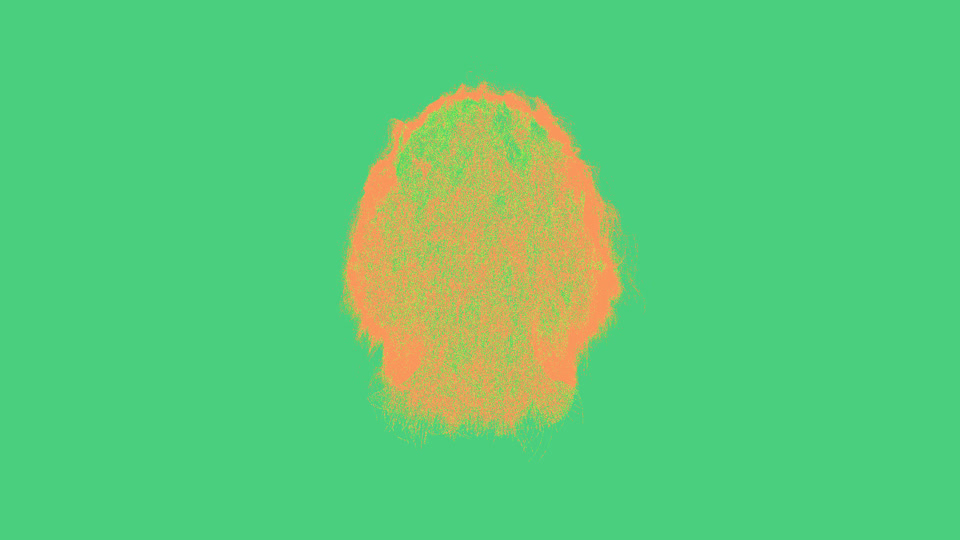}
  {13.70}%
\hspace{0.01\linewidth}%
\colorresult
  {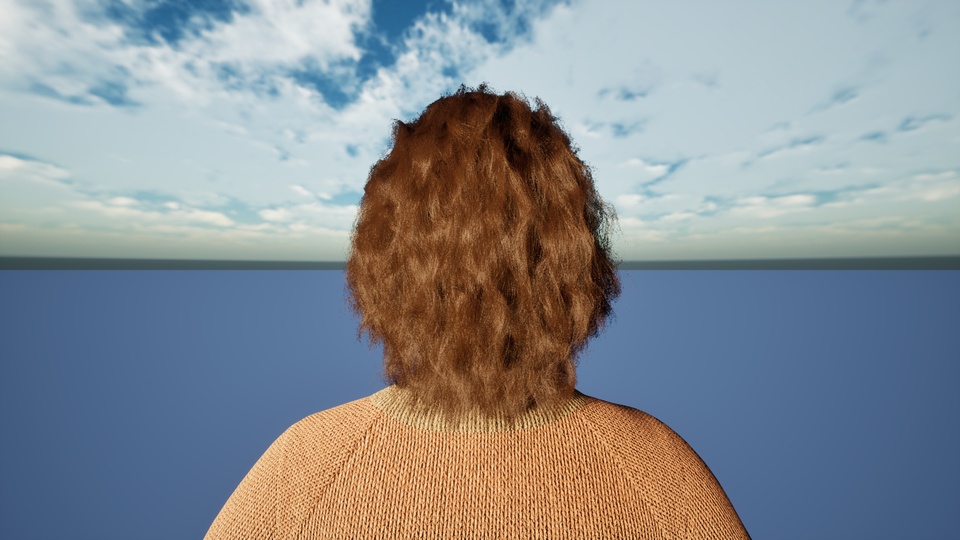}
  {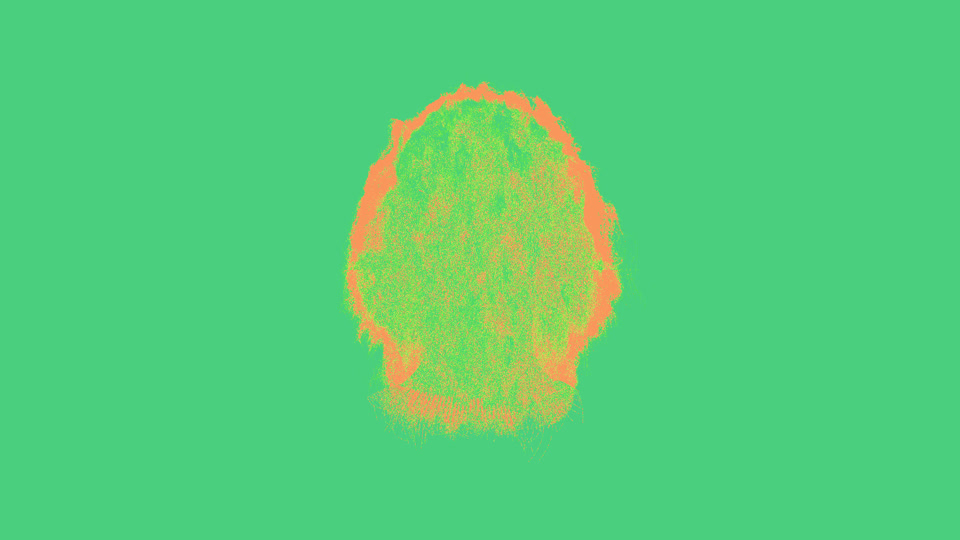}
  {17.86}%
\hspace{0.01\linewidth}%
\colorresult
  {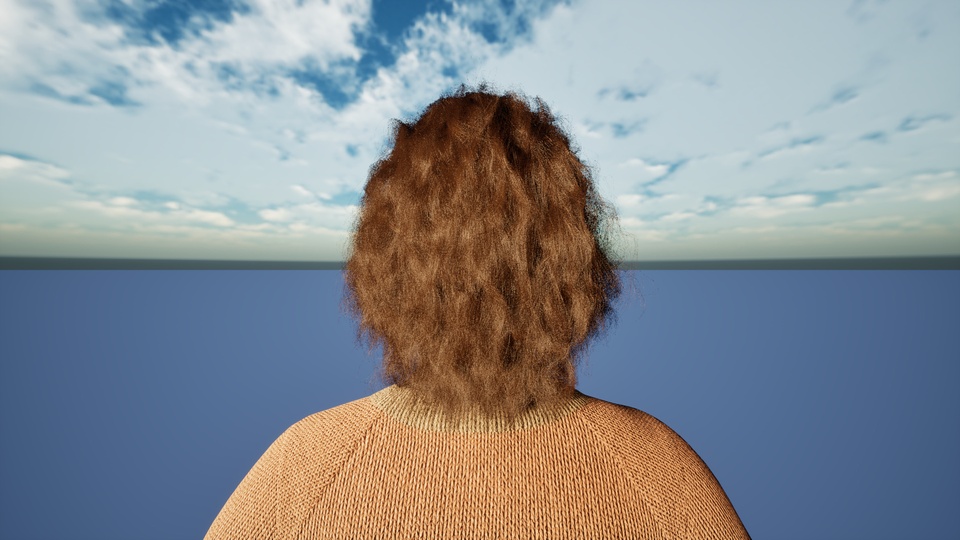}
  {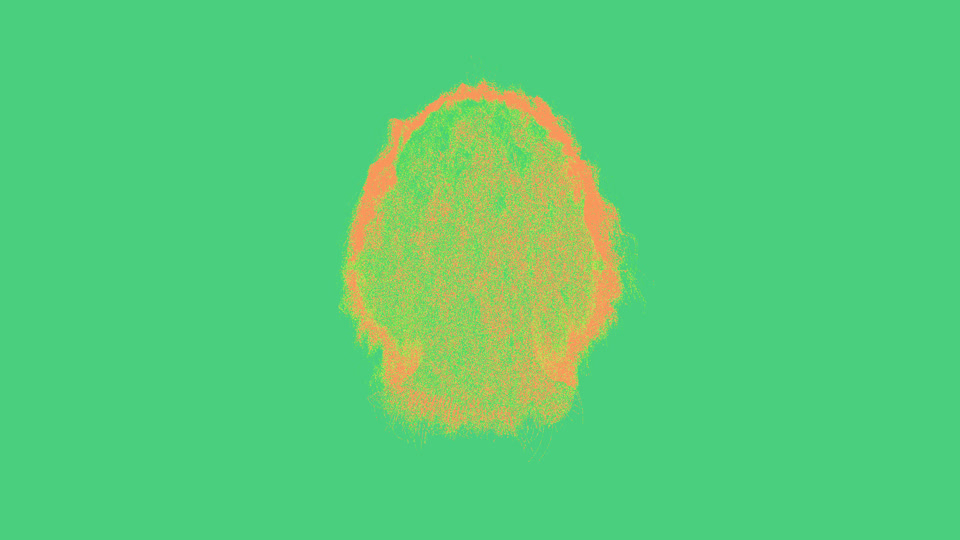}
  {18.21}%
\hspace{0.01\linewidth}%
\colorresult
  {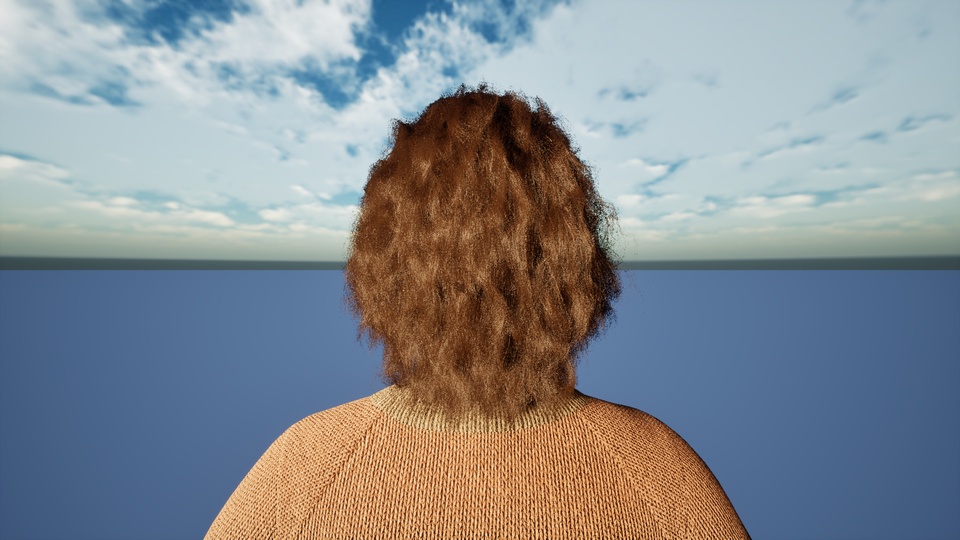}
  {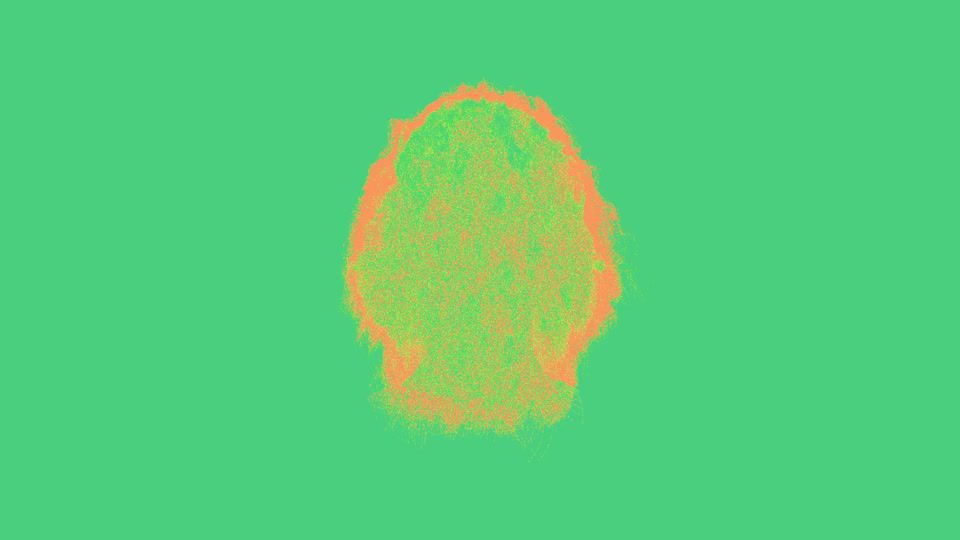}
  {17.36}%
\hspace{0.01\linewidth}%
\colorresult
  {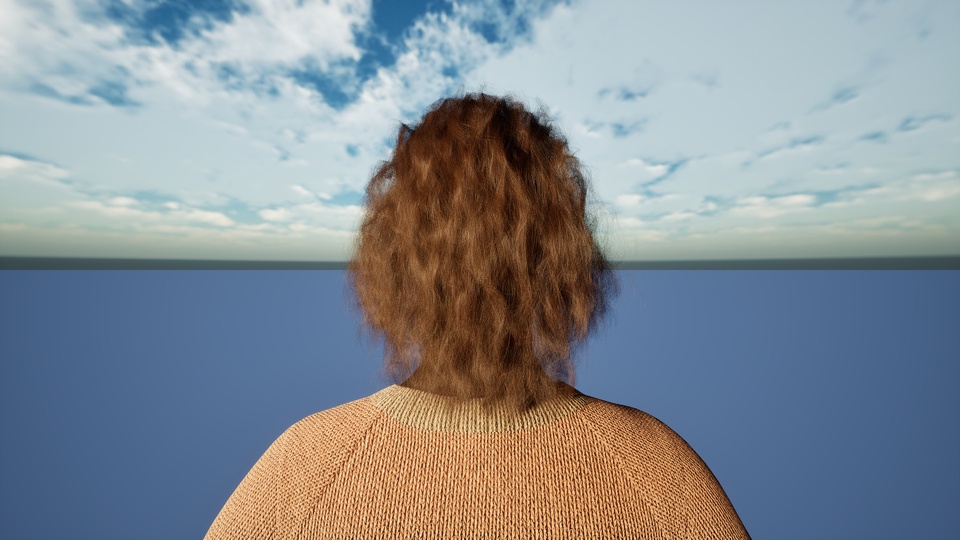}
  {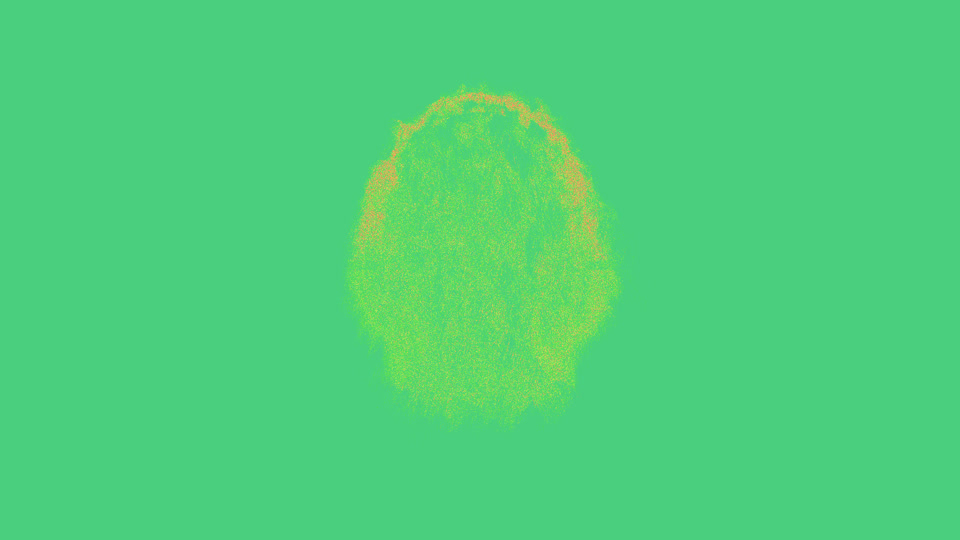}
  {26.49}%
\hspace{0.01\linewidth}%
\colorref{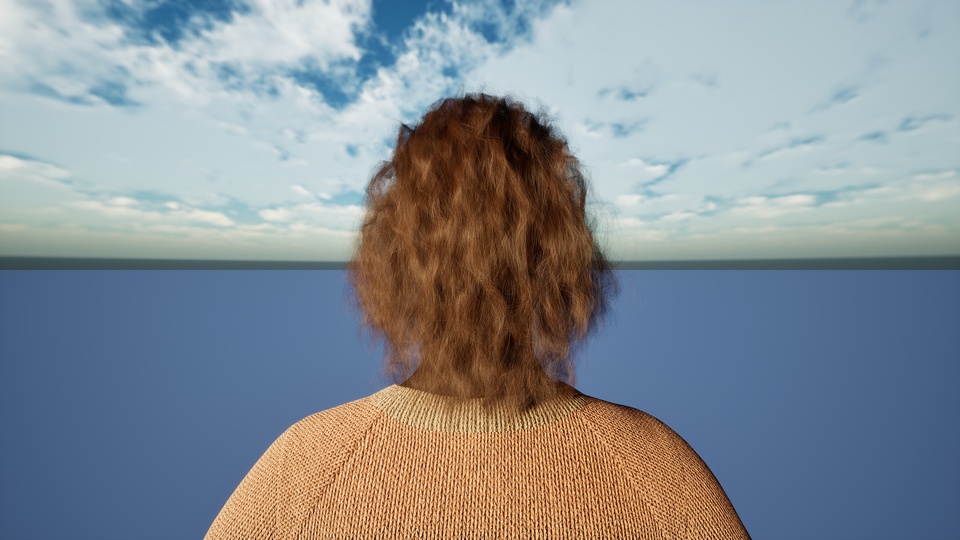}%
\\[2pt]

\colorresult
  {Fig_static/curly/static_human_curly_spp1.jpg}
  {Fig_static/curly/error_static_human_curly_spp1.jpg}
  {14.97}%
\hspace{0.01\linewidth}%
\colorresult
  {Fig_static/curly/static_human_curly_TAA.jpg}
  {Fig_static/curly/error_static_human_curly_TAA.jpg}
  {19.92}%
\hspace{0.01\linewidth}%
\colorresult
  {Fig_static/curly/static_human_curly_DLSS.jpg}
  {Fig_static/curly/error_static_human_curly_DLSS.jpg}
  {20.08}%
\hspace{0.01\linewidth}%
\colorresult
  {Fig_static/curly/static_human_curly_FSR.jpg}
  {Fig_static/curly/error_static_human_curly_FSR.jpg}
  {19.07}%
\hspace{0.01\linewidth}%
\colorresult
  {Fig_static/curly/static_human_curly_pred.jpg}
  {Fig_static/curly/error_static_human_curly_pred.jpg}
  {27.43}%
\hspace{0.01\linewidth}%
\colorref{Fig_static/curly/static_human_curly_spp128.jpg}%
\\[2pt]

\colorresult
  {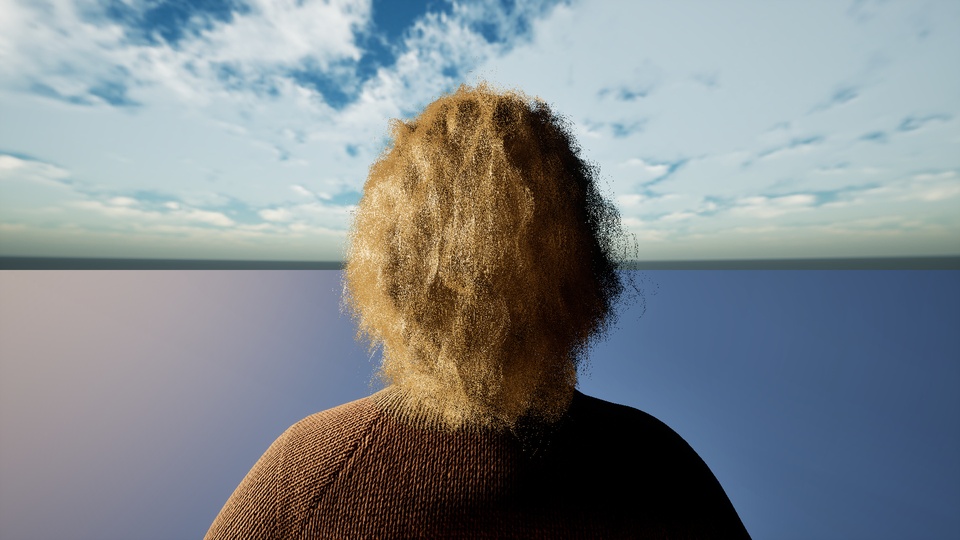}
  {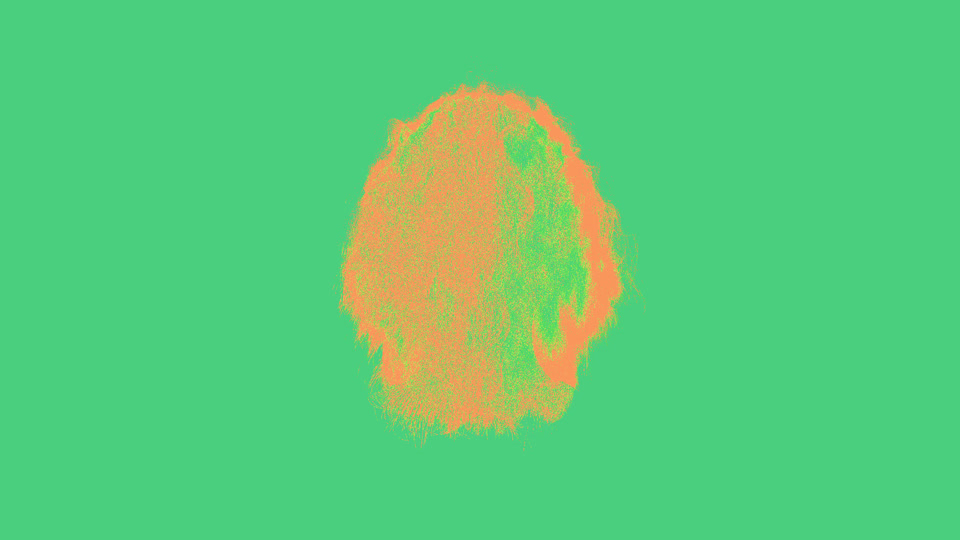}
  {14.19}%
\hspace{0.01\linewidth}%
\colorresult
  {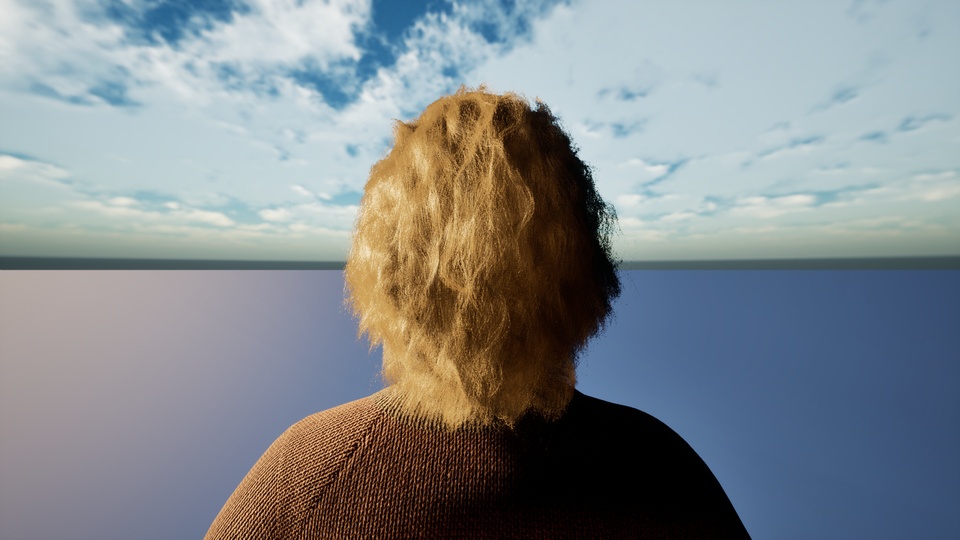}
  {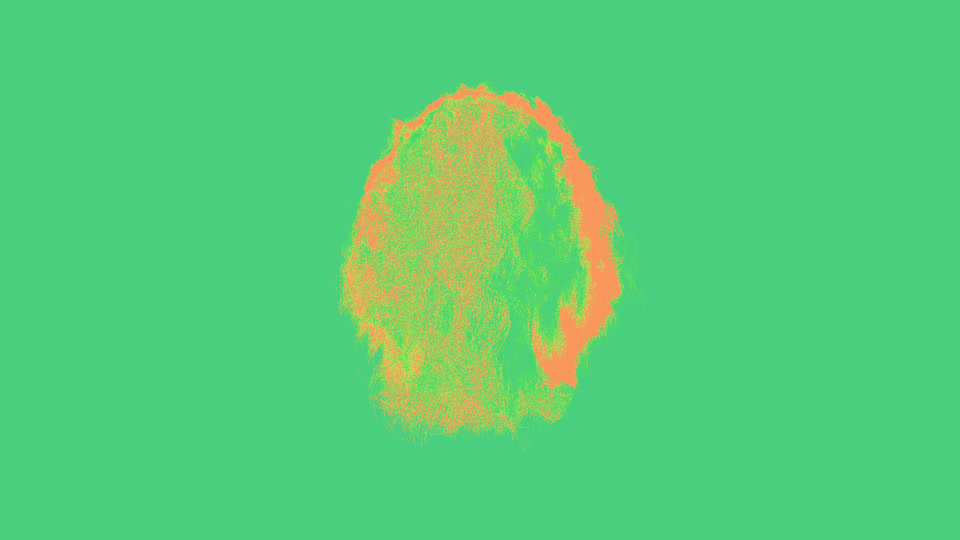}
  {17.81}%
\hspace{0.01\linewidth}%
\colorresult
  {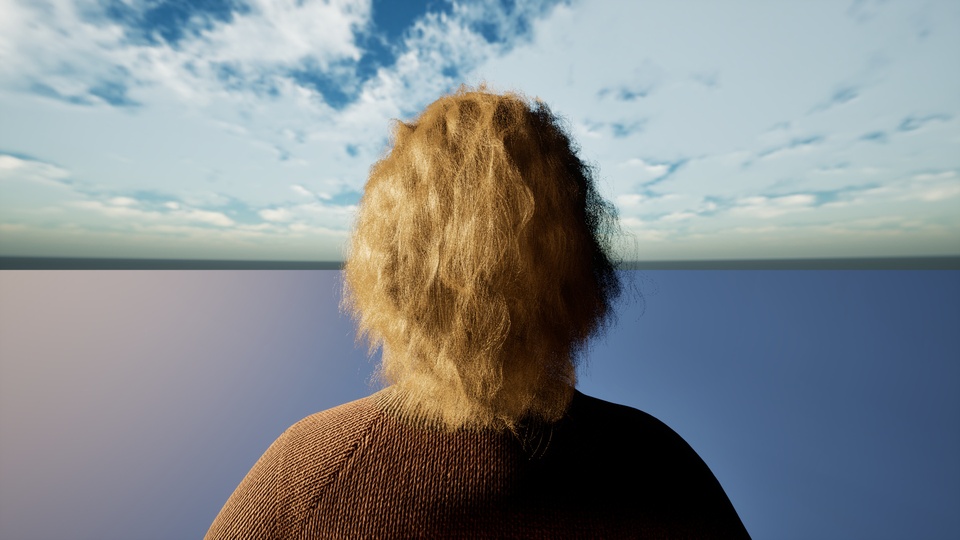}
  {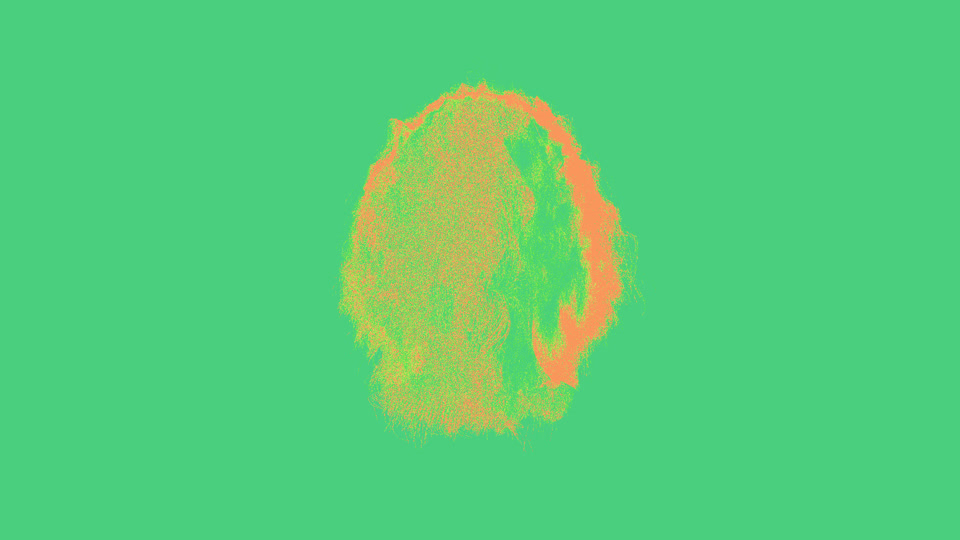}
  {18.47}%
\hspace{0.01\linewidth}%
\colorresult
  {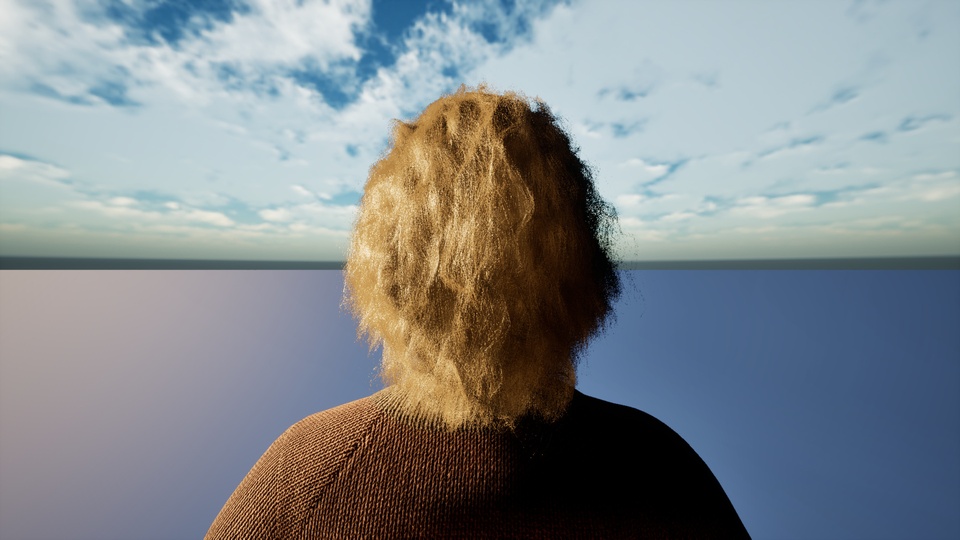}
  {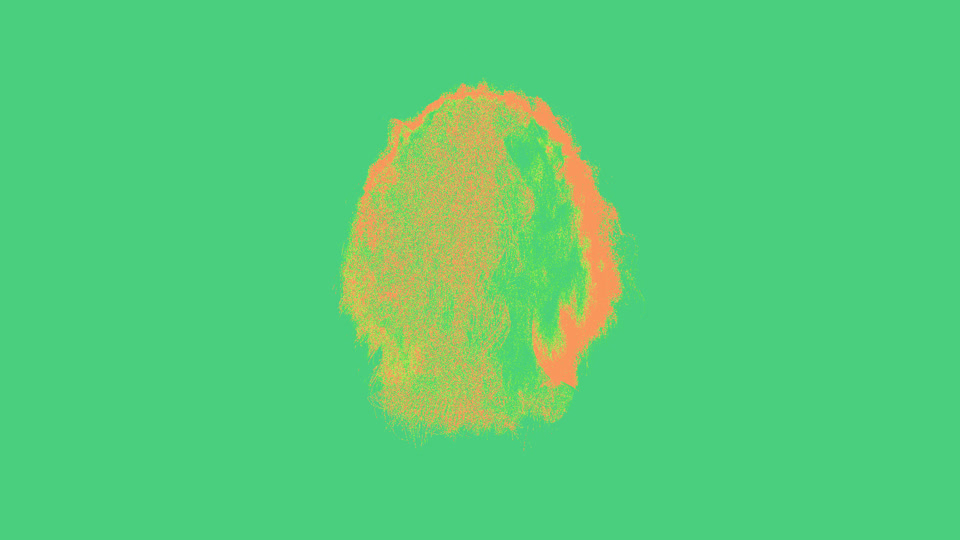}
  {17.65}%
\hspace{0.01\linewidth}%
\colorresult
  {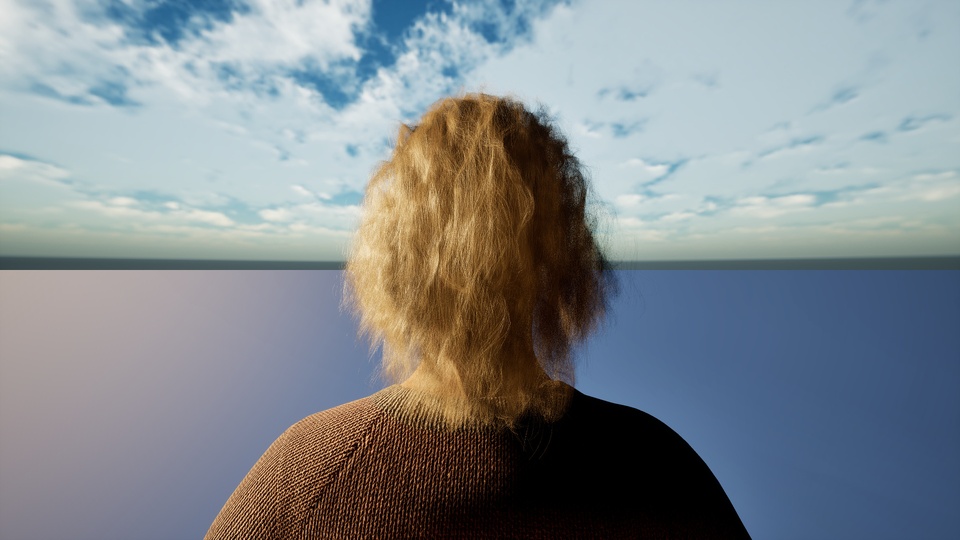}
  {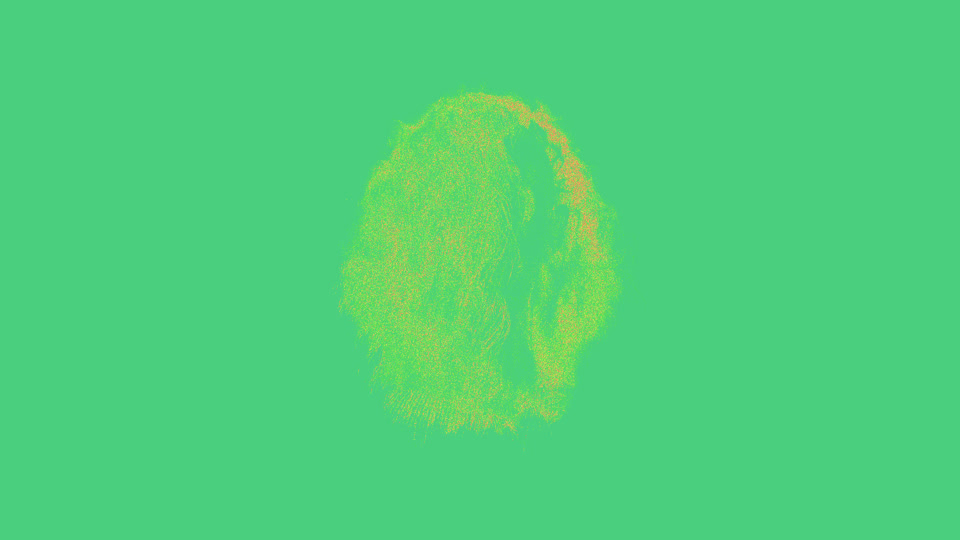}
  {26.18}%
\hspace{0.01\linewidth}%
\colorref{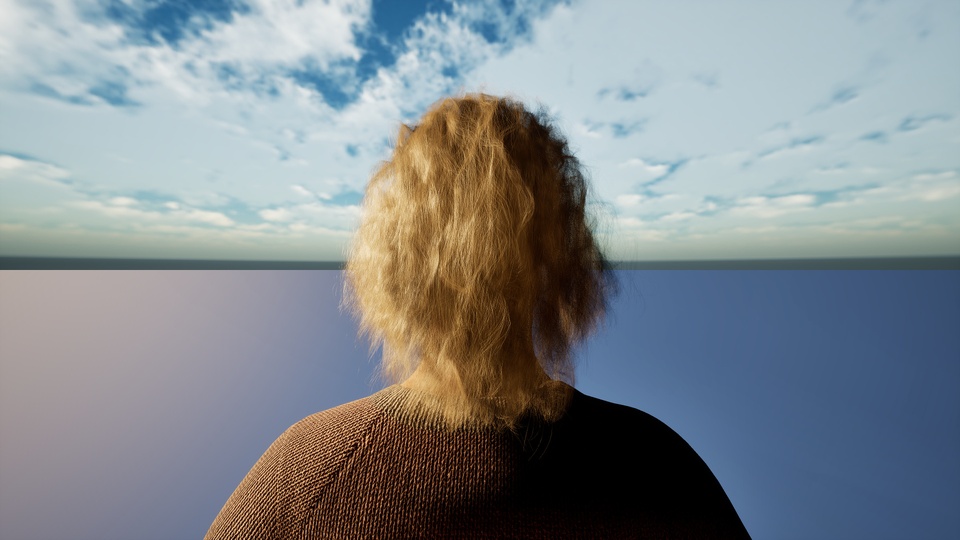}%
\\[2pt]

\colorresult
  {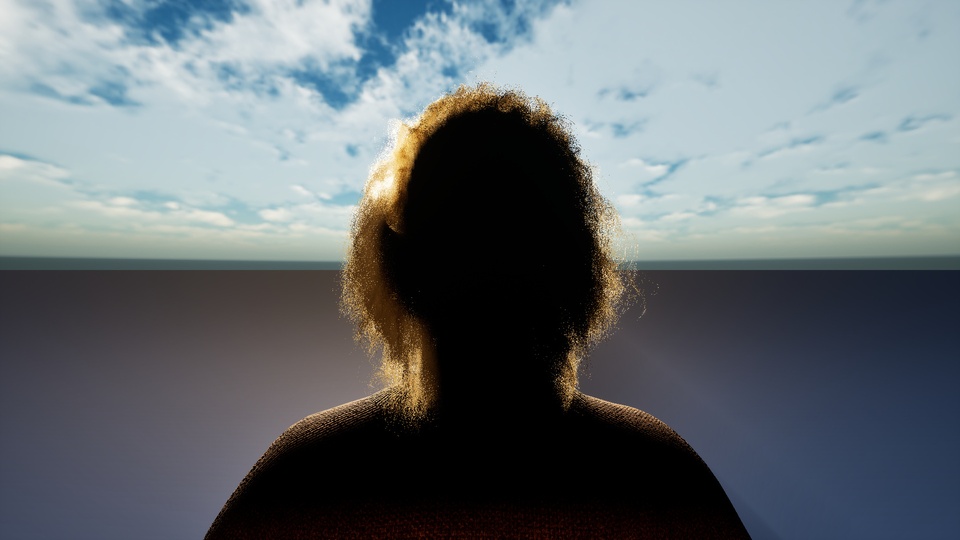}
  {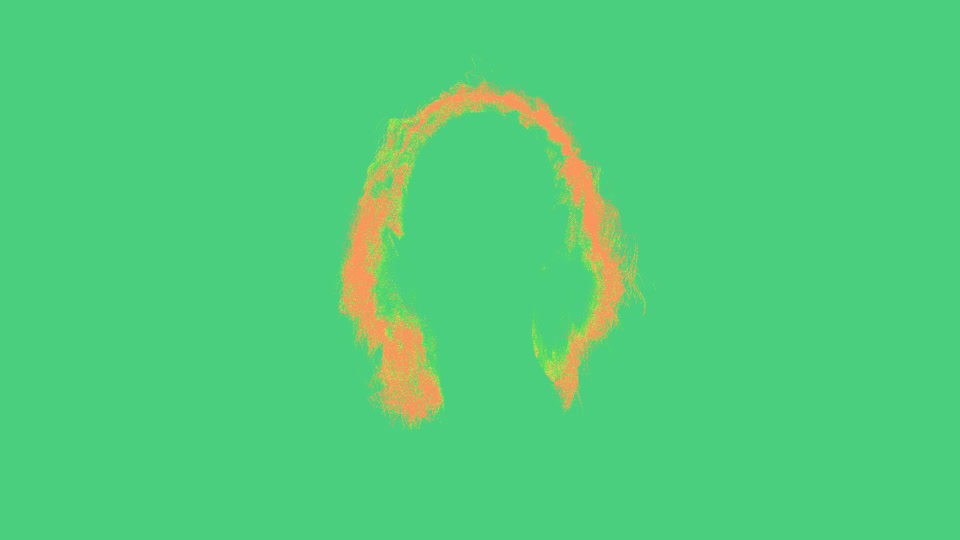}
  {17.54}%
\hspace{0.01\linewidth}%
\colorresult
  {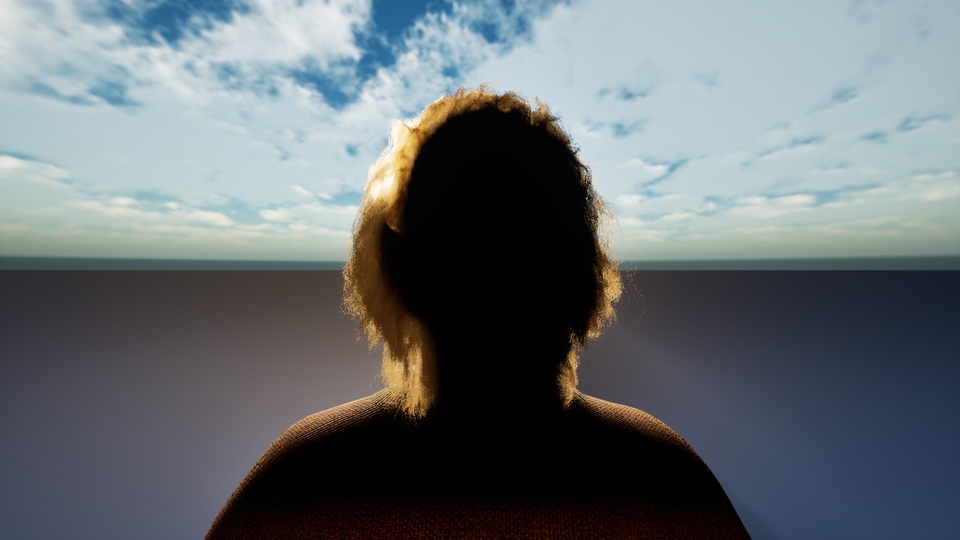}
  {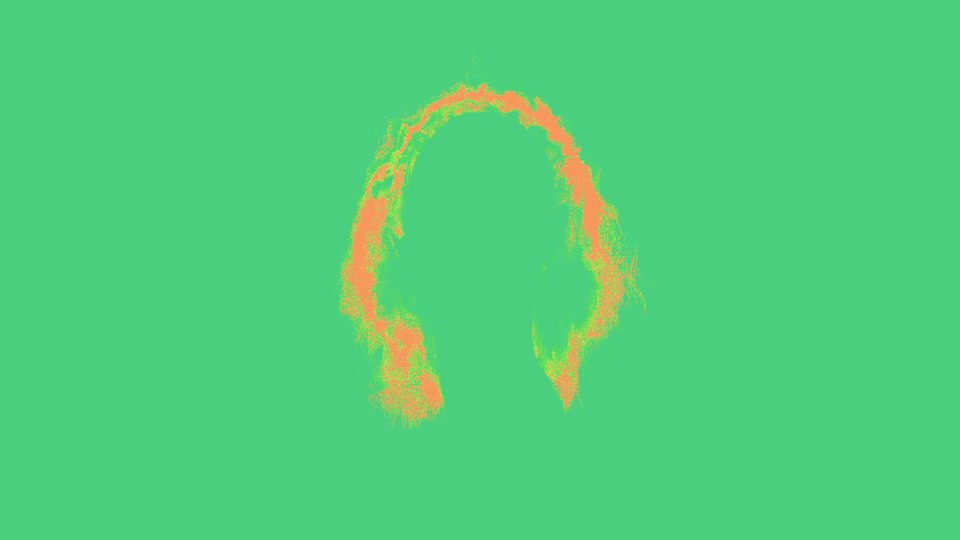}
  {20.20}%
\hspace{0.01\linewidth}%
\colorresult
  {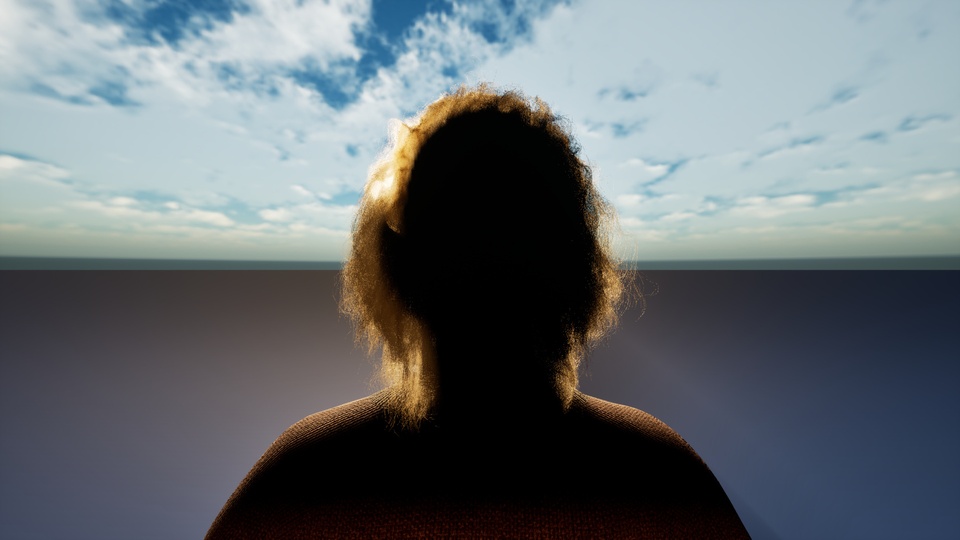}
  {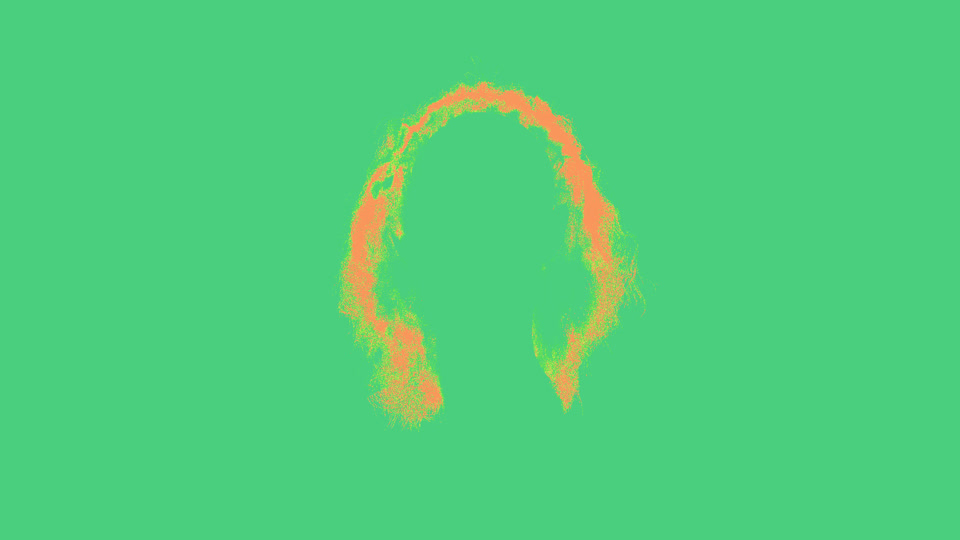}
  {20.14}%
\hspace{0.01\linewidth}%
\colorresult
  {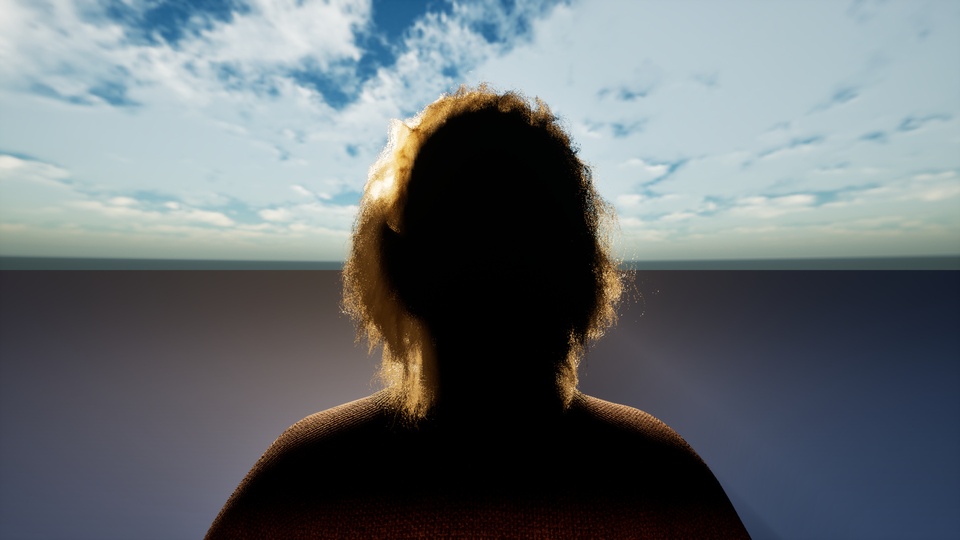}
  {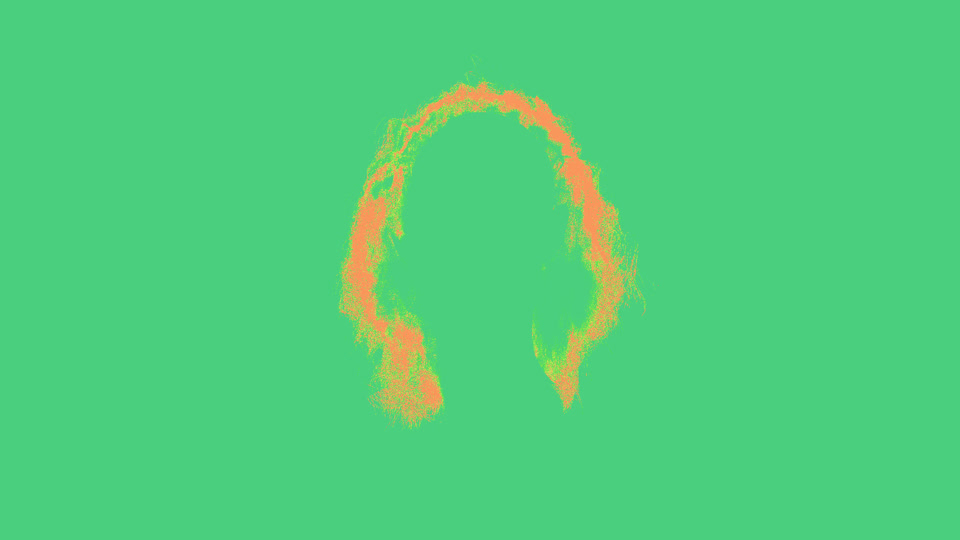}
  {19.76}%
\hspace{0.01\linewidth}%
\colorresult
  {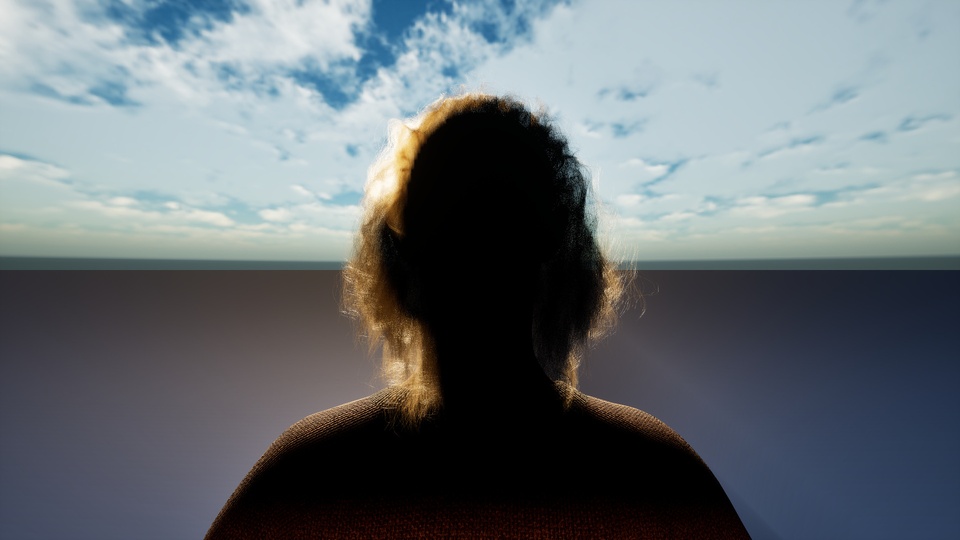}
  {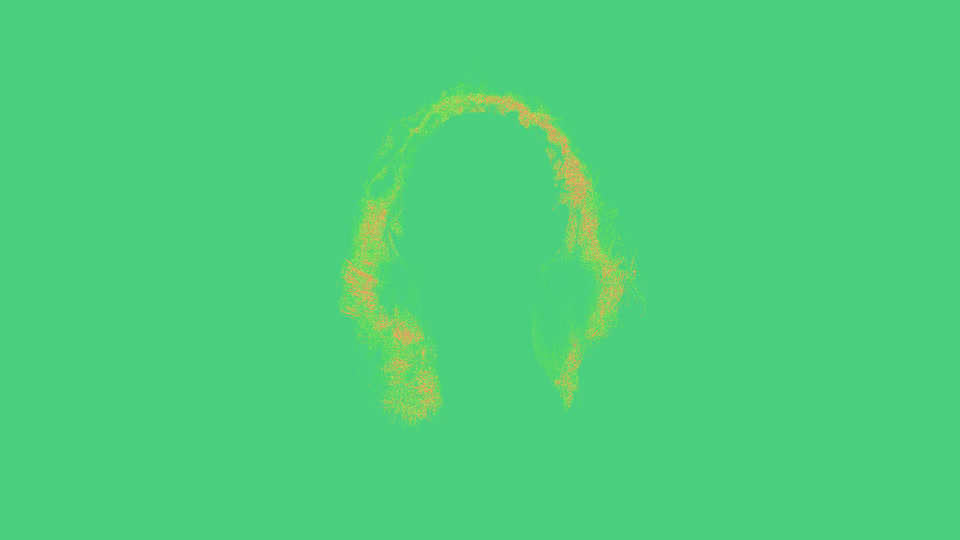}
  {27.27}%
\hspace{0.01\linewidth}%
\colorref{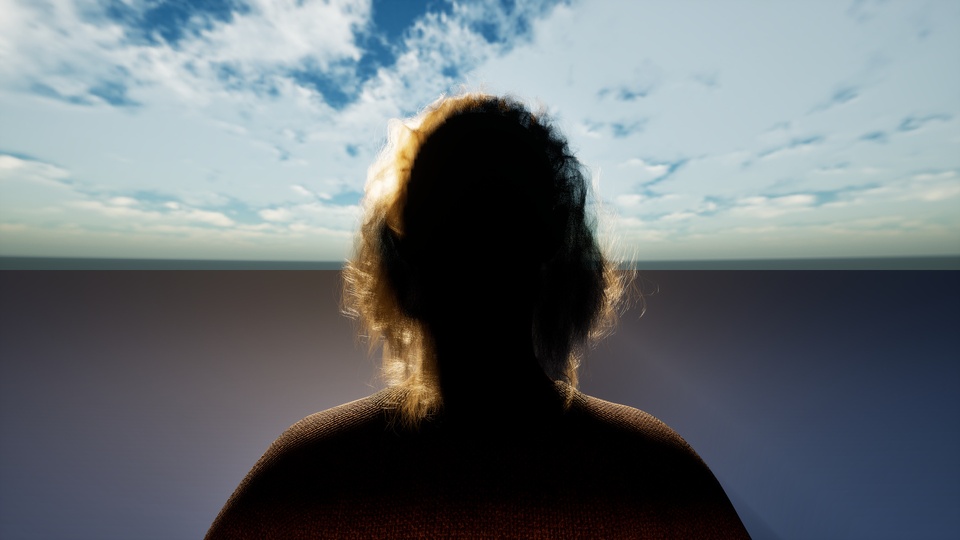}%

\caption{Different colors and lighting. The bottom-left corner shows the error map. Our method remains robust and yields the smallest error across different hair colors and lighting conditions, as it operates on G-buffers. The top-left number indicates the corresponding PSNR relative to the reference.}
\Description{}
\label{fig:color_light}
\end{figure*}
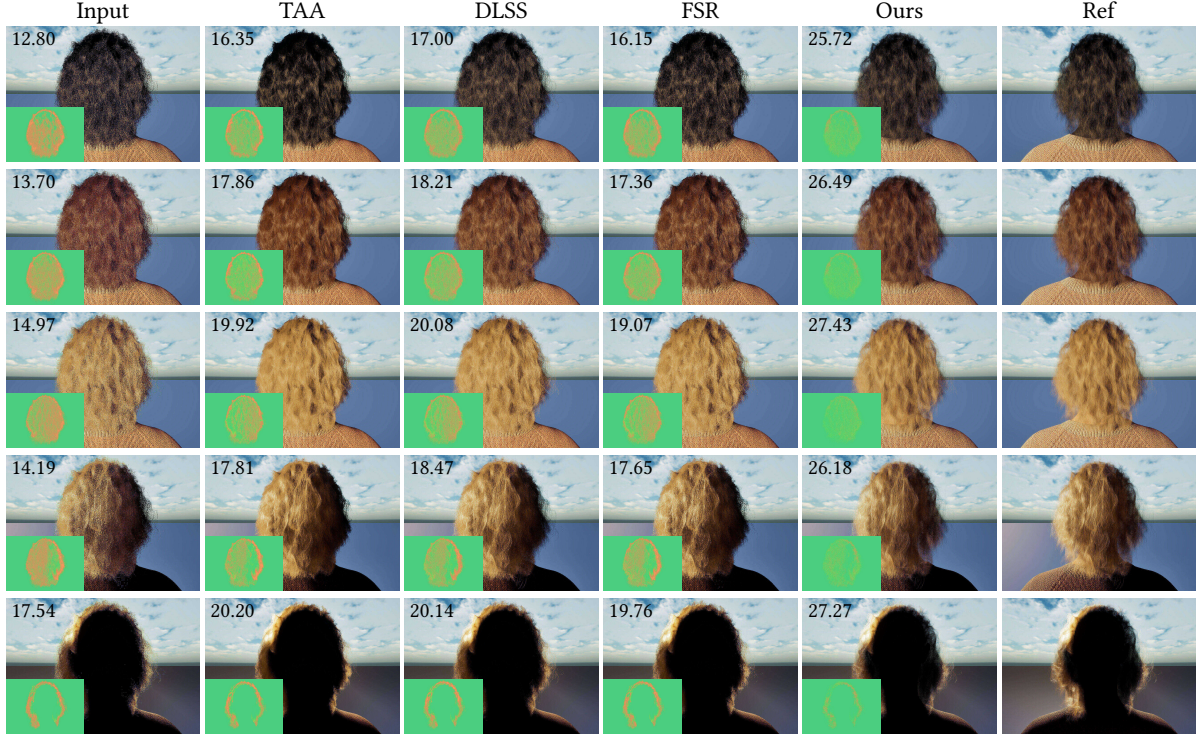

\paragraph{More Evaluations}
We further evaluate our method under different lighting configurations, including front, side, and back lighting, as well as across different hair colors. Because our method reconstructs the rasterized hair G-buffer, it naturally supports such variations, as shown in~\autoref{fig:color_light}. We also compare performance across a variety of hairstyles, including straight, wavy, curly, and afro hair, shown in~\autoref{fig:comparison_static_scene}, which exhibit different strand densities, curvatures, and silhouette complexities. The results show that our method can recover the spatial structure of hair, reduce aliasing, and produce images closer to the high-sample reference. More quantitative comparisons are provided in the supplemental document.

\begin{figure}[htb]
\centering

\newcommand{\cvhead}[1]{%
  \begin{minipage}[b]{0.182\linewidth}
    \centering
    \small #1
  \end{minipage}%
}

\newcommand{\cvrowlabel}[1]{%
  \begin{minipage}[b][0.12\linewidth][c]{0.04\linewidth}
    \centering
    \rotatebox[origin=l]{90}{\small #1}
  \end{minipage}%
}

\begin{minipage}[b]{0.04\linewidth}
\mbox{}
\end{minipage}%
\hspace{0.01\linewidth}%
\cvhead{Straight}%
\hspace{0.01\linewidth}%
\cvhead{Ponytail}%
\hspace{0.01\linewidth}%
\cvhead{Curly}%
\hspace{0.01\linewidth}%
\cvhead{Afro}%
\hspace{0.01\linewidth}%
\cvhead{All}%
\\[0.2em]

\cvrowlabel{Straight}%
\includegraphics[
  trim={150 25 150 25},
  clip,
  width=0.182\linewidth
]{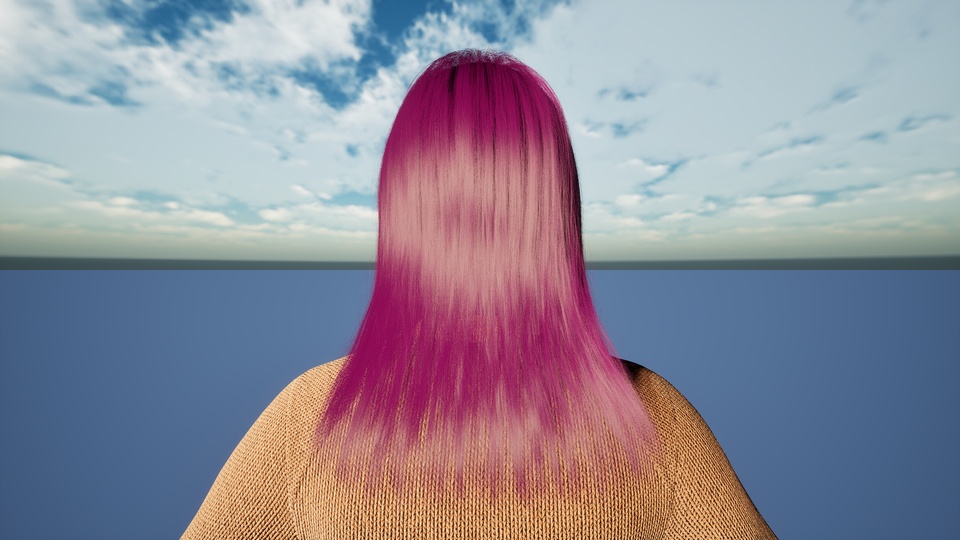}%
\hspace{0.01\linewidth}%
\includegraphics[
  trim={150 25 150 25},
  clip,
  width=0.182\linewidth
]{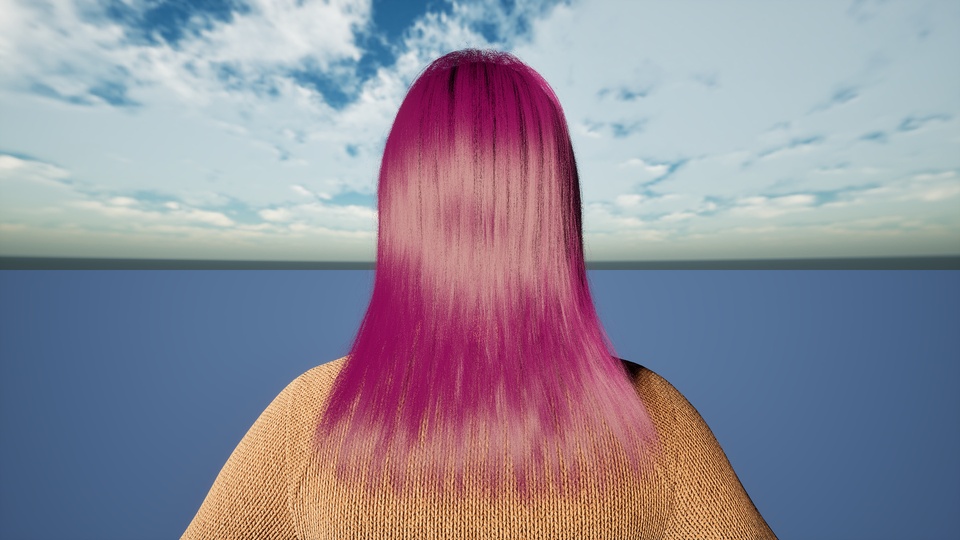}%
\hspace{0.01\linewidth}%
\includegraphics[
  trim={150 25 150 25},
  clip,
  width=0.182\linewidth
]{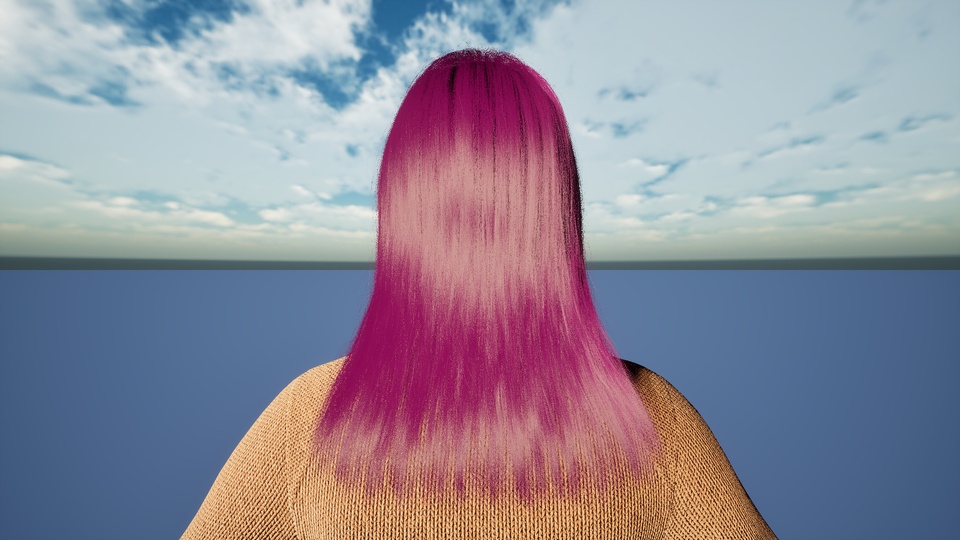}%
\hspace{0.01\linewidth}%
\includegraphics[
  trim={150 25 150 25},
  clip,
  width=0.182\linewidth
]{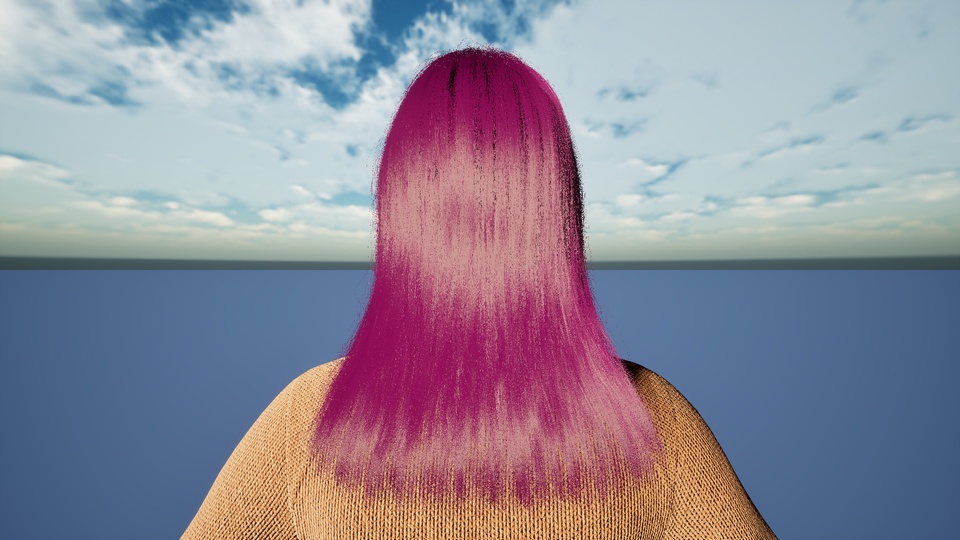}%
\hspace{0.01\linewidth}%
\includegraphics[
  trim={150 25 150 25},
  clip,
  width=0.182\linewidth
]{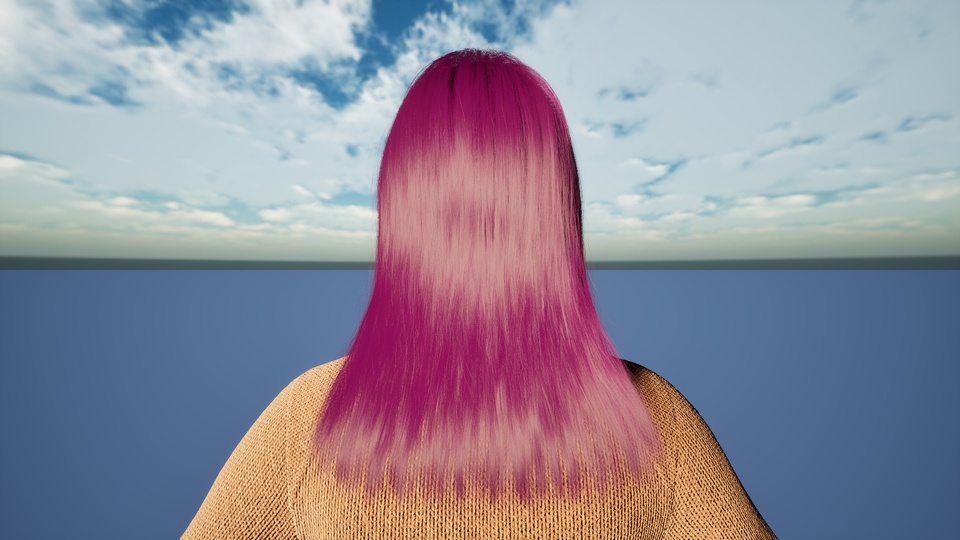}%
\\[3pt]

\cvrowlabel{Ponytail}%
\includegraphics[
  trim={150 25 150 25},
  clip,
  width=0.182\linewidth
]{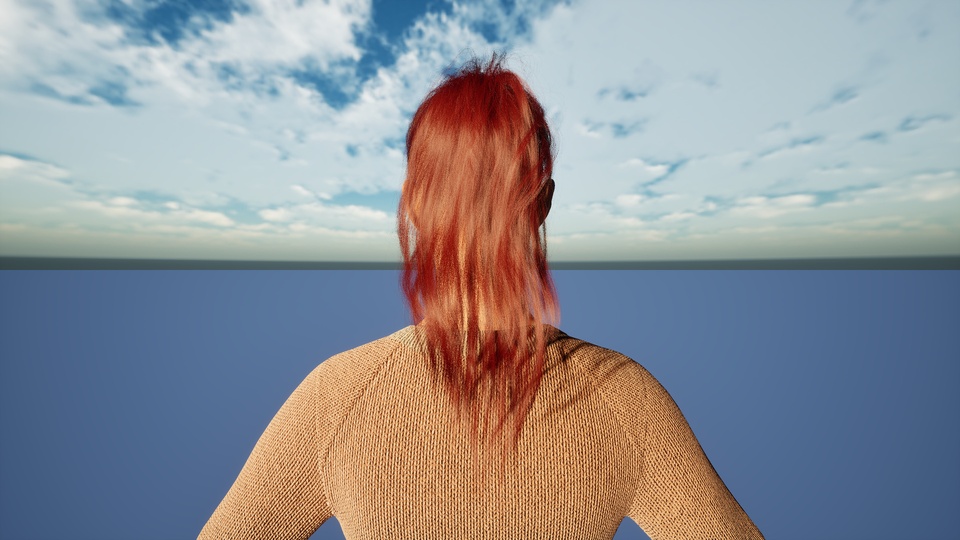}%
\hspace{0.01\linewidth}%
\includegraphics[
  trim={150 25 150 25},
  clip,
  width=0.182\linewidth
]{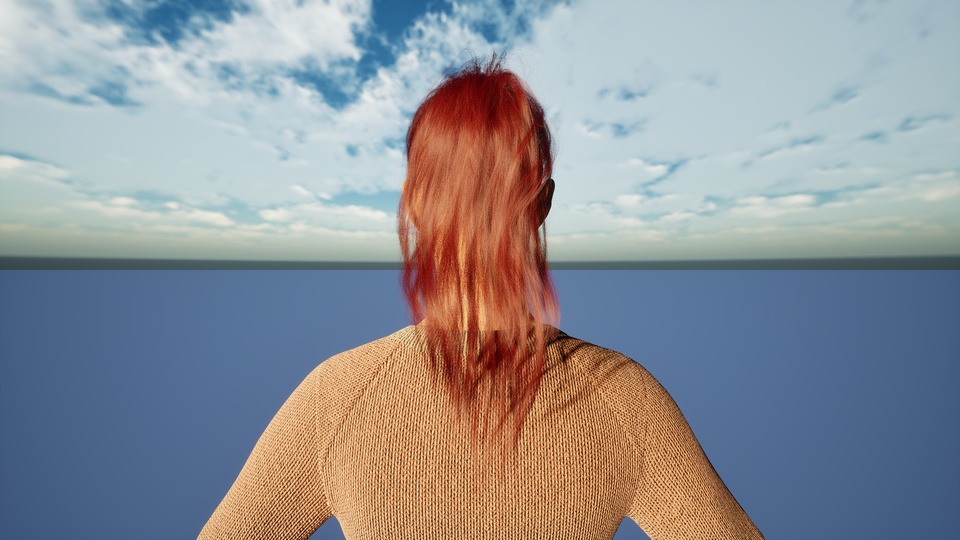}%
\hspace{0.01\linewidth}%
\includegraphics[
  trim={150 25 150 25},
  clip,
  width=0.182\linewidth
]{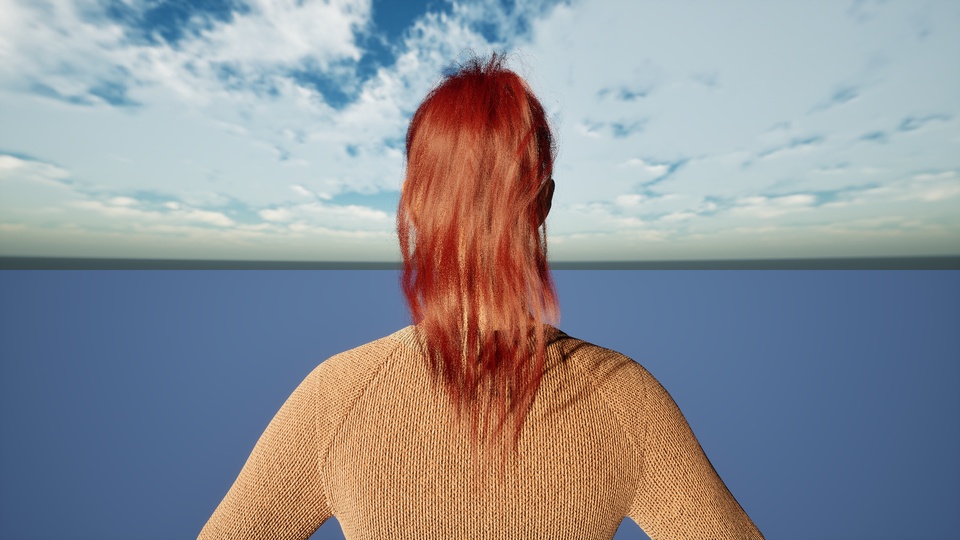}%
\hspace{0.01\linewidth}%
\includegraphics[
  trim={150 25 150 25},
  clip,
  width=0.182\linewidth
]{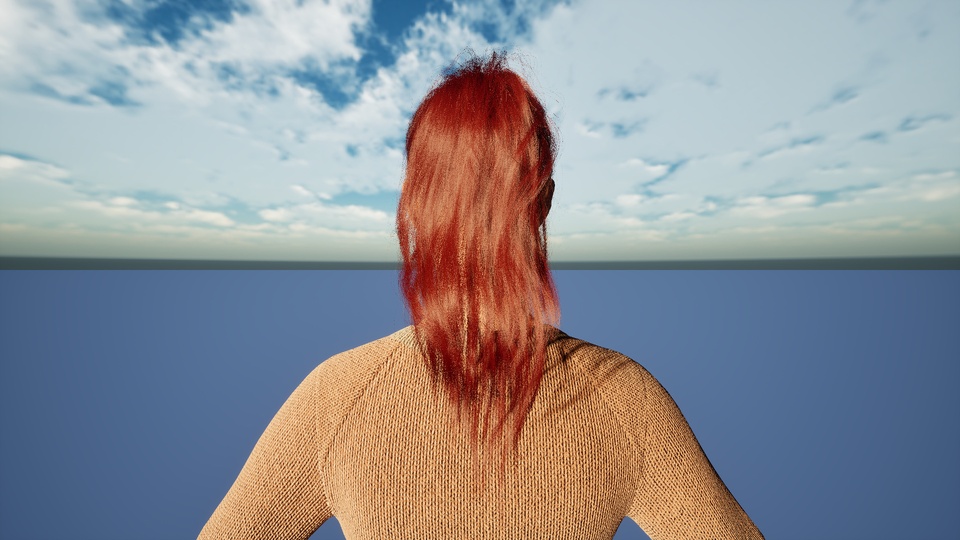}%
\hspace{0.01\linewidth}%
\includegraphics[
  trim={150 25 150 25},
  clip,
  width=0.182\linewidth
]{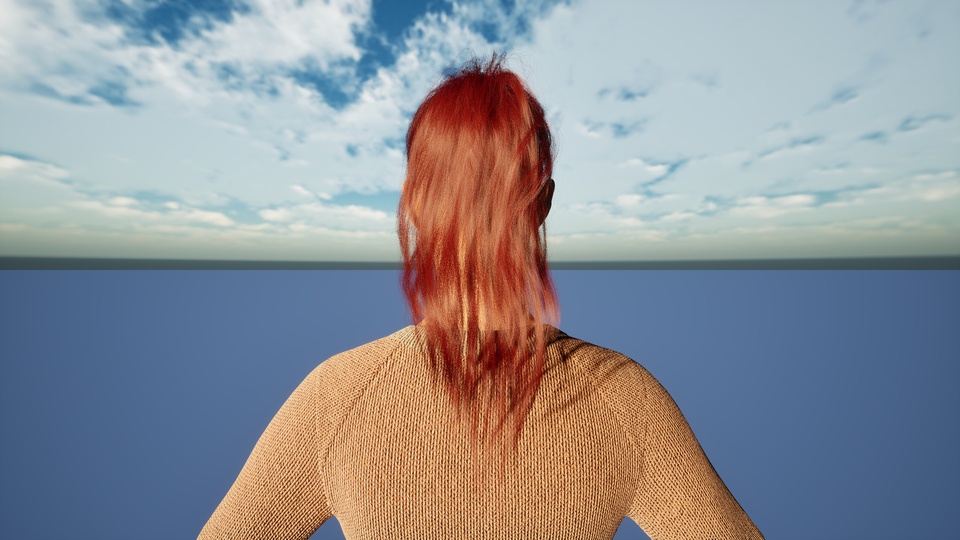}%
\\[3pt]

\cvrowlabel{Curly}%
\includegraphics[
  trim={150 25 150 25},
  clip,
  width=0.182\linewidth
]{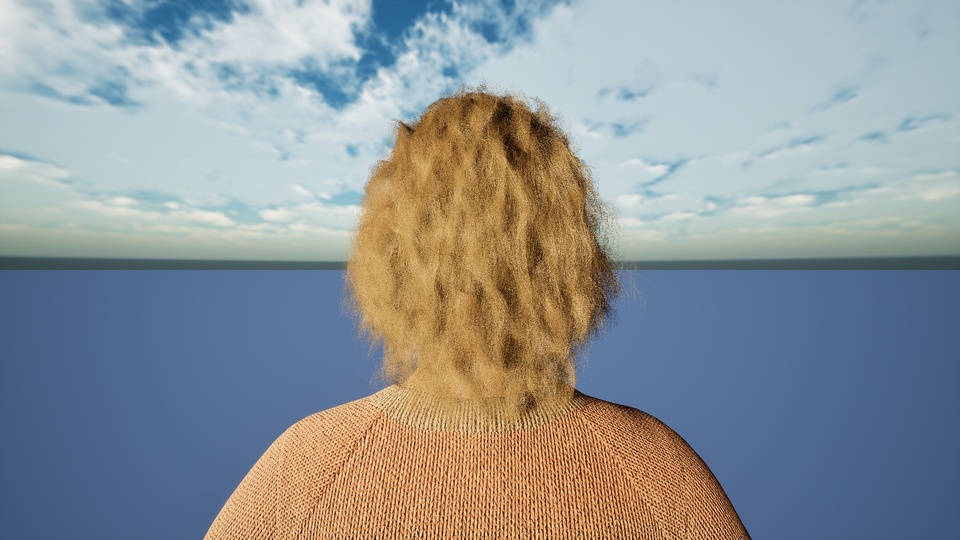}%
\hspace{0.01\linewidth}%
\includegraphics[
  trim={150 25 150 25},
  clip,
  width=0.182\linewidth
]{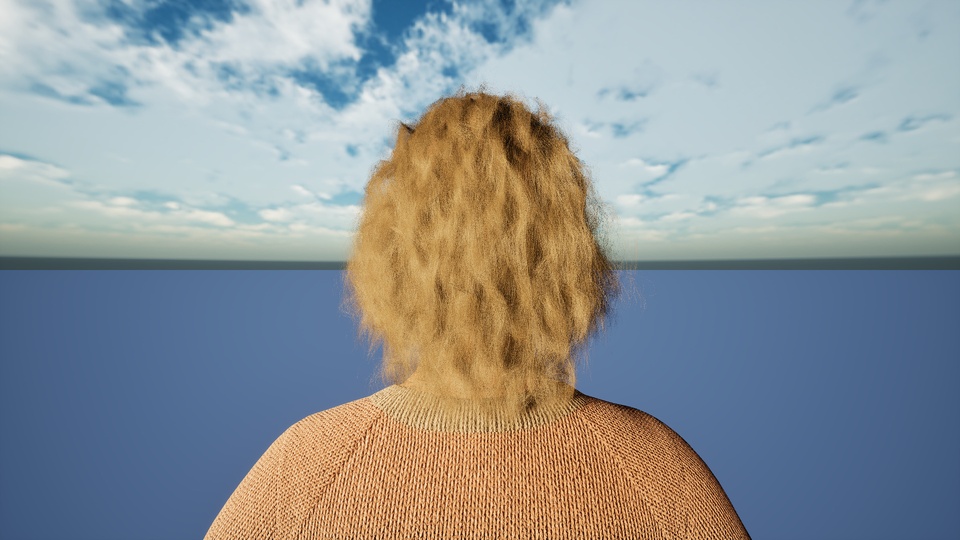}%
\hspace{0.01\linewidth}%
\includegraphics[
  trim={150 25 150 25},
  clip,
  width=0.182\linewidth
]{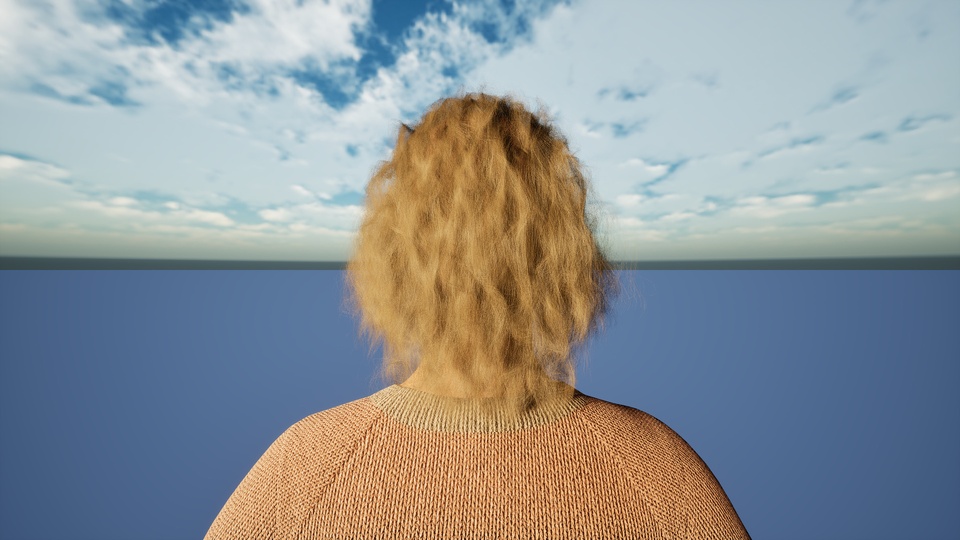}%
\hspace{0.01\linewidth}%
\includegraphics[
  trim={150 25 150 25},
  clip,
  width=0.182\linewidth
]{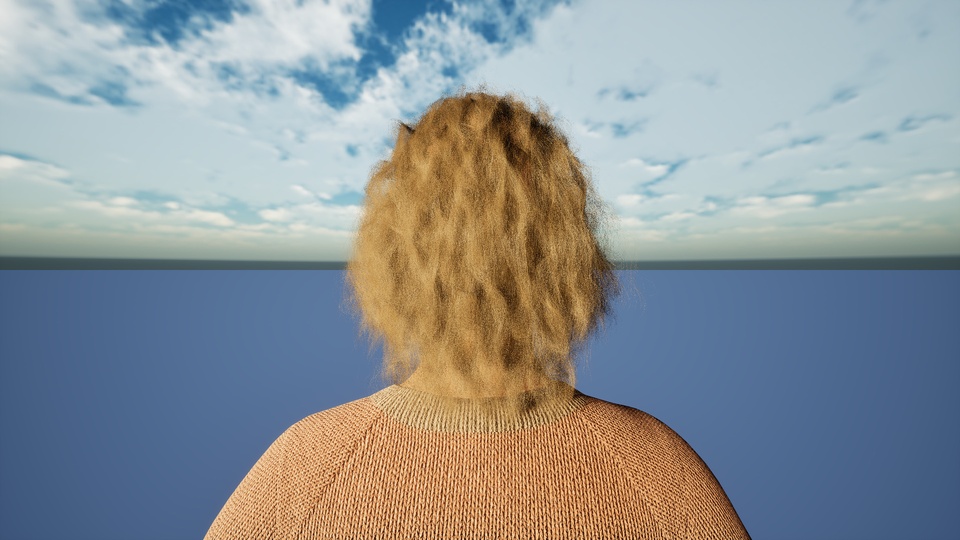}%
\hspace{0.01\linewidth}%
\includegraphics[
  trim={150 25 150 25},
  clip,
  width=0.182\linewidth
]{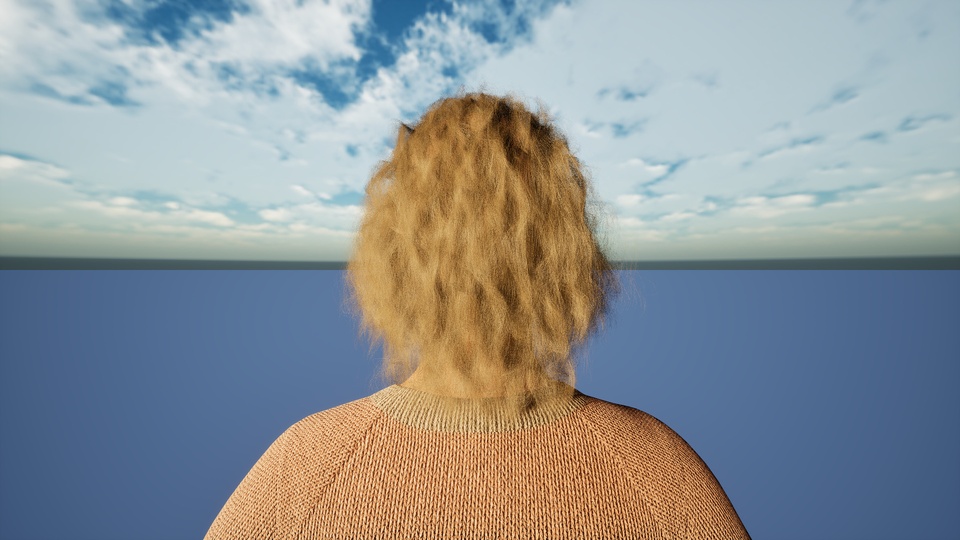}%
\\[3pt]

\cvrowlabel{Afro}%
\includegraphics[
  trim={150 25 150 25},
  clip,
  width=0.182\linewidth
]{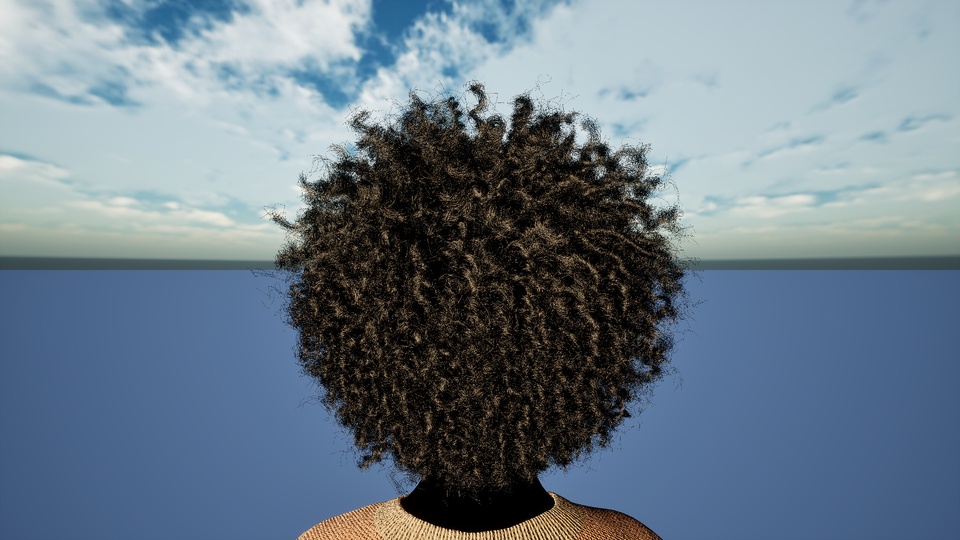}%
\hspace{0.01\linewidth}%
\includegraphics[
  trim={150 25 150 25},
  clip,
  width=0.182\linewidth
]{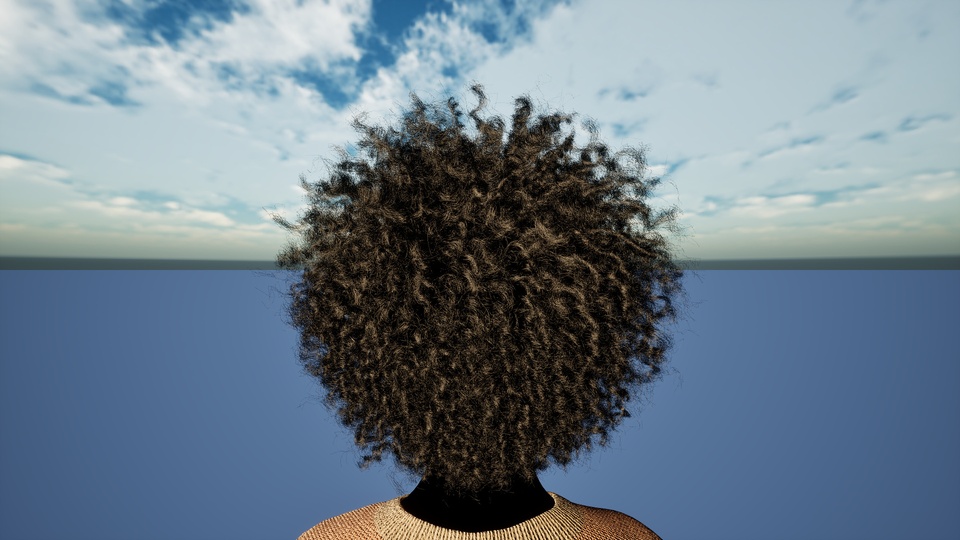}%
\hspace{0.01\linewidth}%
\includegraphics[
  trim={150 25 150 25},
  clip,
  width=0.182\linewidth
]{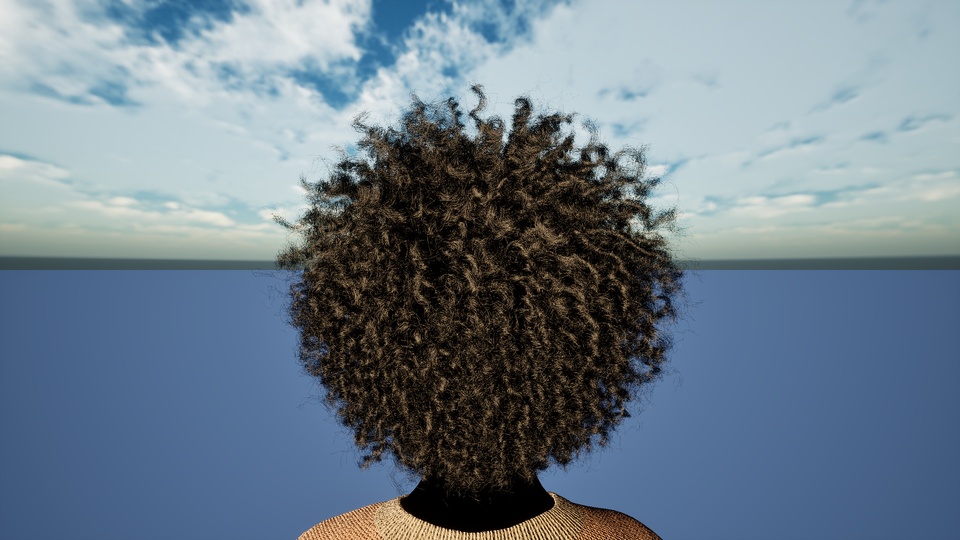}%
\hspace{0.01\linewidth}%
\includegraphics[
  trim={150 25 150 25},
  clip,
  width=0.182\linewidth
]{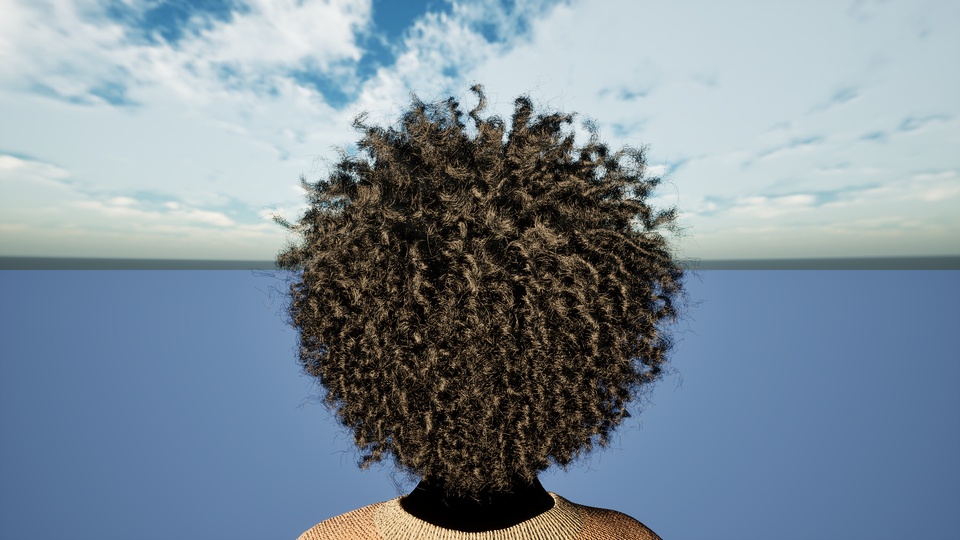}%
\hspace{0.01\linewidth}%
\includegraphics[
  trim={150 25 150 25},
  clip,
  width=0.182\linewidth
]{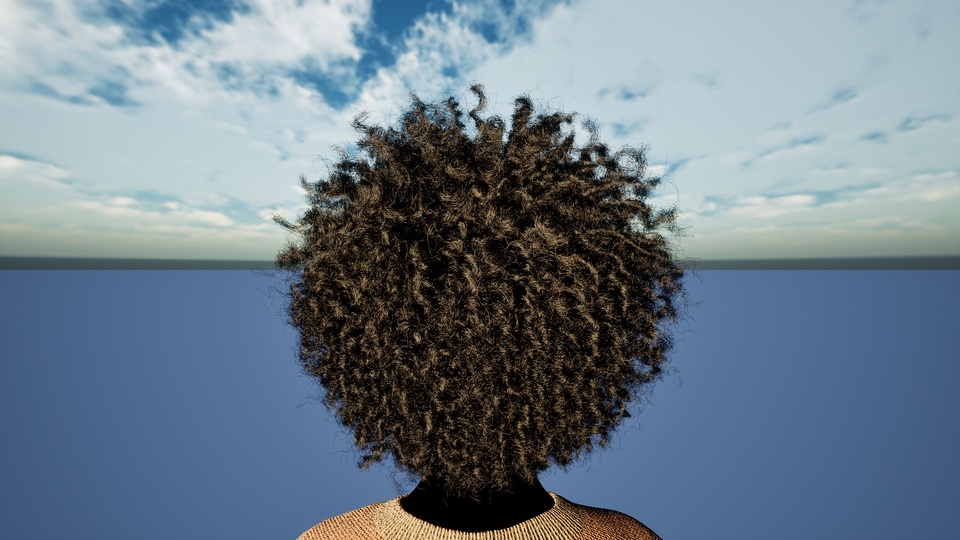}%

\caption{Cross validation. Row label represents the hairstyle used for training; column label represents the hairstyle used for testing. ``All'' refers to training on all four hairstyles.}
\Description{}
\label{fig:cross_validation}
\end{figure}

\paragraph{Cross Validation}
To evaluate the cross-hairstyle robustness of our method, we conducted a cross-validation study over four hairstyles using the same experimental setup as in~\autoref{fig:comparison_static_scene}. These hairstyles cover a broad range of geometric characteristics, from straight hair to highly curly (Afro) hair, as shown in~\autoref{fig:cross_validation}. In~\autoref{tab:cross_validation}, we report two metrics, PSNR and LPIPS. As expected, the diagonal entries correspond to the overfitting case and therefore achieve the best performance. The off-diagonal entries further indicate a certain degree of cross-hairstyle generalization, as each model is evaluated on hairstyles not seen during its training. More importantly, the rightmost column shows that a single model trained jointly on all four hairstyles can robustly handle diverse hairstyle geometries while still outperforming existing methods.

\begin{table}[hbt]
\centering
\small
\caption{Cross-validation across different hairstyles. Row labels indicate the hairstyle used for training, while column labels indicate the hairstyle used for testing. “All” refers to training on all four hairstyles. Results are reported in PSNR (top) and LPIPS (bottom).}
\label{tab:cross_validation}
\renewcommand{\tabcolsep}{0.1cm}


\begin{tabular}{cccccc}
\toprule
Test $\backslash$ Train & Straight  & Ponytail  & Curly     & Afro      & All   \\
Straight      & \textbf{27.21} & 24.62     & 22.20     & 16.10     & 25.31 \\
Ponytail      & 24.34     & \textbf{26.33} & 22.60     & 17.74     & 24.31 \\
Curly         & 20.70     & 21.80     & \textbf{25.78} & 20.62     & 24.91 \\
Afro          & 14.61     & 17.07     & 18.35     & \textbf{23.33} & 23.40 \\
\midrule
Test $\backslash$ Train & Straight        & Ponytail        & Curly           & Afro            & All    \\
Straight       & \textbf{0.0258} & 0.0483          & 0.0692          & 0.1103          & 0.0352 \\
Ponytail       & 0.0761          & \textbf{0.0158} & 0.0883          & 0.0857          & 0.0231 \\
Curly          & 0.1135          & 0.0967          & \textbf{0.0188} & 0.0821          & 0.0511 \\
Afro           & 0.2251          & 0.1942          & 0.1946          & \textbf{0.0324} & 0.0745 \\
\bottomrule

\end{tabular}
\end{table}

\begin{figure*}[htb]
\centering

\newcommand{\ablationhead}[1]{%
  \begin{minipage}{0.143\linewidth}
    \centering
    \small #1
  \end{minipage}%
}

%
%
\newcommand{\ablationrowlabel}[1]{%
  \begin{minipage}[b][0.106\linewidth][c]{0.03\linewidth}
    \centering
    \rotatebox[origin=c]{90}{\small #1}
  \end{minipage}%
}

%
\newcommand{\ablationresult}[2]{%
  \begingroup
  \setlength{\unitlength}{0.143\linewidth}%
  \begin{picture}(1,0.742)

    \put(0,0){%
      \includegraphics[
        trim={150 25 150 25},
        clip,
        width=\unitlength
      ]{#1}%
    }%

    \put(0,0){%
      \includegraphics[
        trim={150 25 150 25},
        clip,
        width=0.40\unitlength
      ]{#2}%
    }%

  \end{picture}%
  \endgroup
}

\newcommand{\ablationref}[1]{%
  \includegraphics[
    trim={150 25 150 25},
    clip,
    width=0.143\linewidth
  ]{#1}%
}

\begin{minipage}{0.03\linewidth}
\mbox{}
\end{minipage}%
\hspace{0.008\linewidth}%
\ablationhead{Input}%
\hspace{0.008\linewidth}%
\ablationhead{w/o spatial}%
\hspace{0.008\linewidth}%
\ablationhead{w/o temporal}%
\hspace{0.008\linewidth}%
\ablationhead{w/o analytic}%
\hspace{0.008\linewidth}%
\ablationhead{FullGym (Ours)}%
\hspace{0.008\linewidth}%
\ablationhead{Ref}%
\\[1pt]

\ablationrowlabel{Frame 0}%
\hspace{0.008\linewidth}%
\ablationresult
  {Fig_dynamic/spp1/Ponytail_Dynamic1.0000.jpg}
  {Fig_dynamic/spp1/error_ponytail_dynamic_spp1_000000.jpg}%
\hspace{0.008\linewidth}%
\ablationresult
  {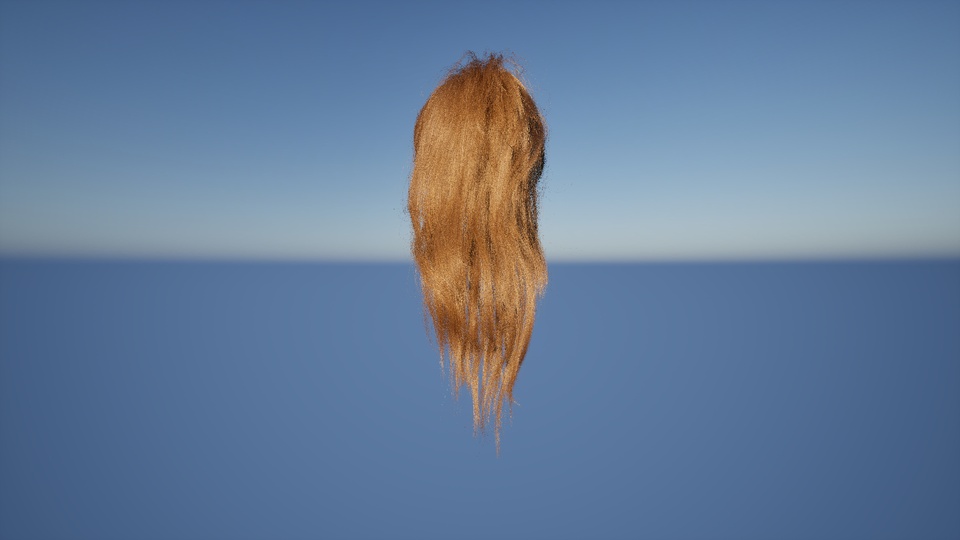}
  {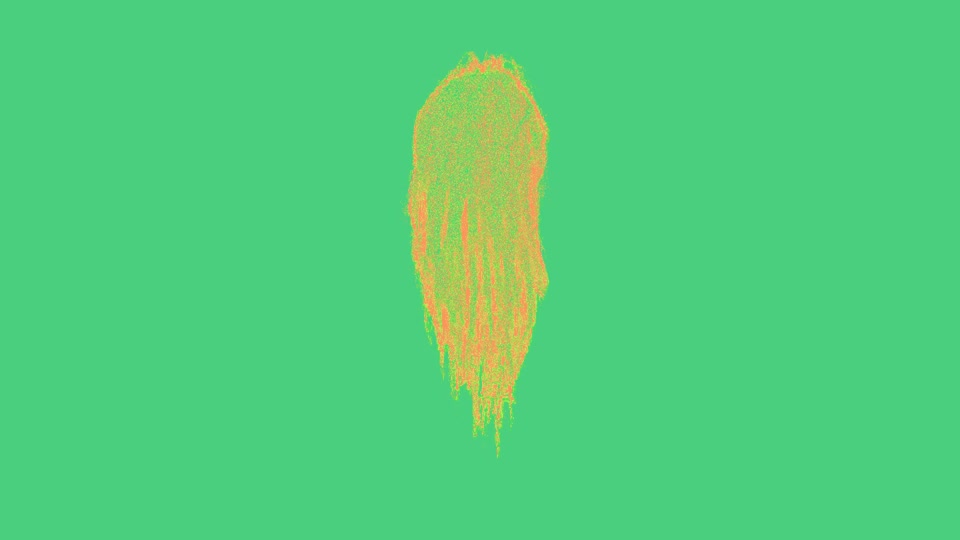}%
\hspace{0.008\linewidth}%
\ablationresult
  {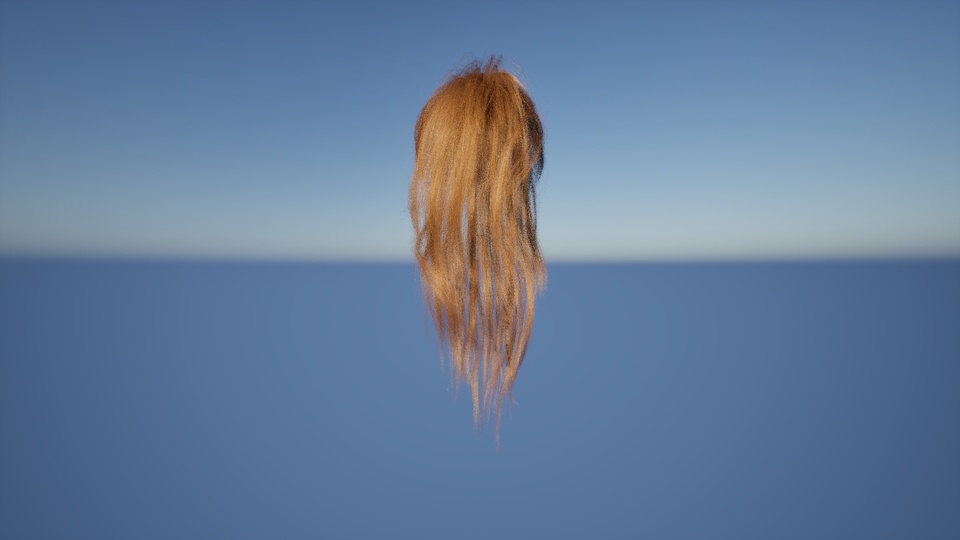}
  {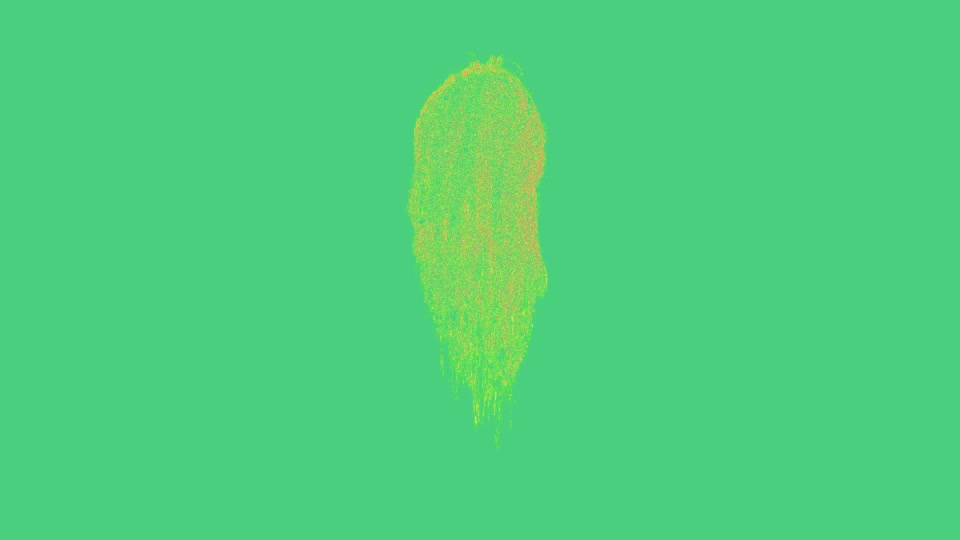}%
\hspace{0.008\linewidth}%
\ablationresult
  {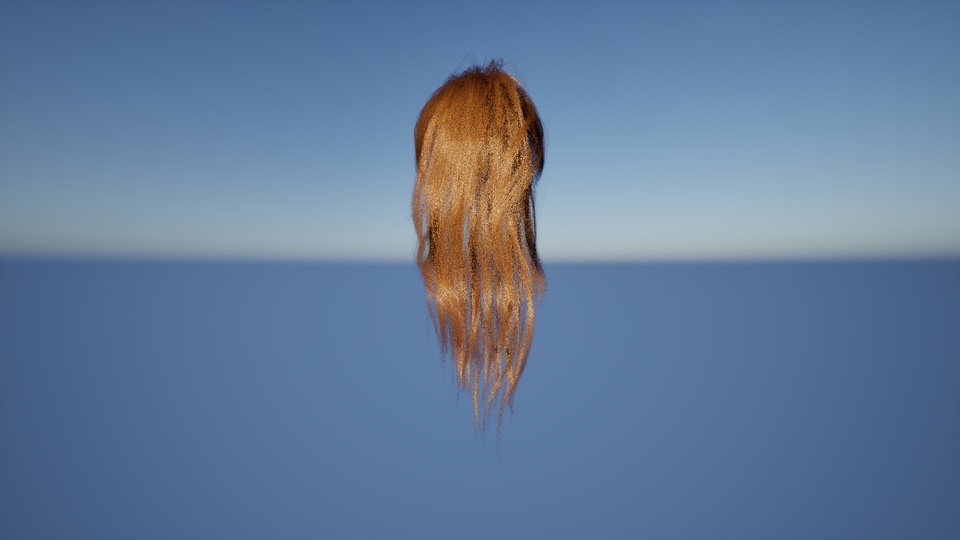}
  {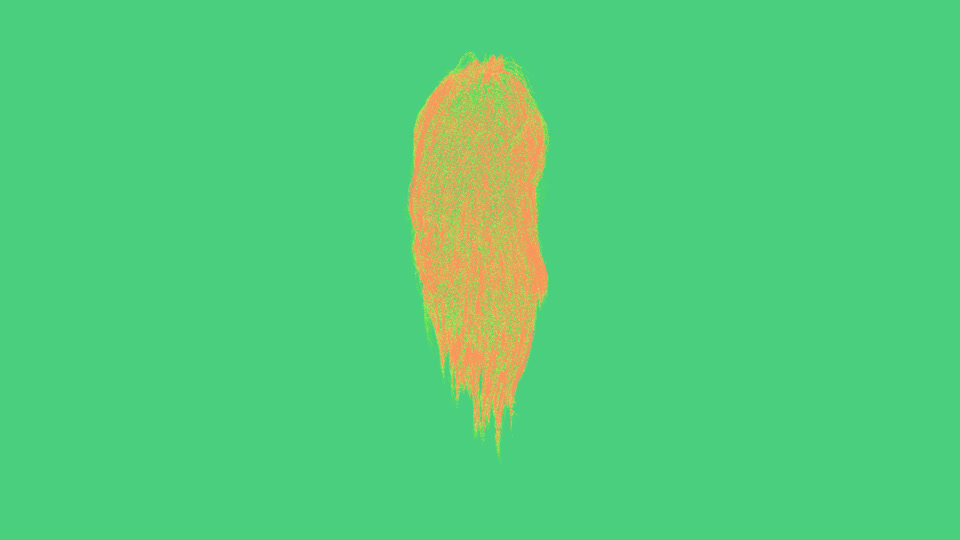}%
\hspace{0.008\linewidth}%
\ablationresult
  {Fig_dynamic/pred/Ponytail_Dynamic1.0000.jpg}
  {Fig_dynamic/pred/error_ponytail_dynamic_pred_000000.jpg}%
\hspace{0.008\linewidth}%
\ablationref
  {Fig_dynamic/spp128/Ponytail_Dynamic1.0000.jpg}%
\\[2pt]

\ablationrowlabel{Frame 50}%
\hspace{0.008\linewidth}%
\ablationresult
  {Fig_dynamic/spp1/Ponytail_Dynamic1.0050.jpg}
  {Fig_dynamic/spp1/error_ponytail_dynamic_spp1_000050.jpg}%
\hspace{0.008\linewidth}%
\ablationresult
  {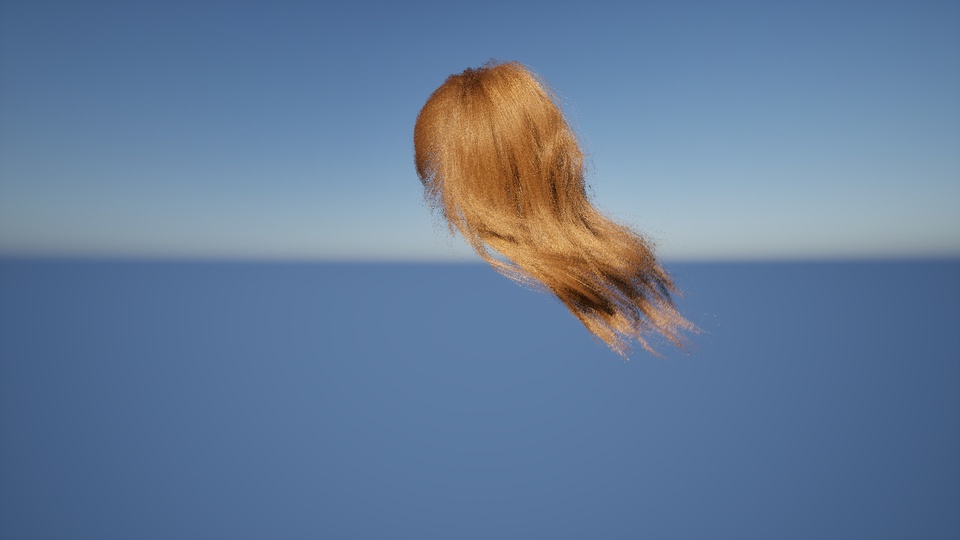}
  {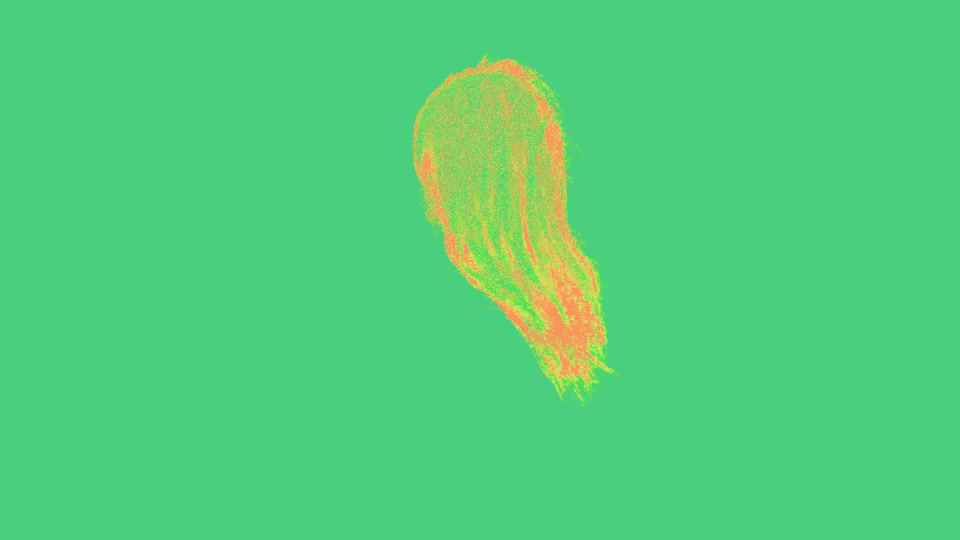}%
\hspace{0.008\linewidth}%
\ablationresult
  {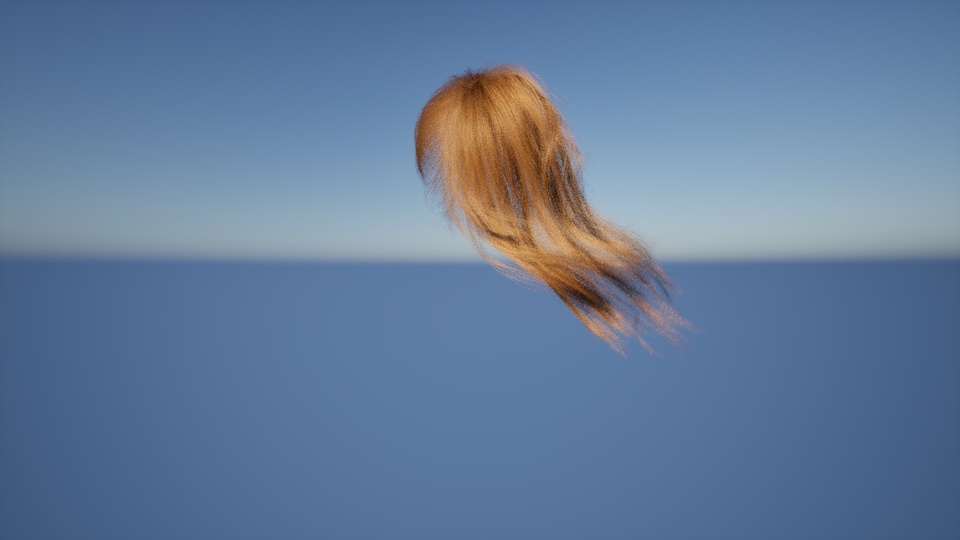}
  {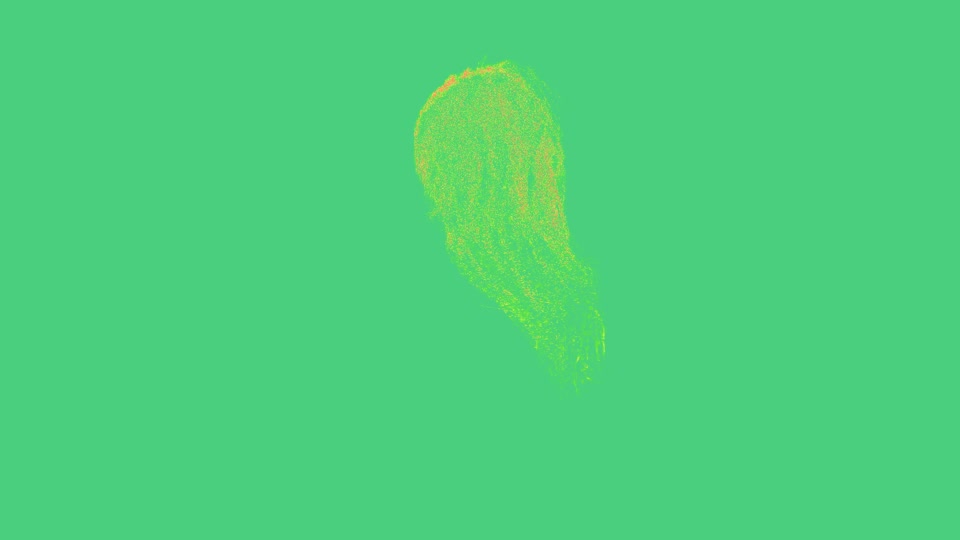}%
\hspace{0.008\linewidth}%
\ablationresult
  {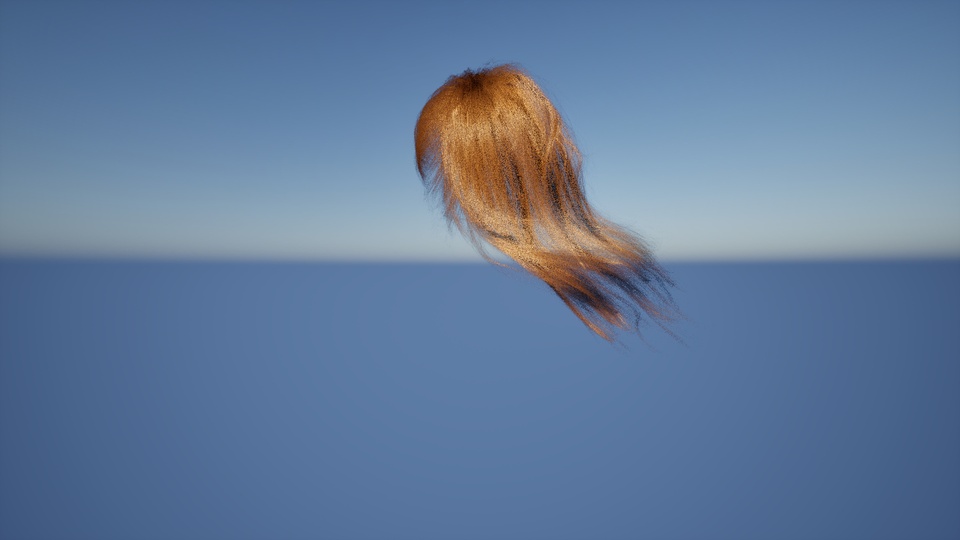}
  {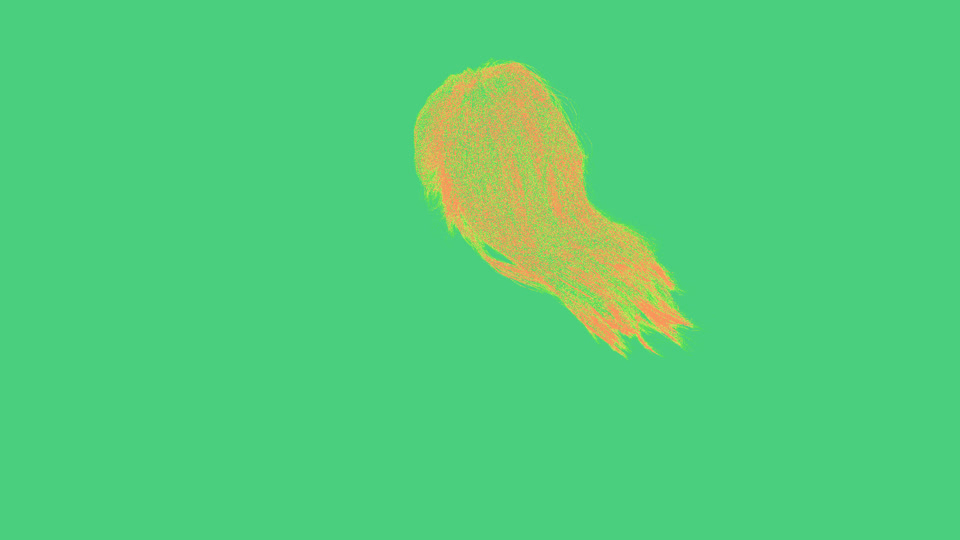}%
\hspace{0.008\linewidth}%
\ablationresult
  {Fig_dynamic/pred/Ponytail_Dynamic1.0050.jpg}
  {Fig_dynamic/pred/error_ponytail_dynamic_pred_000050.jpg}%
\hspace{0.008\linewidth}%
\ablationref
  {Fig_dynamic/spp128/Ponytail_Dynamic1.0050.jpg}%
\\[2pt]

\ablationrowlabel{Frame 150}%
\hspace{0.008\linewidth}%
\ablationresult
  {Fig_dynamic/spp1/Ponytail_Dynamic1.0150.jpg}
  {Fig_dynamic/spp1/error_ponytail_dynamic_spp1_000150.jpg}%
\hspace{0.008\linewidth}%
\ablationresult
  {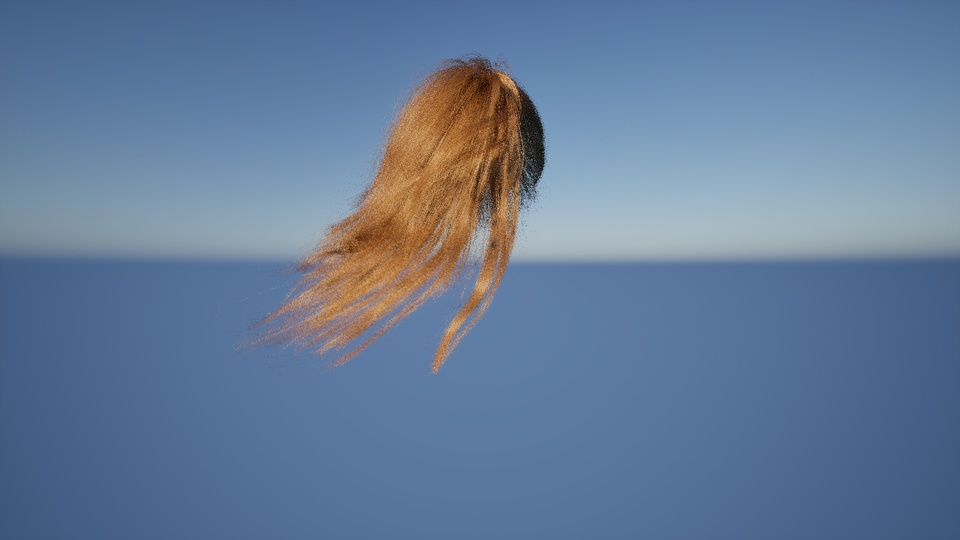}
  {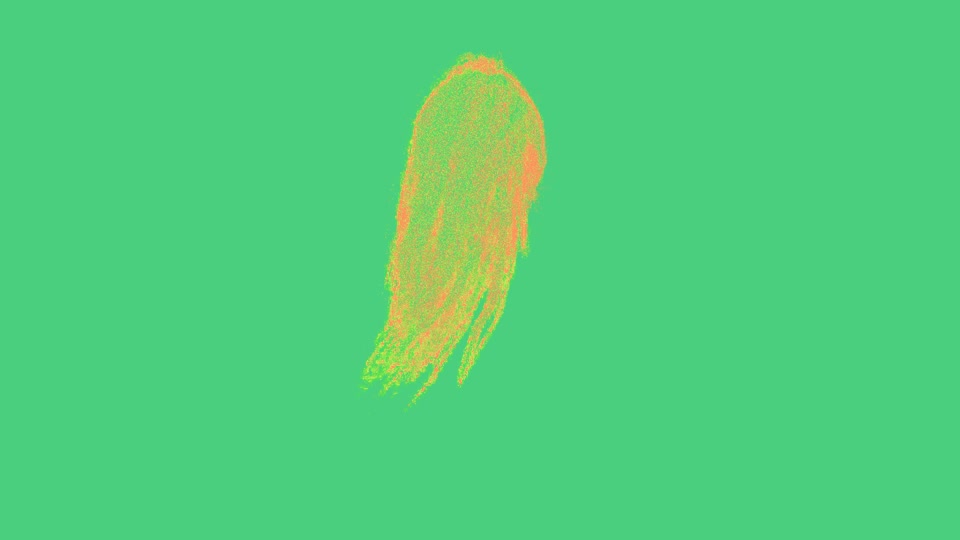}%
\hspace{0.008\linewidth}%
\ablationresult
  {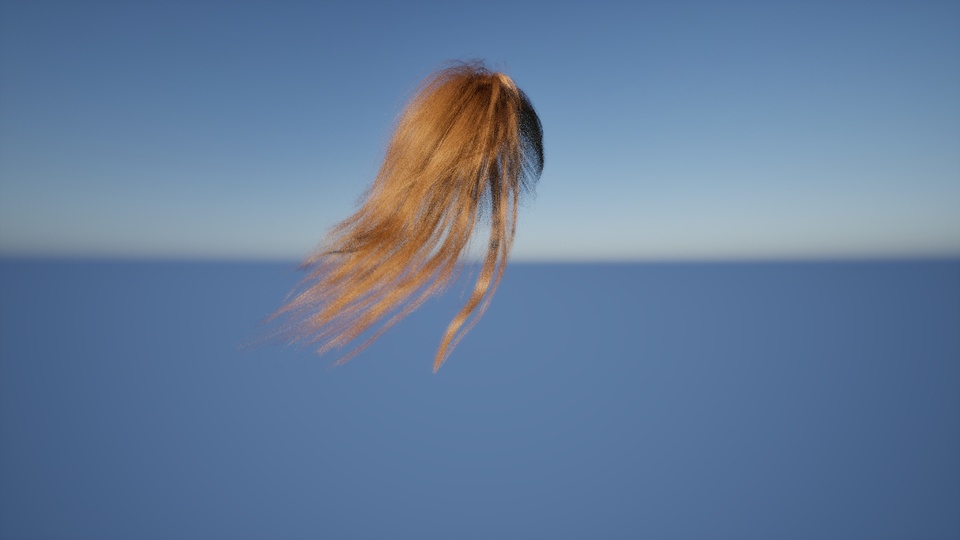}
  {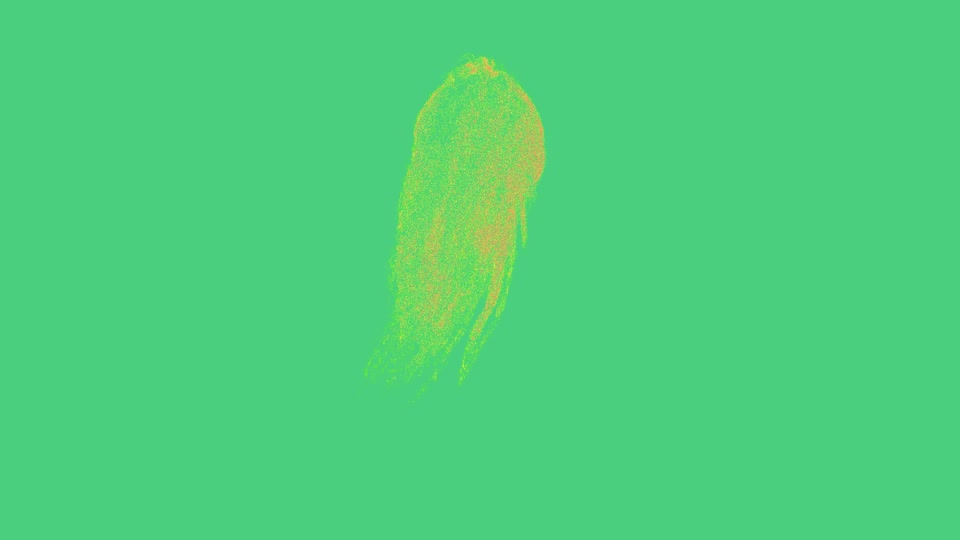}%
\hspace{0.008\linewidth}%
\ablationresult
  {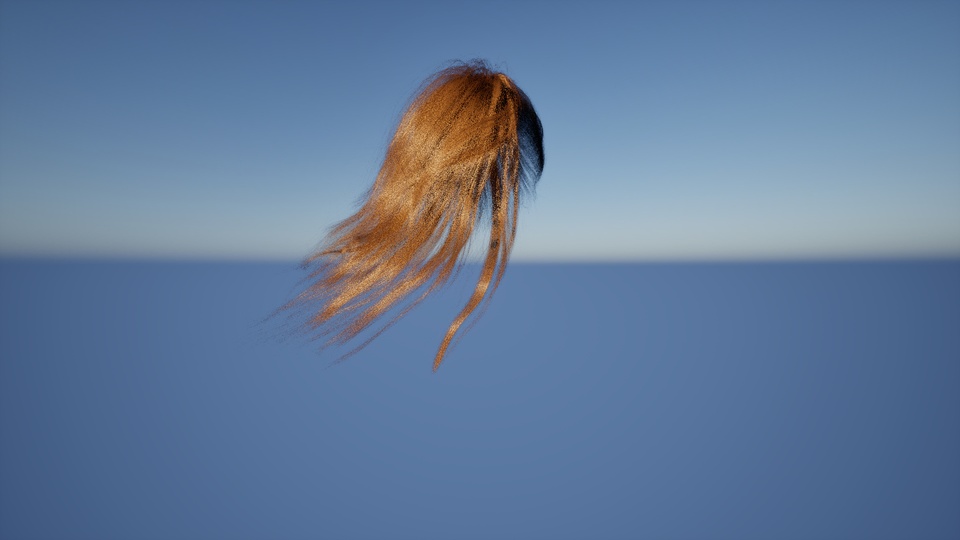}
  {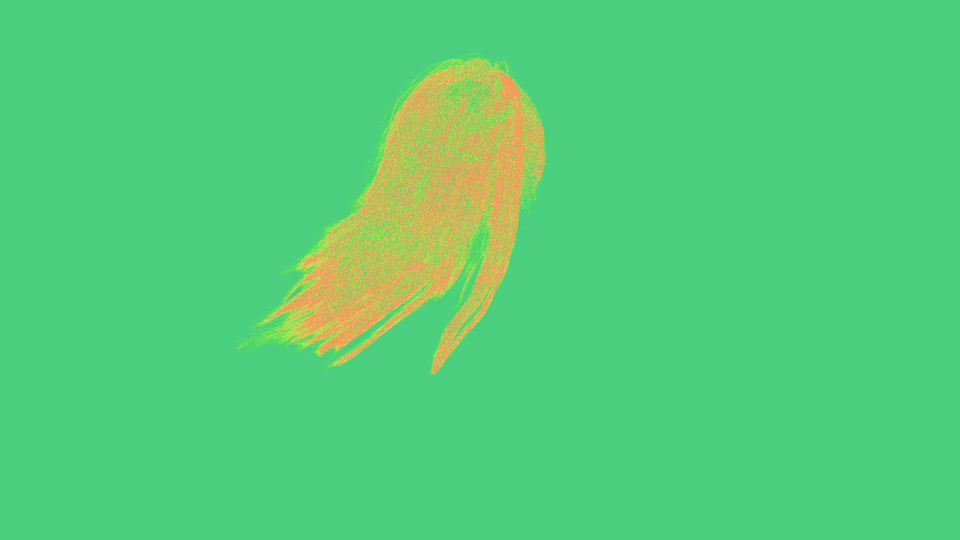}%
\hspace{0.008\linewidth}%
\ablationresult
  {Fig_dynamic/pred/Ponytail_Dynamic1.0150.jpg}
  {Fig_dynamic/pred/error_ponytail_dynamic_pred_000150.jpg}%
\hspace{0.008\linewidth}%
\ablationref
  {Fig_dynamic/spp128/Ponytail_Dynamic1.0150.jpg}%

\caption{Ablation studies on components of our method. The corresponding quantitative comparison is provided in the supplemental document.}
\Description{}
\label{fig:ablation}

\end{figure*}

\paragraph{Ablation Study on Proposed Components.}
To evaluate the contribution of each major component in our pipeline, we perform an ablation study of the spatial module, temporal module, and tangent-guided reconstruction. In summary, as shown in~\autoref{fig:ablation}, removing any component degrades the final rendering quality. Without the spatial module, the result preserves the overall hair structure but contains more residual noise and less accurate fine-strand details. Removing temporal accumulation reduces both reconstruction accuracy and temporal stability under dynamic hair motion. Interestingly, directly shading the reconstructed coverage and tangent produced by our spatiotemporal module while using noisy positions leads to even larger errors because misalignment among coverage, tangent, and position causes incorrect shadows and multi-bounce scattering. Our neural spatiotemporal modules and tangent-guided reconstruction are complementary. The neural modules improve the quality and temporal stability of coverage and tangent, while tangent-guided reconstruction aligns the position with the reconstructed G-buffers, which is essential for robust deferred shading. A quantitative comparison is provided in the supplemental document.

\paragraph{Performance}
We measure the performance of all experiments using NVIDIA Nsight. As an example, our spatial module, temporal accumulation, and analytic filtering take 0.6\,ms, 0.3\,ms, and 0.1\,ms, respectively, for a total of 1.0\,ms. The cost scales proportionally with the screen resolution.

\section{Conclusion}
We have presented a lightweight, real-time neural hair filtering pipeline for reconstructing severely undersampled strand-based hair G-buffers prior to deferred shading. Our method reconstructs the G-buffers by restoring coverage and tangent with neural spatiotemporal modules and completing position with a tangent-guided reconstruction module. This design preserves the geometric consistency required for physically based hair shading and avoids directly correcting errors that are already entangled with visibility, lighting, shadows, and scattering.

\paragraph{Limitations}

Our method has several limitations. 
First, the neural modules may generalize less effectively to hairstyles whose geometric characteristics differ substantially from the training data. For applications that require high-precision reconstruction of a specific hairstyle, fine-tuning the network on that hairstyle or on a closely matched dataset may be beneficial. Second, our tangent-guided position reconstruction relies on local directional consistency, which is the same assumption used by Dual Scattering, and may become unreliable near strand crossings, strong depth discontinuities, or highly complex local structures, where neighboring tangents can correspond to different strands or depth layers. This may cause incorrect position propagation and subsequent shading errors, although we did not observe such failures in the evaluated examples. In addition, our method introduces additional runtime compared with applying highly optimized frame-level methods such as DLSS or FSR alone. Since our method operates only on the hair G-buffer before shading, its output can in principle be composited with the remaining framebuffer and subsequently processed by DLSS or FSR, making the two approaches architecturally compatible. However, we have not systematically evaluated their interaction, as our current implementation prioritizes algorithmic clarity and research validation over low-level engine optimization. Investigating joint integration and optimization is an important direction for future work. For example, kernel fusion and platform-specific acceleration could further reduce runtime. For static camera and hair geometry, the filtered G-buffer can be cached and reused for lighting- or material-only changes.






\bibliographystyle{ACM-Reference-Format}
\bibliography{bibliography}


\end{document}